\documentclass{article}
\usepackage{microtype}
\usepackage{graphicx}
\usepackage{subcaption}
\usepackage{booktabs} 
\usepackage[breaklinks,colorlinks]{hyperref}
\usepackage{url,xurl}

\usepackage[preprint]{icml2026}
\usepackage{array}

\usepackage{amsmath}
\usepackage{amssymb}
\usepackage{mathtools}
\usepackage{amsthm}
\usepackage{tikz}
\usepackage{xcolor}
\usetikzlibrary{arrows.meta,positioning,fit,calc,backgrounds,shapes.multipart,matrix}
\usepackage{pifont}
\usepackage{listings}

\usepackage{multirow}

\usepackage[capitalize,noabbrev]{cleveref}

\theoremstyle{plain}
\newtheorem{theorem}{Theorem}[section]

\theoremstyle{definition}
\newtheorem{definition}[theorem]{Definition}

\theoremstyle{remark}

\newcommand{\equalSupervision}{\textsuperscript{*}Equally contributed insights, supervision, and responsibility for the successful completion of this work.}
\newcommand{\equalAuth}{\textsuperscript{$\dagger$}These authors contributed equally to this work.}
\usepackage[textsize=tiny]{todonotes}

\icmltitlerunning{On the Evidentiary Limits of Membership Inference for Copyright Auditing}

\begin{document}

\twocolumn[
  \icmltitle{On the Evidentiary Limits of Membership Inference for Copyright Auditing}



  \icmlsetsymbol{equal}{*}
  \icmlsetsymbol{eqauth}{$\dagger$}
\begin{icmlauthorlist}
  \icmlauthor{Murat Bilgehan Ertan}{cwi,vu,eqauth}
  \icmlauthor{Emirhan Böge}{ind,eqauth}
  \icmlauthor{Min Chen}{vu}
  \icmlauthor{Kaleel Mahmood}{uri,equal}
  \icmlauthor{Marten van Dijk}{cwi,vu,equal}
\end{icmlauthorlist}

\icmlaffiliation{cwi}{Centrum Wiskunde \& Informatica (CWI), Amsterdam, The Netherlands}
\icmlaffiliation{vu}{Vrije Universiteit Amsterdam, Amsterdam, The Netherlands}
\icmlaffiliation{uri}{University of Rhode Island, Kingston, RI, United States}
\icmlaffiliation{ind}{Independent Researcher}

\icmlcorrespondingauthor{Murat Bilgehan Ertan}{bilgehan.ertan@cwi.nl}

  \icmlkeywords{Machine Learning, ICML, MIA, copyright, membership inference attacks, copyright auiditing, privacy, security}

  \vskip 0.3in
]




\printAffiliationsAndNotice{\equalAuth\equalSupervision}

\begin{abstract}
As large language models (LLMs) are trained on increasingly opaque corpora, membership inference attacks (MIAs) have been proposed to audit whether copyrighted texts were used during training, despite growing concerns about their reliability under realistic conditions. We ask whether MIAs can serve as admissible evidence in adversarial copyright disputes where an accused model developer may obfuscate training data while preserving semantic content, and formalize this setting through a judge-prosecutor-accused communication protocol. To test robustness under this protocol, we introduce SAGE (Structure-Aware SAE-Guided Extraction), a paraphrasing framework guided by Sparse Autoencoders (SAEs) that rewrites training data to alter lexical structure while preserving semantic content and downstream utility. Our experiments show that state-of-the-art MIAs degrade when models are fine-tuned on SAGE-generated paraphrases, indicating that their signals are not robust to semantics-preserving transformations. While some leakage remains in certain fine-tuning regimes, these results suggest that MIAs are brittle in adversarial settings and insufficient, on their own, as a standalone mechanism for copyright auditing of LLMs.
\end{abstract}

\section{Introduction}
\label{sec:intro}
\begin{figure*}[t]
\centering
\begin{tikzpicture}[
  scale=0.72, 
  transform shape,
  font=\normalsize,
  >=Latex,
  line cap=round, line join=round,
  x=1mm,y=1mm,
  prosfill/.style={fill=yellow!14},
  judgefill/.style={fill=purple!10},
  accfill/.style={fill=blue!9},
  techfill/.style={fill=black!6},
  box/.style={
    draw=black!30,
    line width=0.5pt,
    rounded corners=1.6mm,
    inner sep=3.2mm,
    align=left,
    fill=#1
  },
  pill/.style={
    draw=black!30,
    line width=0.5pt,
    rounded corners=3.6mm,
    inner sep=2.0mm,
    align=center,
    fill=#1
  },
  tag/.style={
    draw=none,
    fill=black!8,
    rounded corners=1.0mm,
    inner sep=1.4mm,
    font=\small
  },
  panel/.style={
    draw=black!18,
    rounded corners=3mm,
    inner sep=3.0mm,
    fill=#1
  },
  arrow/.style={->, line width=0.75pt, draw=black!55},
  arrow2/.style={->, line width=0.75pt, draw=black!35},
  dashedarrow/.style={->, dashed, line width=0.75pt, draw=black!55},
  emphasis/.style={draw=black!35, line width=0.9pt},
]

\coordinate (Cpro)   at (0,0);
\coordinate (Cjudge) at (84,0);
\coordinate (Cacc)   at (168,0);

\node[pill=yellow!18, minimum width=58mm,align=center] (pros)  at (Cpro)
{\textbf{Prosecutor}\\[-1mm]\small claim \& evidence};

\node[pill=purple!14, minimum width=58mm,align=center] (judge) at (Cjudge)
{\textbf{Judge}\\[-1mm]\small sets threshold $\tau$ \& decides};

\node[pill=blue!14, minimum width=58mm,align=center] (acc)   at (Cacc)
{\textbf{Accused}\\[-1mm]\small model owner};

\coordinate (RowMain) at ($(pros.south)+(0,-18mm)$);

\node[box=yellow!14, minimum width=58mm,align=center] (xstar)
at (Cpro |- RowMain)
{\textbf{Suspect text $x^\star$}\\[-1mm]\small allegedly from $D$};

\node[box=blue!9, minimum width=58mm,align=center] (model)
at (Cacc |- RowMain)
{\textbf{Trained LLM}\\[-1mm]\small released / audited};

\node[box=blue!9, minimum width=58mm, below=9mm of model,align=center] (Dpara)
{\textbf{Paraphrased dataset $\tilde{D}$}\\[-1mm]\small semantic-preserving variants (\textbf{SAGE})};

\draw[arrow2] (Dpara.west) -- ++(-10mm,0) |- (model.west)
  node[tag, pos=0.52, xshift=-2mm, yshift=-18mm] {train / fine-tune};

\node[
  box=black!6,
  minimum width=58mm,
  align=center
] (audit)
at (Cjudge |- RowMain)
{\textbf{Membership Inference}\\[-1mm]\small compute score $s(x^\star)$; compare to $\tau$};

\node[tag] (tau) at ($(audit.south)+(0,-6mm)$)
{\textbf{Threshold selection:} choose $\tau$};

\draw[arrow] (xstar.east) -- (audit.west)
  node[tag, pos=0.52] {inputs};

\draw[arrow, emphasis] (audit.east) -- (model.west)
  node[tag, fill=black!10, pos=0.56, yshift=0mm, xshift=-5mm, rotate=0]{audit};

\draw[dashedarrow] (pros.east) -- (judge.west)
  node[tag, pos=0.52, yshift=-5.6mm,align=center] {submits claim \& evidence};

\draw[dashedarrow] (acc.west) -- (judge.east)
  node[tag, pos=0.52, yshift=-5.6mm] {provides model access};

\draw[arrow2] (judge.south) -- ++(0,-10mm) -| (audit.north)
  node[tag, pos=0.25, yshift=5.2mm] {sets $\tau$ / decides};

\begin{scope}[on background layer]
  \node[panel=yellow!6,
        fit=(pros) (xstar),
        label={[tag, anchor=west]north west:\textbf{Prosecutor}}] {};
  \node[panel=purple!6,
        fit=(judge) (audit) (tau),
        label={[tag, anchor=west]north west:\textbf{Judge}}] {};
  \node[panel=blue!5,
        fit=(acc) (model) (Dpara),
        label={[tag, anchor=west]north west:\textbf{Accused / Defendant}}] {};
\end{scope}

\end{tikzpicture}
\vspace{-2mm}
\caption{\textbf{Judge--Prosecutor--Accused protocol and artifact flow.}
Party-owned artifacts (prosecutor evidence; accused model/data) are color-coded by role, while
auditing and obfuscation \emph{techniques} (membership inference; paraphrasing/SAGE) are shown in neutral gray.}
\label{fig:protocol-mia}
\end{figure*}
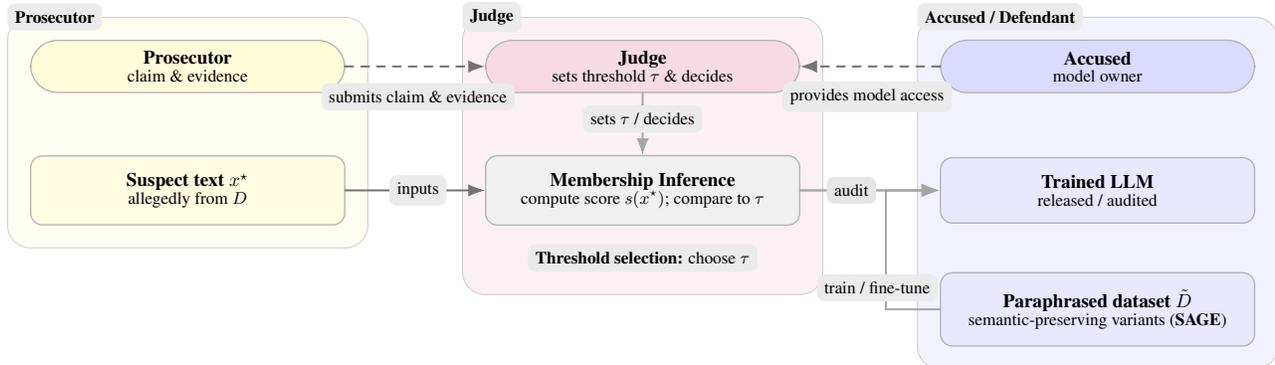

Large language models (LLMs) are increasingly trained and fine-tuned on massive text corpora whose precise composition is rarely disclosed~\citep{longpre2023data, touvronllama}, raising concern that training datasets may contain copyrighted material used without permission. These concerns are amplified by evidence that LLMs can memorize and reproduce training data~\citep{DBLP:conf/uss/CarliniTWJHLRBS21, DBLP:conf/emnlp/MireshghallahGU22, nasr2025extracting, ahmed2026extracting}, and by demonstrated vulnerabilities to extraction and inference attacks~\citep{DBLP:conf/uss/CarliniTWJHLRBS21,DBLP:conf/emnlp/ChuS0024,DBLP:conf/nips/Guo0J000S024}. In practice, fine-tuning is widely used in sensitive domains such as medical~\citep{Singhal2023clinical}, legal~\citep{guha2023legal}, financial~\citep{DBLP:journals/corr/abs-2303-17564}, and code-generation~\citep{DBLP:journals/tmlr/LiAZMKMMALCLZZW23} applications, where it often relies on private, regulated, or copyrighted data. As courts begin to confront these issues, a central question emerges: \emph{what technical evidence, if any, can reliably demonstrate that a particular copyrighted work was used to train a model?} 

A recent U.S.\ federal ruling highlights the stakes of this question. In \emph{Bartz v.\ Anthropic PBC} (2025), the court held that Anthropic’s use of books for model training did not violate copyright law\footnote[2]{See \emph{Bartz v.\ Anthropic PBC}, No.\ 24-05417 (N.D.\ Cal.\ 2025).}. Notably, the decision relied entirely on documentary evidence and company admissions, without considering any technical analysis of model behavior. While such evidence may suffice in cases involving voluntary disclosures, future disputes are likely to arise in settings where factual admissions are absent, adversarial incentives are strong~\citep{henderson2023fair}, and credible technical evidence would be required to substantiate claims of unauthorized data use.

Membership inference attacks (MIAs) \citep{DBLP:conf/sp/ShokriSSS17,DBLP:conf/uss/CarliniTWJHLRBS21} are a natural candidate for providing such technical evidence. Originally developed to detect whether a record belonged to the training set of a classifier, MIAs for LLMs typically rely on loss reduction, likelihood ratios, or calibration signals \citep{DBLP:conf/uss/MeeusJRM24,DBLP:conf/nips/MainiJPD24}. Prior work has shown that LLMs can be vulnerable to MIAs under certain fine-tuning regimes \citep{DBLP:conf/ccs/ChenTZYJWSZWT24,DBLP:conf/uss/00020GCSA0FK0L25}. Recent studies reveal substantial limitations: high reported AUCs often arise from temporal or distributional artifacts rather than genuine memorization~\citep{DBLP:journals/corr/abs-2402-07841,DBLP:conf/icml/MeeusSFM24,DBLP:conf/satml/MeeusSJFRM25, DBLP:conf/nips/MainiJPD24}, and MIAs against pre-training data are largely ineffective because each pre-training sample is typically seen only once \citep{DBLP:journals/corr/abs-1906-06669,kandpal2022deduplicating,DBLP:conf/nips/MuennighoffRBST23,DBLP:conf/sp/DasZT25}.

Any method introduced in court must remain robust to manipulation by an accused party, interpretable to non-experts, and stable under distribution shifts. Existing MIAs fail to guarantee these properties \citep{DBLP:conf/uss/00020GCSA0FK0L25,DBLP:conf/satml/MeeusSJFRM25,DBLP:conf/sp/DuZ0ZSC0025}. In particular, an accused can paraphrase or otherwise obfuscate text prior to training \citep{krishna2023evades, sadasivan2025reliable, DBLP:conf/uss/00020GCSA0FK0L25}. Because judges cannot observe the original training corpus, any useful technical evidence must withstand such adversarial transformations.

In this work, we examine whether MIAs can reliably serve as evidence that a specific copyrighted text was used during LLM training, both conceptually and empirically. We formalize this question through a communication protocol between three parties—\emph{judge}, \emph{prosecutor}, and \emph{accused}—that captures the evidentiary structure of copyright disputes and highlights a key requirement: any MIA-based claim must remain valid even when the accused can paraphrase or otherwise obfuscate the suspected text. Guided by this protocol, we systematically evaluate state-of-the-art MIAs on fine-tuned LLMs under paraphrasing-based transformations that preserve semantic meaning while substantially altering surface form, using a new paraphrasing pipeline we introduce, \emph{SAGE} (Structure-Aware SAE-Guided Extraction). By leveraging Sparse Autoencoders (SAEs)~\citep{bricken2023towards, huben2024sae}, SAGE operates independently of the target model’s internals.

SAGE\footnote{{\url{https://github.com/kiraz-ai/sage-sps-mia}}} enforces semantic preservation while suppressing surface-level token overlap by guiding paraphrasing with a semantic similarity signal derived from SAEs. This signal is summarized by the \emph{Semantic Persistence Score} ($\mathrm{SPS}$), which measures whether paraphrased text retains underlying semantic content despite substantial changes in phrasing. Using this framework, our analysis reveals a structural limitation of existing membership inference attacks: their signals are dominated by lexical and distributional artifacts rather than robust evidence that a model has internalized a specific copyrighted work. Because this vulnerability persists under semantics-preserving obfuscation, current MIAs struggle to meet the robustness requirements implied by our communication protocol, limiting their suitability as technical evidence in adversarial copyright auditing.

\textbf{Contributions.}
\textbf{(i)}, we formalize a \emph{judge--prosecutor--accused} \textbf{communication protocol} that makes explicit the evidentiary assumptions, adversarial incentives, and admissibility requirements in copyright auditing of LLM training data. \textbf{(ii)}, we introduce \textbf{SAGE} (Structure-Aware SAE-Guided Extraction), a semantic paraphrasing framework guided by {Sparse Autoencoders} and a \textbf{Semantic Persistence  Score} ($\mathrm{SPS}$), which preserves meaning and document structure while enabling ablations that isolate structural and factual sources of membership signal. \textbf{(iii)}, through systematic experiments, we show that state-of-the-art membership inference attacks degrade substantially under realistic, semantics-preserving obfuscation even when the model retains the underlying knowledge, and, grounded in our protocol and prior work, argue that under adversarial settings, in terms of robustness, \textbf{MIAs are insufficient, on their own, as evidentiary tools for copyright enforcement}.

\textbf{Paper outline.} Section~\ref{sec:communication} establishes the judge--prosecutor--accused communication protocol, formalizing the robustness and admissibility requirements for technical evidence in copyright disputes. Section~\ref{sec:methodology} introduces our framework, SAGE, and the $\mathrm{SPS}$ used to quantify semantic preservation. Section~\ref{sec:results} presents our experimental evaluation, demonstrating the sensitivity of MIAs to paraphrasing and validating the downstream utility of audited models. Section~\ref{sec:discussion} discusses the implications for copyright enforcement and the limitations of our study. Section~\ref{sec:rworks} reviews prior work on membership inference and auditing. Finally, Section~\ref{sec:conc} concludes with a summary.

\section{Copyright Auditing}
\label{sec:communication}
Under \citet{daubert1993daubert}, admissibility of technical evidence hinges on reliability.  In adversarial auditing settings, we focus on one necessary aspect of this requirement: \textbf{robustness} to semantics-preserving transformations available to an accused party. If a membership inference claim can be invalidated by such transformations, it cannot constitute reliable evidence regardless of its stability or explanatory framing.

We study the problem of \emph{copyright auditing} for LLMs: given a suspect text $x \in \mathcal{X}$ and black-box access to a deployed model $f_\theta$, can one determine whether $x$, or a semantically equivalent variant, was used during training? Because copyright concerns protected expression rather than exact duplication, any auditing procedure must reason about semantic content under meaning-preserving transformations. Here, black-box access refers to query-only access to the model, allowing the auditor to observe outputs such as generated text, token probabilities, or likelihoods, but not internal states.

We formalize copyright auditing as a \textbf{communication protocol} between three parties: a \emph{prosecutor}, an \emph{accused}, and a \emph{judge} (see Figure~\ref{fig:protocol-mia}). The \textbf{prosecutor} provides a suspect text $x$ and black-box access to the model $f_\theta$, and claims that $x$ appeared in the training data of $f_\theta$. The prosecutor’s objective is to supply technical evidence, typically via a membership inference attack, that the model’s behavior on $x$ is inconsistent with non-membership. The \textbf{accused} is the model owner or developer. Their objective is to minimize legal, financial, or regulatory exposure while protecting proprietary datasets and complying with privacy or contractual obligations. In pursuit of this objective, the accused may apply transformations to training data, including paraphrasing, summarization, or other semantic-preserving defenses. The \textbf{judge} must evaluate the prosecutor’s claim without access to the training data. This reflects practical and legal realities: training corpora are often protected by intellectual property law, privacy regulation, or contractual constraints, and may be unavailable even under litigation.

\textbf{Operational constraints.} 
This protocol operates under three realistic constraints. First, \emph{training datasets are opaque}: neither party can be assumed to disclose the full corpus, even in cooperative settings (See Appendix~\ref{app:dataset} for a discussion on dataset-level comparisons). Second, \emph{semantic-preserving transformations are allowed}: the accused may legally or strategically obfuscate any training text by applying transformations that preserve meaning, such as paraphrasing, summarization, or stylistic rewriting. Third, the judge must fix a decision rule to determine whether a specific input $x$ was used during training, typically by thresholding a membership inference attack score $s(x)$ at some value $\tau_{mia}$, such that $s(x) > \tau_{mia}$ implies membership. In practice, the appropriate choice of $\tau_{mia}$ depends on the model, dataset, and input distribution, as membership scores are sensitive to the intrinsic difficulty of the input text~\citep{DBLP:conf/iclr/WatsonGCS22,DBLP:conf/sp/CarliniCN0TT22} and to the model's generalization behavior. Consequently, thresholds calibrated on one distribution may fail to control false positive rates under distribution shift~\citep{DBLP:journals/corr/abs-2402-07841,DBLP:conf/icml/MeeusSFM24,DBLP:conf/satml/MeeusSJFRM25,DBLP:conf/acl/MatternMJSSB23}.

\begin{definition}[Semantic Equivalence]
Two texts $x, \tilde{x} \in \mathcal{X}$ are \emph{semantically equivalent}, denoted $x \sim \tilde{x}$, if \textbf{(i)} their high-level semantic representations are close under a semantic similarity metric, instantiated later in Section~\ref{sec:methodology} as the Semantic Persistence Score (SPS) with $\mathrm{SPS}(x, \tilde{x}) \ge \tau_{\mathrm{sps}}$, and \textbf{(ii)} replacing $x$ with $\tilde{x}$ in the training data does not materially degrade downstream utility of the resulting fine-tuned model, i.e.,
$|U(f_\theta^{(x)}) - U(f_\theta^{(\tilde{x})})| \le \varepsilon_{\mathrm{util}}$, where $f_\theta^{(x)}$ denotes the model $f_\theta$ fine-tuned on data containing $x$.
\end{definition}

\textbf{Formal evidentiary criteria.   }
Let $A(x; f_\theta) \in \mathbb{R}$ denote an auditing statistic (for example, an MIA score), where larger values indicate stronger evidence of membership. Let $\tau_{mia} \in \mathbb{R}$ be a fixed decision threshold, and let $\mathbb{I}[\cdot]$ denote the indicator function that outputs $1$ if its argument is true and $0$ otherwise. Let $\mathcal{T}$ denote a family of semantic-preserving transformations on $\mathcal{X}$. Paraphrasing is a concrete and practically important subclass of $\mathcal{T}$, but our definitions apply more broadly.

\begin{definition}[Robustness]
\label{def:robustness}
An auditing method with score $A(x;f_\theta)$ and threshold $\tau_{\mathrm{mia}}$
is \emph{robust} to a family of semantic-preserving transformations $\mathcal{T}$
if there exists $\varepsilon_{\mathrm{rob}}>0$ such that for any $T\in\mathcal{T}$
with $x\sim T(x)$,
\[
\lvert A(x;f_\theta)-A(T(x);f_\theta)\rvert \le \varepsilon_{\mathrm{rob}}.
\]
In particular, if $A(x;f_\theta)$ is \emph{non-ambiguous}\footnote{
If $A(x;f_\theta)\ge\tau_{\mathrm{mia}}$ and
$|A(x;f_\theta)-\tau_{\mathrm{mia}}|\ge\varepsilon_{\mathrm{rob}}$, then
$A(x;f_\theta)\ge\tau_{\mathrm{mia}}+\varepsilon_{\mathrm{rob}}$, which implies
$A(T(x);f_\theta)\ge\tau_{\mathrm{mia}}$.
The case $A(x;f_\theta)\le\tau_{\mathrm{mia}}$ follows analogously.
} in the sense that 
$|A(x;f_\theta)-\tau_{\mathrm{mia}}|\ge\varepsilon_{\mathrm{rob}},$
then induced membership decisions agree: 
$\mathbb{I}[A(x;f_\theta)\ge\tau_{\mathrm{mia}}]=\mathbb{I}[A(T(x);f_\theta)\ge\tau_{\mathrm{mia}}].$
\end{definition}

Accordingly, the remainder of this work focuses exclusively on evaluating whether existing MIAs satisfy this robustness requirement. We emphasize that this section does not propose a legal standard; rather, it formalizes a technical threat model that captures information asymmetries and adversarial incentives that may arise in litigation.

\section{Structure Aware SAE-Guided Extraction}
\label{sec:methodology}

\newcommand{\ph}[1]{{\textbf{\ttfamily\detokenize{<<#1>>}}}}

\definecolor{sageBlueFill}{RGB}{233,245,255}
\definecolor{sageBlueLine}{RGB}{125,182,240}
\definecolor{sageGreen}{RGB}{34,139,34}
\definecolor{sageRed}{RGB}{200,60,60}
\definecolor{sageGray}{RGB}{90,90,90}

\newcommand{\okmark}{\textcolor{white}{\ding{51}}}
\newcommand{\badmark}{\textcolor{white}{\ding{55}}}

\definecolor{sageBlueFill}{RGB}{233,245,255}
\definecolor{sageBlueLine}{RGB}{125,182,240}
\definecolor{sageGreen}{RGB}{34,139,34}
\definecolor{sageRed}{RGB}{200,60,60}
\definecolor{sageGray}{RGB}{90,90,90}

\begin{figure*}[htbp]
\centering
{\resizebox{\textwidth}{!}{%
\begin{tikzpicture}[
  font=\normalsize,
  >=Latex,
  line cap=round, line join=round,
  coltitle/.style={font=\bfseries\Large, text=black},
  bigbox/.style={
    draw=sageBlueLine, fill=sageBlueFill,
    rounded corners=6mm,
    line width=1.2pt,
    inner sep=7mm,
    align=center,
    text width=7.1cm,
    anchor=center,
    font=\normalsize
  },
  box/.style={
    draw=sageBlueLine, fill=sageBlueFill,
    rounded corners=2.2mm,
    line width=1.0pt,
    inner sep=2.0mm,
    align=center,
    text width=5.2cm,
    anchor=north,
    font=\normalsize
  },
  boxS/.style={
    draw=sageBlueLine, fill=sageBlueFill,
    rounded corners=2.2mm,
    line width=1.0pt,
    inner sep=2.0mm,
    align=center,
    text width=4.4cm,
    anchor=center,
    font=\normalsize
  },
  removed/.style={
    draw=sageBlueLine, fill=sageBlueFill,
    rounded corners=2.2mm,
    line width=1.0pt,
    inner sep=2.0mm,
    align=center,
    text width=4.4cm,
    font=\bfseries\Large,
    text=sageGray,
    anchor=center
  },
  midarrow/.style={-Latex, line width=2.2pt, draw=sageGray},
  rowarrow/.style={-Latex, line width=1.6pt, draw=sageGray},
  ok/.style={circle, fill=sageGreen, draw=sageGreen, minimum size=6.5mm, inner sep=0pt},
  bad/.style={circle, fill=sageRed, draw=sageRed, minimum size=6.5mm, inner sep=0pt}
]

\pgfmathsetmacro{\xOrig}{0}
\pgfmathsetmacro{\xPara}{8.8}
\pgfmathsetmacro{\xSage}{15.9}
\pgfmathsetmacro{\xR}{22.5}

\pgfmathsetmacro{\yTop}{3.25}
\pgfmathsetmacro{\vsep}{0.2mm}

\node[coltitle] at (\xOrig,4.35) {Original Document};
\node[coltitle] at (\xPara,4.35) {Paraphrase};
\node[coltitle] at (\xSage,4.35) {SAGE};
\node[coltitle] at (\xR,4.35) {SAGE-R};

\node[box] (p1) at (\xPara,\yTop) {%
  \textbf{\texttt{<section type="structure">}}\\
  Description\\
  \textbf{\texttt{</section>}}
};
\node[box] (p2) [below=\vsep of p1.south, anchor=north] {%
  \textbf{\texttt{<section type="narrative">}}\\
  Using pdfbox, I produced\\
  a PDF/A-2b document\\
  and ...\\
  \textbf{\texttt{</section>}}
};
\node[box] (p3) [below=\vsep of p2.south, anchor=north] {%
  \textbf{\texttt{<section type="structure">}}\\
  Specification: ISO 19005-\\
  2:2011, Clause: 6.2.11.4,\\
  Test number: 4\\
  \textbf{\texttt{</section>}}
};
\node[box] (p4) [below=\vsep of p3.south, anchor=north] {%
  \textbf{\texttt{<section type="narrative">}}\\
  Specifically, certain CIDs\\
  included in the\\
  CIDToGidMap are absent\\
  from the CIDSet.\\
  \textbf{\texttt{</section>}}
};

\coordinate (y1) at ($(0,0)!(p1.center)!(0,1)$);
\coordinate (y2) at ($(0,0)!(p2.center)!(0,1)$);
\coordinate (y3) at ($(0,0)!(p3.center)!(0,1)$);
\coordinate (y4) at ($(0,0)!(p4.center)!(0,1)$);

\coordinate (spos1) at ($(\xSage,0) + (y1)$);
\coordinate (spos2) at ($(\xSage,0) + (y2)$);
\coordinate (spos3) at ($(\xSage,0) + (y3)$);
\coordinate (spos4) at ($(\xSage,0) + (y4)$);

\coordinate (rpos1) at ($(\xR,0) + (y1)$);
\coordinate (rpos2) at ($(\xR,0) + (y2)$);
\coordinate (rpos3) at ($(\xR,0) + (y3)$);
\coordinate (rpos4) at ($(\xR,0) + (y4)$);

\node[boxS] (s1) at (spos1) {Description};
\node[boxS] (s2) at (spos2) {%
  Using pdfbox, I\\
  produced a PDF/A-\\
  2b document and ...
};
\node[boxS] (s3) at (spos3) {%
  \detokenize{Specification: ISO}
  \\ \detokenize{19005-2:2011,}
  \\ \detokenize{Clause: 6.2.11.4, Test}
  \\ \detokenize{number: 4}
};
\node[boxS] (s4) at (spos4) {%
  Specifically, certain\\
  CIDs included in the\\
  CIDToGidMap are\\
  absent from the\\
  CIDSet.
};

\node[removed] (r1) at (rpos1) {REMOVED};
\node[boxS]    (r2) at (rpos2) {%
  Using \ph{FACT_1}, I\\
  produced \ph{FACT_2}\\
  document and
};
\node[removed] (r3) at (rpos3) {REMOVED};
\node[boxS]    (r4) at (rpos4) {%
  Specifically, certain\\
  CIDs included in the\\
  \ph{FACT_3} are\\
  absent from the\\
  \ph{FACT_4}
};

\coordinate (stackMid)  at ($(p1.north)!0.5!(p4.south)$);
\coordinate (stackMidY) at ($(0,0)!(stackMid)!(0,1)$); 
\coordinate (origpos)   at ($(\xOrig,0.55) + (stackMidY)$);

\node[bigbox] (orig) at (origpos) {
  {Description}\\[2mm]
  I have created PDF/A-2b document with\\
  pdfbox and I have found that the font\\
  ...\\[1mm]
  Specification: ISO 19005-2:2011, Clause:\\
  6.2.11.4, Test number: 4\\
  ...\\[1mm]
  That is: some CIDs which are in the\\
  CIDToGidMap don't appear in the CIDSet.
};

\coordinate (p23west) at ($(p2.west)!0.5!(p3.west)$);
\draw[midarrow] (orig.east) -- (p23west);

\foreach \a/\b in {p1/s1,p2/s2,p3/s3,p4/s4}{
  \draw[rowarrow] ([xshift=2mm]\a.east) -- ([xshift=-2mm]\b.west);
  \path ([xshift=2mm]\a.east) -- ([xshift=-2mm]\b.west)
    node[pos=0.37, ok] {\okmark};
}

\foreach \src/\dst/\sym/\sty in {s1/r1/\badmark/bad, s2/r2/\okmark/ok, s3/r3/\badmark/bad, s4/r4/\okmark/ok}{
  \draw[rowarrow] ([xshift=2mm]\src.east) -- ([xshift=-2mm]\dst.west);
  \path ([xshift=2mm]\src.east) -- ([xshift=-2mm]\dst.west)
    node[pos=0.37, \sty] {\sym};
}

\end{tikzpicture}}
}
\caption{\textbf{SAGE qualitative example.}
Structure-aware dataset generation in SAGE and SAGE-R. The original document (left) is decomposed into structural and narrative sections. Paraphrasing preserves structural sections verbatim while rewriting narrative content. SAGE retains factual anchors and document structure, whereas SAGE-R removes structural sections and replaces factual entities with placeholders (\texttt{<<FACT\_i>>}). Green and red markers indicate preserved and removed components, respectively.}

\label{fig:sage-qual}
\end{figure*}
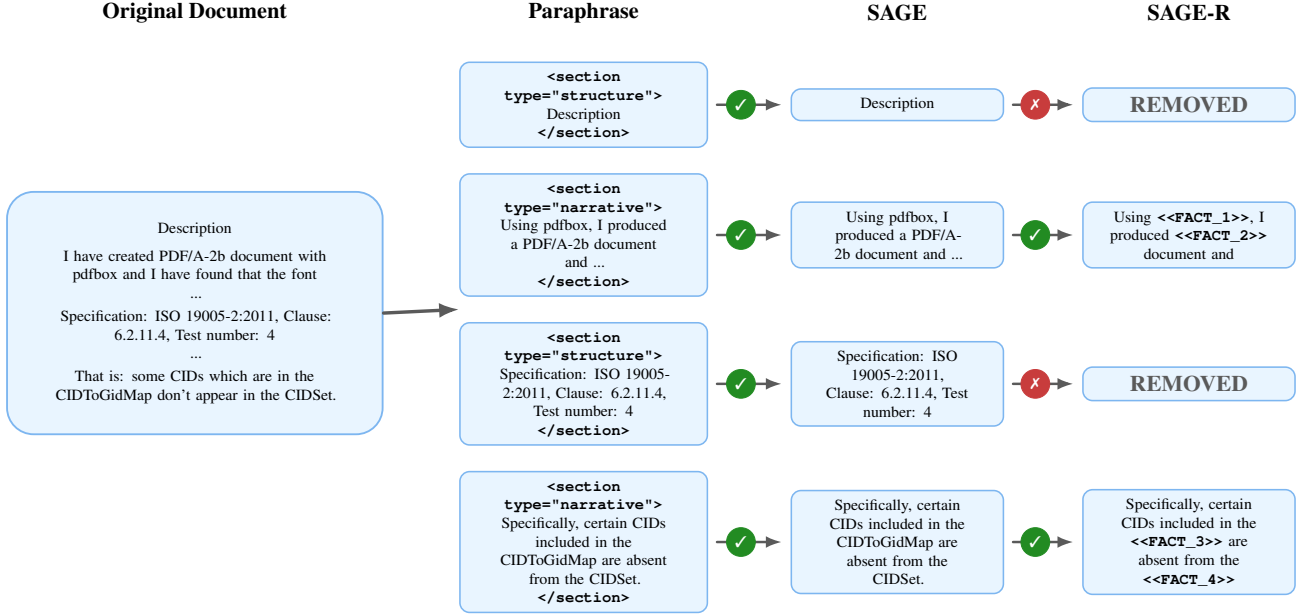

We propose \textbf{S}tructure-\textbf{A}ware \textbf{S}parse \textbf{A}uto\textbf{E}ncoder-\textbf{G}uided \textbf{E}xtraction (SAGE), a structure-aware, metric-guided paraphrasing pipeline that generates obfuscated training data while preserving semantic content. Given an input document $x$, SAGE decomposes the text into an ordered sequence of sections $\{s_k\}_{k=1}^{K}$  using a fixed LLM prompt. Each section is classified as either \emph{structural} (e.g., headers, citations, code blocks, identifiers) or \emph{narrative} (prose and explanations). Structural sections are preserved verbatim, while $N=3$ paraphrasing candidates are generated for narrative sections and evaluated under explicit semantic and surface-form constraints. The final paraphrase $\tilde{x}$ is selected by maximizing a trade-off between semantic preservation and lexical divergence.

\begin{equation}
\vspace{-0.1in}
\label{eq:sage_simplified}
\tilde{x}=
\arg\max_{x'\in\mathcal{C}(x)} \Big[
\underbrace{\mathrm{SPS}(x, x')}_{\text{semantic persistence}} - \underbrace{\mathrm{WordSim}(x, x')}_{\text{surface overlap}}
\Big]
\end{equation}
where $\mathcal{C}(x)$ is the finite set of candidates. This objective ensures $\tilde{x}$ retains the original meaning while minimizing surface-form overlap. The  generation prompts and full algorithm are provided in Appendix~\ref{app:prompts} and~\ref{app:sage-generation}, respectively.

\subsection{Semantic Persistence}
We define \textbf{semantic persistence} as invariance of a fixed semantic observer's feature activations under paraphrasing. Let $f(x)\in\mathbb{R}^M$ be the sparse feature vector extracted from a semantic observer given input $x$. We define active feature indices as $\mathcal{I}_x=\{i \mid f(x)_i > 0\}$.

\subsubsection{The Semantic Oracle}
In all experiments, we instantiate the semantic observer using a Sparse Autoencoder (SAE) attached to an intermediate layer of a separate, fixed language model (the \textbf{semantic oracle}). The SAE induces a function $f:\mathcal{X}\rightarrow\mathbb{R}^M$ that maps a text sequence to a sparse vector of monosemantic feature activations~\citep{bricken2023towards, templeton2024scaling}. To prevent formatting artifacts from influencing the oracle, we compute SAE features only on narrative text spans, excluding structural sections such as headers, citations, and formatting markers.

\begin{definition}[Semantic Persistence Score ($\mathrm{SPS}$)]
\label{def:sps}
Given SAE feature representations, we define the \emph{Semantic Persistence Score} ($\mathrm{SPS}$) as a measure of semantic alignment between an original text $x$ and its paraphrase $\tilde{x}$. Let $\{s_k\}_{k=1}^{K}$ and $\{\tilde{s}_k\}_{k=1}^{K}$ denote the corresponding narrative spans extracted from $x$ and $\tilde{x}$. The SPS is defined as the average cosine similarity between SAE feature vectors computed on each span:
\[
    \mathrm{SPS}(x,\tilde{x})
    = \frac{1}{K}\sum_{k=1}^{K}
    \frac{f(s_k)\cdot f(\tilde{s}_k)}{\lVert f(s_k)\rVert\,\lVert f(\tilde{s}_k)\rVert}.
\]
$\mathrm{SPS}$ evaluates similarity in a sparse feature space where individual activations correspond to specific semantic factors~\citep{bricken2023towards}, allowing it to distinguish true semantic preservation from superficial similarity.
\end{definition}

\subsection{Word Similarity}
In addition to semantic preservation, SAGE enforces surface-form divergence at the word level. Lower values of $\mathrm{WordSim}(x,\tilde{x})$ indicate less surface-form reuse (overlap) between the original and paraphrased text.

\begin{definition}[Word Similarity ($\mathrm{WordSim}$)]
\label{def:wordsim}
Let $M=\{\mathrm{Jacc}_{\mathrm{word}},\, \mathrm{Ov}_{\mathrm{w\text{-}tri}},\, \mathrm{Ov}_{\mathrm{c5}}\}$, where $\mathrm{Jacc}_{\mathrm{word}}(x,\tilde{x})$ denotes the Jaccard similarity between sets of lowercase alphanumeric word tokens, $\mathrm{Ov}_{\mathrm{w\text{-}tri}}(x,\tilde{x})$ denotes the overlap ratio of word trigrams in $x$ that also appear in $\tilde{x}$ (normalized by the number of trigrams in $x$), and $\mathrm{Ov}_{\mathrm{c5}}(x,\tilde{x})$ denotes the overlap ratio of character 5-grams in $x$ that also appear in $\tilde{x}$ (normalized by the number of 5-grams in $x$).  
The combined surface similarity score is defined as:
\[
\mathrm{WordSim}(x,\tilde{x})
= \frac{1}{|M|}\sum_{m \in M} m(x,\tilde{x}).
\]
\end{definition}

\section{Results}
\label{sec:results}

We evaluate whether MIAs satisfy the robustness requirement introduced in Def.~\ref{def:robustness}. Our experiments center on the following question: \textbf{Do MIAs remain robust when the training data is paraphrased without changing its meaning?} Across all evaluated settings, we observe a consistent pattern: MIAs exhibit detectable signals on unmodified text, but their performance degrades sharply under paraphrasing, even though the model’s underlying semantic representations remain stable.

\subsection{Membership Inference Attack Evaluation}
\label{sec:mia_results}

Our experiments follow the MIMIR benchmark introduced by \citet{DBLP:journals/corr/abs-2402-07841}, using the same 13-gram (13\_0.8) split as in prior work \citep{DBLP:conf/uss/00020GCSA0FK0L25}. In contrast to the full benchmark, we evaluate on five representative datasets spanning academic, web, and curated corpora: \texttt{arXiv}, \texttt{HackerNews (HN)}, \texttt{PubMed}, \texttt{Pile-CC}, and \texttt{Wikipedia (Wiki)}. 

Unless otherwise stated, all experiments use \emph{LLaMA-3.2-3B}~\citep{llama3models2024} as the fine-tuned target model (i.e., the defendant model in Fig.~\ref{fig:protocol-mia}). Additional results on \emph{Pythia-7B}~\citep{pythia2023} are reported in Appendix~\ref{app:pythia}. For the SAGE paraphrasing pipeline, we employ three independent paraphrasing models: \emph{DeepSeek-V3.2}~\citep{deepseek2024}, \emph{Gemini-2.5-Flash}~\citep{gemini2025}, and \emph{Grok-4.1-Fast}~\citep{xai2025grok4}. For each paraphraser, we generate a separate paraphrased training dataset, fine-tune an independent target model, and compute all evaluation metrics. Reported SAGE results are obtained by averaging metrics across these independently fine-tuned models, mitigating paraphraser-specific artifacts. Semantic persistence during SAGE optimization is evaluated using a fixed, pretrained SAE~\citep{bloom2024saelens}, trained on \texttt{gemma-2b} and attached to layer~12 (residual stream, post-activation). This SAE serves as an external semantic oracle that is independent of the audited model and is not trained or adapted as part of our experiments.

We consider two fine-tuning regimes for the target model: end-to-end full fine-tuning and parameter-efficient fine-tuning for three epochs on the same MIMIR training split. Low-Rank Adaptation (LoRA)~\citep{DBLP:conf/iclr/HuSWALWWC22} fine-tunes large models by injecting a small number of trainable low-rank matrices into existing weights while keeping the base parameters frozen, substantially reducing memory and compute overhead. We include LoRA because it is a widely adopted fine-tuning paradigm in practice and prior work has shown that parameter-efficient fine-tuning can exhibit different memorization and privacy behaviors than full fine-tuning~\citep{biderman2024learns,shuttleworth2024illusion,DBLP:conf/uss/00020GCSA0FK0L25}. Implementation details and hyperparameters are in Appendix~\ref{app:full_implementation}, with per-model results in Appendix~\ref{app:detailed-result}.

We evaluate \textbf{nine} MIAs proposed for large language models~\citep{DBLP:conf/uss/CarliniTWJHLRBS21,DBLP:conf/uss/00020GCSA0FK0L25}: \emph{Loss}~\citep{DBLP:conf/csfw/YeomGFJ18}, \emph{Zlib}~\citep{DBLP:conf/uss/CarliniTWJHLRBS21}, \emph{Lowercase}~\citep{DBLP:conf/uss/CarliniTWJHLRBS21}, \emph{Min-K\% Prob}~\citep{Shi2023DetectingPD}, \emph{Min-K\%++}~\citep{DBLP:journals/corr/abs-2404-02936}, \emph{ReCall}~\citep{DBLP:conf/emnlp/XieWHZGP0D24}, \emph{CON-ReCall}~\citep{DBLP:conf/coling/WangWHC0C25}, \emph{Bag-of-Words}~\citep{DBLP:conf/satml/MeeusSJFRM25,DBLP:conf/sp/DasZT25}\footnote{\emph{Bag-of-Words} infers membership from lexical features of the dataset only, without accessing the target model, and thus serves as a control for dataset-specific artifacts rather than memorization.}, and the reference-based \emph{Ratio}~\citep{DBLP:conf/uss/CarliniTWJHLRBS21} attack. For reference-based attacks, we follow SOFT~\citep{DBLP:conf/uss/00020GCSA0FK0L25} and use \texttt{OpenLLaMA-7B}~\citep{openlm2023openllama} as the shadow model. In all settings, we evaluate attacks using queries sampled from the original distribution.

\textbf{Training regimes.   }
For each fine-tuning track, we compare the following regimes:
We consider the following training regimes: \emph{Pretrained (PT)}, the base model without additional fine-tuning; \emph{Fine-tuned (FT)}, standard fine-tuning on the original MIMIR training split (reported separately for full and LoRA fine-tuning); \emph{SOFT}, reproduced from the selective obfuscation pipeline of~\citep{DBLP:conf/uss/00020GCSA0FK0L25}, which modifies the training corpus during fine-tuning by replacing influential examples with paraphrased variants generated by an auxiliary language model, using targeted paraphrasing conditioned on access to model internals. \emph{SAGE (structure-aware)}, fine-tuning on SAGE-generated paraphrases obtained via structure-aware extraction and recombination, where narrative sections are paraphrased independently and reinserted into the original document skeleton while preserving section boundaries, metadata, and formatting; and \emph{SAGE-R (structure- and factual-removed)}, a reduced variant in which extracted narrative sections are flattened into a single continuous text stream and factual anchors (e.g., names, numbers, and dates) are removed, eliminating both structural and factual sources of membership signal (see Fig.~\ref{fig:sage-qual})\footnote{Tokens corresponding to removed facts (e.g., \texttt{FACT1}) are masked during training.}. Factual anchors are identified using a fixed LLM-prompting procedure that tags entity mentions, numeric values, and date expressions (see App.~\ref{app:factual-anchor-method}). Qualitative examples are provided in Appendix~\ref{app:qualitative-sage}.

\noindent
We emphasize that both \emph{SAGE} and \emph{SAGE-R} variants rely only on a semantic oracle to guide paraphrasing and do not require access to the internals of the model in question, whereas SOFT explicitly depends on such access. We also observe a discrepancy between our reproduced SOFT results and those reported in the original paper. All SOFT results presented in this work are obtained using a corrected implementation of the SOFT methodology that resolves inconsistencies in the official codebase and follows the intended pipeline; detailed discussion is provided in Appendix~\ref{app:soft_reprod}.

\begin{table*}[t]
\centering
\caption{\textbf{AUC:} Average performance of MIAs across datasets and methods, averaged over three independently fine-tuned models, one per paraphraser-generated dataset (LoRA).}
\resizebox{\textwidth}{!}{
\begin{tabular}{lccccccccccccccc}
\toprule
& \multicolumn{3}{c}{\textbf{\texttt{arxiv}}} & \multicolumn{3}{c}{\textbf{\texttt{wikipedia}}} & \multicolumn{3}{c}{\textbf{\texttt{pile\_cc}}} & \multicolumn{3}{c}{\textbf{\texttt{hackernews}}} & \multicolumn{3}{c}{\textbf{\texttt{pubmed}}} \\
\cmidrule(lr){2-4}
\cmidrule(lr){5-7}
\cmidrule(lr){8-10}
\cmidrule(lr){11-13}
\cmidrule(lr){14-16}
& {{FT}} & {{SAGE}} & {{SAGE-R}} & {{FT}} & {{SAGE}} & {{SAGE-R}} & {{FT}} & {{SAGE}} & {{SAGE-R}} & {{FT}} & {{SAGE}} & {{SAGE-R}} & {{FT}} & {{SAGE}} & {{SAGE-R}} \\
\midrule
Loss & $0.685 $ & $0.602 $ & $0.559 $ & $0.658 $ & $0.582 $ & $0.528 $ & $0.645 $ & $0.547 $ & $0.515 $ & $0.710 $ & $0.588 $ & $0.535 $ & $0.645 $ & $0.549 $ & $0.530 $ \\
Zlib & $0.682 $ & $0.602 $ & $0.562 $ & $0.663 $ & $0.586 $ & $0.529 $ & $0.667 $ & $0.562 $ & $0.527 $ & $0.724 $ & $0.589 $ & $0.532 $ & $0.645 $ & $0.550 $ & $0.530 $ \\
Lowercase & $0.667 $ & $0.593 $ & $0.516 $ & $0.699 $ & $0.607 $ & $0.521 $ & $0.687 $ & $0.572 $ & $0.499 $ & $0.657 $ & $0.573 $ & $0.519 $ & $0.669 $ & $0.582 $ & $0.542 $ \\
Min-K\% & $0.579 $ & $0.546 $ & $0.524 $ & $0.561 $ & $0.533 $ & $0.510 $ & $0.568 $ & $0.531 $ & $0.516 $ & $0.548 $ & $0.512 $ & $0.492 $ & $0.566 $ & $0.523 $ & $0.516 $ \\
Min-K\%++ & $0.634 $ & $0.565 $ & $0.533 $ & $0.657 $ & $0.578 $ & $0.526 $ & $0.624 $ & $0.537 $ & $0.510 $ & $0.628 $ & $0.549 $ & $0.523 $ & $0.626 $ & $0.532 $ & $0.519 $ \\
ReCall & $0.698 $ & $0.607 $ & $0.559 $ & $0.666 $ & $0.588 $ & $0.532 $ & $0.661 $ & $0.559 $ & $0.524 $ & $0.719 $ & $0.590 $ & $0.531 $ & $0.666 $ & $0.561 $ & $0.537 $ \\
CON-ReCall & $0.650 $ & $0.579 $ & $0.541 $ & $0.636 $ & $0.572 $ & $0.531 $ & $0.593 $ & $0.530 $ & $0.513 $ & $0.603 $ & $0.546 $ & $0.526 $ & $0.628 $ & $0.549 $ & $0.534 $ \\
Ratio & $0.842 $ & $0.693 $ & $0.632 $ & $0.903 $ & $0.752 $ & $0.572 $ & $0.916 $ & $0.681 $ & $0.569 $ & $0.894 $ & $0.727 $ & $0.583 $ & $0.857 $ & $0.674 $ & $0.624 $ \\
Bag-of-Words & $0.498 $ & $0.498 $ & $0.498 $ & $0.510 $ & $0.510 $ & $0.510 $ & $0.498 $ & $0.498 $ & $0.498 $ & $0.524 $ & $0.524 $ & $0.524 $ & $0.528 $ & $0.528 $ & $0.528 $ \\
\midrule
Average & $0.68$ & $0.60$ & $0.55$ & $0.68$ & $0.60$ & $0.53$ & $0.67$ & $0.56$ & $0.52$ & $0.69$ & $0.58$ & $0.53$ & $0.66$ & $0.57$ & $0.54$ \\
\bottomrule
\end{tabular}}

\label{tab:results-lora-ensemble-auc-datasets-defenses}
\end{table*}

\begin{table*}[htbp]
\centering
\caption{\textbf{TPR@1\%FPR:} Average performance of MIAs across datasets and methods, averaged over three independently fine-tuned models, one per paraphraser-generated dataset (LoRA).}
\resizebox{\textwidth}{!}{
\begin{tabular}{lccccccccccccccc}
\toprule
& \multicolumn{3}{c}{\textbf{\texttt{arxiv}}} & \multicolumn{3}{c}{\textbf{\texttt{wikipedia}}} & \multicolumn{3}{c}{\textbf{\texttt{pile\_cc}}} & \multicolumn{3}{c}{\textbf{\texttt{hackernews}}} & \multicolumn{3}{c}{\textbf{\texttt{pubmed}}} \\
\cmidrule(lr){2-4}
\cmidrule(lr){5-7}
\cmidrule(lr){8-10}
\cmidrule(lr){11-13}
\cmidrule(lr){14-16}
& {{FT}} & {{SAGE}} & {{SAGE-R}} & {{FT}} & {{SAGE}} & {{SAGE-R}} & {{FT}} & {{SAGE}} & {{SAGE-R}} & {{FT}} & {{SAGE}} & {{SAGE-R}} & {{FT}} & {{SAGE}} & {{SAGE-R}} \\
\midrule
Loss & $0.028 $ & $0.013 $ & $0.006 $ & $0.041 $ & $0.026 $ & $0.014 $ & $0.022 $ & $0.015 $ & $0.012 $ & $0.036 $ & $0.016 $ & $0.008 $ & $0.036 $ & $0.014 $ & $0.011 $ \\
Zlib & $0.031 $ & $0.018 $ & $0.010 $ & $0.051 $ & $0.034 $ & $0.016 $ & $0.039 $ & $0.020 $ & $0.019 $ & $0.036 $ & $0.015 $ & $0.005 $ & $0.037 $ & $0.015 $ & $0.009 $ \\
Lowercase & $0.058 $ & $0.033 $ & $0.008 $ & $0.048 $ & $0.031 $ & $0.015 $ & $0.036 $ & $0.022 $ & $0.013 $ & $0.028 $ & $0.016 $ & $0.008 $ & $0.045 $ & $0.022 $ & $0.008 $ \\
Min-K\% & $0.048 $ & $0.024 $ & $0.015 $ & $0.047 $ & $0.027 $ & $0.017 $ & $0.030 $ & $0.014 $ & $0.011 $ & $0.042 $ & $0.012 $ & $0.015 $ & $0.078 $ & $0.020 $ & $0.017 $ \\
Min-K\%++ & $0.018 $ & $0.009 $ & $0.004 $ & $0.057 $ & $0.029 $ & $0.015 $ & $0.024 $ & $0.015 $ & $0.012 $ & $0.032 $ & $0.020 $ & $0.012 $ & $0.039 $ & $0.018 $ & $0.011 $ \\
ReCall & $0.030 $ & $0.016 $ & $0.007 $ & $0.047 $ & $0.028 $ & $0.016 $ & $0.021 $ & $0.011 $ & $0.009 $ & $0.037 $ & $0.014 $ & $0.007 $ & $0.041 $ & $0.012 $ & $0.011 $ \\
CON-ReCall & $0.030 $ & $0.018 $ & $0.007 $ & $0.044 $ & $0.026 $ & $0.015 $ & $0.028 $ & $0.013 $ & $0.011 $ & $0.022 $ & $0.012 $ & $0.008 $ & $0.027 $ & $0.012 $ & $0.009 $ \\
Ratio & $0.200 $ & $0.077 $ & $0.013 $ & $0.164 $ & $0.073 $ & $0.020 $ & $0.185 $ & $0.034 $ & $0.015 $ & $0.122 $ & $0.032 $ & $0.016 $ & $0.190 $ & $0.072 $ & $0.058 $ \\
Bag-of-Words & $0.007 $ & $0.007 $ & $0.007 $ & $0.004 $ & $0.004 $ & $0.004 $ & $0.010 $ & $0.010 $ & $0.010 $ & $0.011 $ & $0.011 $ & $0.011 $ & $0.011 $ & $0.011 $ & $0.011 $ \\
\midrule
Average & $0.06$ & $0.03$ & $0.01$ & $0.06$ & $0.03$ & $0.02$ & $0.05$ & $0.02$ & $0.01$ & $0.04$ & $0.02$ & $0.01$ & $0.06$ & $0.02$ & $0.02$ \\
\bottomrule
\end{tabular}}
\label{tab:results-lora-ensemble-TPR@FPRdatasets-defenses-None-True}
\end{table*}

\subsection{Impact of Paraphrasing Defenses on MIAs}
\begin{figure}[ht]
    \centering
    \includegraphics[width=\columnwidth]{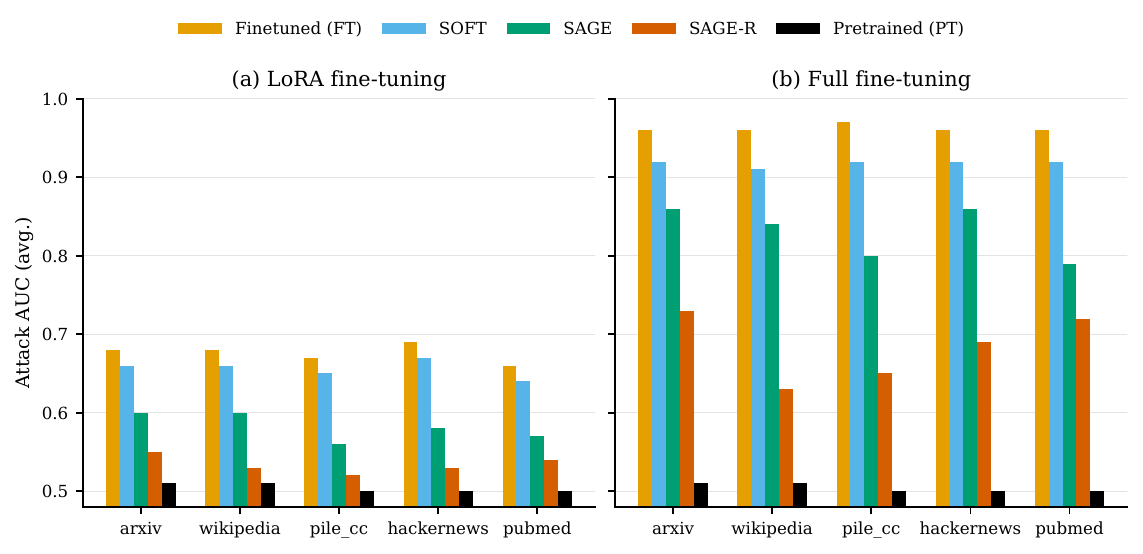}
    \caption{
    Average performance of MIAs across datasets and methods, averaged over three independently fine-tuned models, one per paraphraser-generated dataset. (a) LoRA fine-tuning. (b) Full fine-tuning. Lower is better. In both regimes, the same qualitative ordering holds: FT $>$ SOFT $>$ SAGE $>$ SAGE-R $>$ PT.
    }
    \label{fig:avg-auc-full}
\end{figure}

Tables~\ref{tab:results-lora-ensemble-auc-datasets-defenses} and~\ref{tab:results-lora-ensemble-TPR@FPRdatasets-defenses-None-True} report MIA performance, showing average\footnote{Bag-of-Words is omitted from the average calculation, as it does not query the target model and serves only as a control baseline.} and per-attack AUC across attacks and TPR@1\% FPR, which we use as complementary measures of attack effectiveness. As a sanity check, we also evaluate attacks on the pretrained (PT) model, which has not been exposed to downstream data, and observe random-guessing performance across datasets (AUC $\approx 0.5$, Fig.~\ref{fig:avg-auc-full}). Corresponding TPR@FPR results for the pretrained model, which align with random-guessing behavior, are reported in App.~\ref{app:detailed-result}.

\textbf{Under LoRA fine-tuning, paraphrasing substantially suppresses MIAs.}
Under LoRA fine-tuning, MIAs achieve strong performance on the original data (FT), confirming the presence of exploitable membership signals in the absence of defenses (Tables~\ref{tab:results-lora-ensemble-auc-datasets-defenses} and~\ref{tab:results-lora-ensemble-TPR@FPRdatasets-defenses-None-True}). Introducing paraphrasing consistently degrades attack effectiveness across attacks and datasets: SAGE leads to a substantial reduction in both AUC and low-FPR metrics, while SAGE-R further suppresses performance to levels close to the pretrained baseline. This degradation is systematic and observed across all evaluated datasets, indicating that paraphrasing weakens membership signals under parameter-efficient fine-tuning.

\textbf{Full fine-tuning amplifies leakage but preserves the ordering of defenses.}
Figure~\ref{fig:avg-auc-full} presents the corresponding results under full fine-tuning. While the qualitative ordering of regimes remains consistent, meaning that SAGE variants outperform SOFT and substantially reduce leakage relative to FT,  the absolute level of attack success is uniformly higher than in the LoRA setting. Stronger memorization effects induced by full fine-tuning are therefore more difficult to eliminate entirely via paraphrasing alone. Nevertheless, even in this more challenging regime, SAGE and SAGE-R continue to provide a clear reduction in attack performance compared to standard fine-tuning; detailed per-attack results for full fine-tuning are reported in Appendix~\ref{app:detailed-result}.

\textbf{Stability across regimes.} Across both LoRA and full fine-tuning, the relative ordering remains stable (FT $\!>\!$ SOFT $\!>\!$ SAGE $\!>\!$ SAGE-R), suggesting that existing MIAs rely primarily on surface-level regularities rather than invariant semantic representations (see Fig.~\ref{fig:avg-auc-full}). Accordingly, SOFT~\citep{DBLP:conf/uss/00020GCSA0FK0L25} reduces leakage relative to FT but remains consistently weaker than SAGE and SAGE-R, despite requiring access to model internals, while SAGE achieves stronger suppression without such access. Although SAGE-R removes approximately 9.8\% of tokens on average, these correspond specifically to extracted factual anchors; an ablation removing facts without paraphrasing yields only limited MIA suppression (Appendix~\ref{app:facts-ablation}).

\subsection{Utility Evaluation}
We evaluate utility using two complementary criteria. For paraphrase quality we use SPS which measures semantic consistency beyond surface overlap. For downstream utility, we follow prior work and adopt an LLM-as-a-Judge framework~\citep{DBLP:conf/uss/00020GCSA0FK0L25}. Although perplexity is a commonly used metric, it is sensitive to surface-level variation and is difficult to interpret under paraphrasing, particularly in datasets such as MIMIR. Human evaluation, while informative, is expensive and does not scale. We therefore rely on SPS and LLM-based judgments as practical alternatives.

\label{sec:utility}
\subsubsection{Paraphrase Quality}

\begin{figure}[htbp]
    \centering
    \includegraphics[width=\columnwidth]{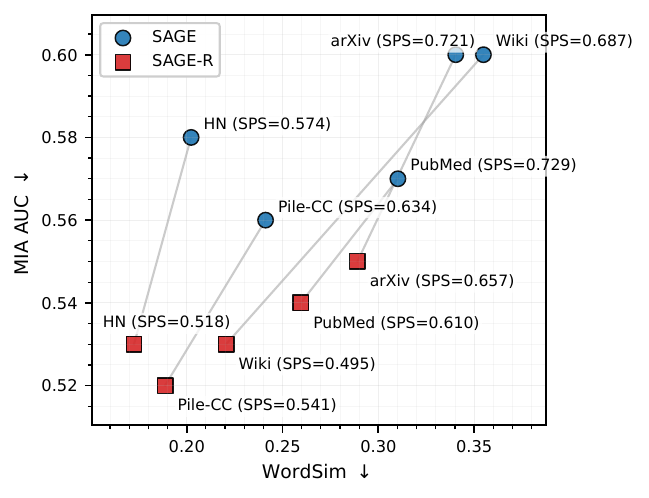}
    \caption{AUC--WordSim tradeoff: Average MIA AUC ($\downarrow$ means stronger privacy) vs. WordSim ($\downarrow$ indicates greater divergence from original). Each point represents a dataset-level average over paraphraser models; gray segments connect SAGE to SAGE-R within each dataset. Labels indicate SPS.
}
    \label{fig:auc_vs_wordsim}
    \vspace{-0.1in}
\end{figure}

Using the $\mathrm{SPS}$ and $\mathrm{WordSim}$ metrics introduced in Definitions~\ref{def:sps} and~\ref{def:wordsim}, we evaluate the quality of our paraphrasing. Figure~\ref{fig:auc_vs_wordsim} shows that, SAGE retains the conceptual meaning within the original datasets, as reported with $\mathrm{SPS}$ results. Furthermore, low $\mathrm{WordSim}$ scores confirm that the paraphrasing process substantially reduces surface overlap with the original text while preserving semantic content. Full results and detailed analysis of $\mathrm{SPS}$ are in Appendix~\ref{app:sps-anal}. This combination supports the suppression of MIA signals without degrading downstream fine-tuning utility. We ablate over SAE layers and models and find that $\mathrm{SPS}$ is stable across layers and consistently separates SAGE and SAGE-R, though absolute values vary by model (Appendix~\ref{app:sae_ablation}).

\subsubsection{LLM-as-a-Judge}
\begin{table}[ht]
\centering
\small
\caption{Utility evaluation using an LLM-as-a-Judge under LoRA fine-tuning. Higher scores indicate better knowledge retention. Results are averaged over three independently fine-tuned models, each fine-tuned on a dataset generated by a different paraphraser.}
\label{tab:utility-judge-lora}
\begin{tabular}{lcccc}
\toprule
\textbf{Dataset} & \textbf{PT} & \textbf{FT} & \textbf{SAGE} & \textbf{SAGE-R} \\
\midrule
ArXiv & 45.7 & 47.0 & 53.1 & 52.8 \\
Wikipedia & 46.6 & 48.3 & 54.6 & 49.0 \\
Pile-CC & 45.1 & 51.6 & 49.4 & 45.2 \\
PubMed & 53.9 & 55.9 & 60.9 & 60.0 \\
HackerNews & 42.1 & 45.0 & 45.9 & 39.3 \\
\midrule
\textbf{Average} & 47.8 & 49.5 & 52.8 & 49.1\\
\bottomrule
\end{tabular}
\end{table}

The collapse of MIA performance under paraphrasing does not imply that fine-tuned models forget the underlying training content. We therefore evaluate downstream utility using an \emph{LLM-as-a-Judge} framework~\citep{zheng2023judging}, following prior work~\citep{DBLP:conf/uss/00020GCSA0FK0L25}. To mitigate the inherent noise and variance of LLM evaluators~\citep{chenghan2023alternative, wang2024evaluators}, we employ a jury approach: we query three independent LLM judges three times per example (325 questions) and report the averaged scores~\citep{vega2024replacing}. As shown in Table~\ref{tab:utility-judge-lora}, models fine-tuned on SAGE-paraphrased data achieve utility comparable to standard fine-tuning and consistently outperform the pre-trained baseline, with no utility degradation under SAGE or SAGE-R. These results show that both SAGE and SAGE-R satisfy the semantic equivalence criterion defined in Definition~\ref{def:sps}: high SPS establishes semantic persistence, while the LLM-as-a-Judge evaluation confirms utility preservation. Full evaluation details and discussion are in Appendix~\ref{app:llm-judge}.
\section{Discussion}
\label{sec:discussion}
Framed by our communication protocol (Section~\ref{sec:communication}), our results highlight a tension between the behavior of existing MIAs and the robustness requirements implied by copyright auditing (see Def.~\ref{def:robustness}). While prior work attributes strong MIA performance to overfitting or distributional artifacts~\citep{DBLP:conf/csfw/YeomGFJ18,DBLP:conf/nips/MainiJPD24,DBLP:journals/corr/abs-2402-07841,DBLP:conf/icml/MeeusSFM24,DBLP:conf/satml/MeeusSJFRM25}, our results address a distinct robustness limitation. We do not claim that MIAs are universally ineffective; rather, we observe that even under such favorable conditions, membership scores can change substantially under semantics-preserving transformations. This sensitivity suggests that current MIAs depend strongly on surface-level lexical regularities, rather than features that are invariant to meaning-preserving rewrites, undermining their evidentiary value in adversarial auditing settings.

\textbf{Limitations.}
We acknowledge that paraphrasing does not fully eliminate memorization: MIAs retain performance substantially above chance (AUC $0.7$--$0.8$) under full fine-tuning. However, the presence of a statistical signal does not directly translate into evidentiary reliability for copyright auditing. As noted in Definition~\ref{def:robustness}, admissible evidence requires a decision threshold $\tau_{\text{mia}}$. Our results show that semantics-preserving transformations can shift score distributions sufficiently to make such thresholds unstable, even when aggregate performance remains high. As a result, the remaining signal appears sensitive to surface-level artifacts and difficult to interpret consistently in adversarial settings. Finally, we note that our findings are specific to the text modality considered here; other domains, such as image or code generation, may exhibit different trade-offs.

\textbf{Future work.} Promising directions include auditing mechanisms that target semantic or functional behavior rather than surface-level memorization, as well as protocol-aware and game-theoretic formulations that explicitly model strategic behavior, incentives, and evidentiary thresholds between the parties involved. An important extension is to study adaptive auditing strategies in which the prosecutor generates and aggregates membership signals across multiple semantically equivalent paraphrases of a suspect text, and to characterize whether such ensemble-based procedures yield stable evidence under adversarial, semantics-preserving obfuscation.

\section{Related Works}
\label{sec:rworks}
A prominent line of work proposes \emph{model-centric} defenses based on differential privacy. Techniques such as DP-SGD~\citep{DBLP:conf/ccs/AbadiCGMMT016} and parameter-efficient variants including DP-LoRA~\citep{DBLP:journals/tmis/LiuZZGZWQ25,DBLP:conf/iclr/YuNBGI0KLMWYZ22,DBLP:conf/iclr/LiTLH22} bound the influence of individual training samples by injecting noise during optimization. While these methods provide formal privacy guarantees, they typically incur nontrivial utility degradation, require careful hyperparameter tuning, and often struggle to scale to large models or low-noise regimes relevant in practice. Recent work has proposed alternative membership inference attacks for fine-tuned LLMs, including SPV-MIA~\citep{DBLP:conf/nips/0005W0L0J24}, which leverages self-prompted reference models and probabilistic variation signals. Our evaluation focuses on attacks whose decision rules operate directly on the queried text and model outputs under a fixed auditing protocol. We exclude adaptive methods like SPV-MIA, as their reliance on dynamic calibration mechanisms introduces distinct questions about robustness and admissibility, which we leave to future work.%

A parallel line of research focuses on dataset copyright auditing. Existing approaches include intrusive watermarking, dataset inference, and document-level MIAs \citep{DBLP:conf/icml/MeeusSFM24, DBLP:conf/sp/DuZ0ZSC0025}. For instance, DE-COP~\citep{duarte2024decop} proposes detecting copyrighted content by probing whether a model can distinguish verbatim text from paraphrased variations. Dataset inference methods~\citep{DBLP:conf/nips/MainiJPD24} address aggregate dataset-level inclusion rather than record-level membership, and are thus orthogonal to the individual-instance auditing setting considered in this work. Watermarking-based audits operate under different assumptions, typically relying on proactive watermark insertion, though recent work has shown they are vulnerable to evasion via recursive paraphrasing~\citep{sadasivan2025reliable}. In contrast, we consider post hoc auditing without assuming such instrumentation.

\section{Conclusion}
\label{sec:conc}
We evaluated membership inference attacks (MIAs) through a formal judge--prosecutor--accused communication protocol and showed that they face significant challenges in providing robust evidence of copyright infringement under realistic adversarial conditions. Even when MIAs exhibit non-trivial performance under favorable settings, semantic-preserving paraphrasing destabilizes their signals while preserving meaning and downstream utility. This indicates that MIAs largely depend on surface-level artifacts rather than semantic use, whereas copyright claims fundamentally concern the use of protected expression. This suggests that reliable copyright auditing cannot rely on membership inference alone and instead requires protocol-aware approaches aligned with legal standards of evidence.

\section*{Impact Statement}
\label{sec:impact}
This paper membership inference attacks as technical evidence for copyright auditing of large language models under adversarial conditions. Our results show that the effectiveness of existing MIAs degrades under semantics-preserving transformations, raising concerns about their robustness in adversarial settings.

This work helps clarify the limitations of current auditing signals in legal, regulatory, and policy settings, and encourages the development of threat-model-aware alternatives. While the paraphrasing techniques studied here could be misused to obfuscate training data, they reflect realistic capabilities already available to model developers rather than introducing new risks.

Overall, this work aims to promote more responsible use and interpretation of machine learning security tools in high-stakes auditing contexts.

\section*{Acknowledgments}
\label{sec:acknow}
The contribution of Marten van Dijk, Murat Bilgehan Ertan and Min Chen to this publication is part of the project CiCS of the research program Gravitation which is (partly) financed by the Dutch Research Council (NWO) under the grant 024.006.037. We acknowledge the use of the DAS-6 High-Performance Computing cluster at Vrije Universiteit Amsterdam for GPU-based experiments~\citep{bal2016medium}.
\newpage
\bibliography{bib}
\bibliographystyle{icml2026}

\newpage
\appendix
\onecolumn

\section{Appendix Organization}
\label{app:app-org}
The appendix is organized as follows:
\begin{itemize}
    \item \textbf{Appendix~\ref{app:dataset}} discusses the theoretical and practical limitations of relying on dataset-level comparisons for copyright auditing.
    \item \textbf{Appendix~\ref{app:soft_reprod}} details the reproduction issues identified in the SOFT framework, documenting specific implementation bugs regarding training regimes, model checkpointing, and model loading.
    \item \textbf{Appendix~\ref{app:full_implementation}} provides comprehensive implementation details for SAGE, including algorithmic pseudocode, generation procedures, prompt strategies, and the methodology for the LLM-as-a-Judge utility evaluation.
    \item \textbf{Appendix~\ref{app:sps-anal}} presents ablation studies and sensitivity analyses for the Semantic Persistence Score (SPS) and Word Similarity (WordSim) metrics, examining their stability across different probe models and layers.
    \item \textbf{Appendix~\ref{app:qualitative-sage}} showcases qualitative examples of original documents alongside their SAGE and SAGE-R paraphrased variants across representative datasets.
    \item \textbf{Appendix~\ref{app:facts-ablation}} briefly addresses ablations regarding the removal of factual anchors in the original dataset.
    \item Finally, \textbf{Appendix~\ref{app:detailed-result}} and \textbf{Appendix~\ref{app:pythia}} report the extensive results for Membership Inference Attack (MIA) performance on \texttt{Llama-3.2-3B} and \texttt{EleutherAI/pythia-6.9b} respectively, covering all evaluated fine-tuning regimes, defense mechanisms, and evaluator models.
\end{itemize}

\section{On Dataset-level Comparisons}
\label{app:dataset}
\paragraph{Why dataset-level comparisons fail.} One might attempt to bypass MIAs by asking the accused to disclose a dataset and then comparing the suspect text $x$ against disclosed documents using token embeddings or nearest-neighbor matching. However, in realistic disputes this approach is not evidentially reliable. First, the accused can omit relevant items, disclose only partial corpora, or disclose a transformed corpus whose provenance cannot be independently verified. Second, adding large volumes of unrelated or near-duplicate documents can change nearest-neighbor structure and undermine any fixed decision threshold. Third, without access to the true training corpus, and its preprocessing, the judge cannot calibrate what similarity level constitutes meaningful evidence of use. Consistent with this, prior work notes that most copyright auditing methods already operate under black-box access assumptions~\cite{DBLP:conf/sp/DuZ0ZSC0025}. As a result, dataset-level similarity may be suggestive in cooporative settings, but it does not provide a robust test of infringement under the communication protocol in Section~\ref{sec:communication}.
\section{Reproduction Issues in SOFT Framework}
\label{app:soft_reprod}
During our attempt to reproduce the results reported in~\cite{DBLP:conf/uss/00020GCSA0FK0L25}, we identified several critical implementation issues in the official \href{https://github.com/KaiyuanZh/SOFT/}{SOFT codebase}\footnote{Accessed January 19, 2026, latest commit was: \texttt{7ca2b7b3b44e352e8ade6bea62889156fe1bff94}.\\ \url{https://github.com/KaiyuanZh/SOFT/}}\textsuperscript{,}\footnote{\url{https://web.archive.org/web/20251211132711/https://github.com/KaiyuanZh/SOFT}} that prevent faithful reproduction of the reported results. We document these issues below for transparency and to justify our corrected reimplementation.

  \subsection{Bug 1: Inconsistent Training Regime}
  \label{appendix:bug1}

  In \texttt{finetune.py} (lines 254--265), the implementation applies a LoRA (Low-Rank Adaptation) wrapper to the model, then immediately enables gradient computation for all parameters:

  \begin{lstlisting}[language=Python, basicstyle=\small\ttfamily]
  from peft import get_peft_model, LoraConfig
  lora_config = LoraConfig(
      r=128, lora_alpha=256, lora_dropout=0.1,
      bias="none", task_type="CAUSAL_LM",
  )
  model = get_peft_model(model, lora_config)

  for param in model.parameters():
      param.requires_grad = True
  \end{lstlisting}

  This creates an undefined training regime that is neither standard LoRA fine-tuning (where only adapter parameters are trained) nor full fine-tuning (where all parameters of the \emph{original} architecture are trained). Instead, it trains all parameters of a \emph{modified} architecture that includes additional LoRA layers. Notably, the original paper reports results using full fine-tuning without LoRA (Tables 1 and 2), making the presence of this snippet unexplained. This ambiguity makes the reported MIA results difficult to interpret and weakens claims that the results correspond to ``full fine-tuning'' baselines.

  \subsection{Bug 2: Incorrect Model Checkpointing}
  \label{appendix:bug2}

  The training regime inconsistency in Bug 1 causes a cascading failure in model saving. When \texttt{trainer.save\_model()} is called (lines 103 and 351) in \texttt{finetune.py}, the HuggingFace \texttt{SFTTrainer} detects that the model is a \texttt{PeftModel} instance and consequently saves \emph{only the LoRA adapter weights}, not the full model. Since all base model parameters were also trained (due to the \texttt{requires\_grad=True} loop), these trained weights are discarded and never persisted to disk.

  The correct approach for full fine-tuning would be to not wrap the model with LoRA at all. While calling \texttt{model.merge\_and\_unload()} before saving would merge the adapter weights into the base model, this would not resolve the fundamental issue: the model was trained in an undefined state where both LoRA adapters \emph{and} the base model parameters were simultaneously trained. This optimization of redundant parameter sets (the base weights and their low-rank corrections) does not correspond to any standard fine-tuning methodology and produces a model in an ill-defined state. Thus, even after merging, the resulting weights reflect this anomalous training dynamic rather than either pure full fine-tuning or pure LoRA adaptation. Neither a proper fix nor any workaround is implemented in the original codebase.
  
  \subsection{Bug 3: Model Loading Mismatch}
  \label{appendix:bug3}

  In \texttt{main.py} (lines 29--32), where MIA attacks are executed, the model is loaded using standard HuggingFace utilities:

  \begin{lstlisting}[language=Python, basicstyle=\small\ttfamily]
  model = AutoModelForCausalLM.from_pretrained(
      model_name, torch_dtype=torch.float16
  )
  \end{lstlisting}

  This loading mechanism expects a complete model checkpoint. However, due to Bug 2, the saved checkpoint contains only LoRA adapter weights. When \texttt{AutoModelForCausalLM} encounters this, it fails to locate full model weights and instead loads the \textbf{base pretrained model} without applying any fine-tuning adaptations\footnote{\url{https://huggingface.co/docs/transformers/v5.0.0rc2/en/main\_classes/model\#transformers.PreTrainedModel.from\_pretrained}}. No PEFT-specific loading mechanism (e.g., \texttt{PeftModel.from\_pretrained()}) is implemented. We note that even if such PEFT-aware loading were implemented, it would not resolve the underlying issues: the loaded adapter would still reflect the ill-defined training state from Bug 1, and the jointly-trained base model weights would remain lost due to Bug 2.

  \subsection{Impact on Reproducibility}

  The combined effect of these bugs is that all MIA attacks in the evaluation pipeline are executed against the \textbf{original pretrained model} rather than the fine-tuned model. This explains why our initial reproduction attempts yielded near-random-guessing performance for SOFT (AUC $\approx$ 0.50), as the model has no membership-specific information to leak.

  Using the provided codebase with these bugs intact, we were able to reproduce the SOFT defense results reported in their Table 1 and 2 (obtaining near-random-guessing AUC scores). However, we were unable to reproduce the baseline fine-tuning results (approximately 80\% AUC) using the same flawed pipeline, as the evaluation necessarily runs against the pretrained model regardless of whether SOFT's obfuscation is applied. It remains unclear how the original authors obtained their reported baseline results with this implementation.

  To ensure a fair and accurate comparison, we developed our own training and evaluation pipeline with the following corrections:
  \begin{enumerate}
      \item Removed the LoRA wrapper for full fine-tuning experiments
      \item Ensured complete model weights are saved after training
      \item Verified model loading correctly restores fine-tuned weights
  \end{enumerate}

  We retained the original SOFT data preparation and obfuscation methodology to ensure the defense mechanism itself is faithfully reproduced. Specifically, we preserved: (i) their dataset obfuscation pipeline, (ii) the loss-based sample selection strategy that identifies high-risk training samples, and (iii) the dynamic swapping mechanism that replaces original samples with paraphrased counterparts during training. Using this corrected pipeline, we re-trained the models and report our own results in the main tables to ensure correctness and reproducibility. For dataset obfuscation, we adhere to the original SOFT setup and generate paraphrased datasets for SOFT using the \texttt{gpt-4o-mini-2024-07-18} model specified in their repository.

\section{Implementation Details}
\label{app:full_implementation}

\subsection{Prompts and Hyperparameters Used}
\label{app:prompts}
Prompts and hyperparameters used can be found in our code repository in supplementary material or \href{https://github.com/kiraz-ai/sage-sps-mia}{here}\footnote{{\url{https://github.com/kiraz-ai/sage-sps-mia}}}. 

\subsection{Additional Details on SAGE Generation}
\label{app:sage-generation}
We provide here full pseudocode for SAGE, the paraphrasing pipeline used throughout our experiments to generate semantic-preserving variants of training documents. The algorithm formalizes the structure-aware rewriting procedure described in Section~\ref{sec:methodology}.

\begin{algorithm}[t]
\caption{SAGE {\small(Structure-Aware SAE-Guided Extraction)}}
\label{alg:sageapp}
\begin{algorithmic}[1]
\STATE \textbf{Input:} document $x$; max attempts $N$; thresholds $\tau_{\mathrm{sps}},\tau_{\mathrm{ov}}$
\STATE \textbf{Output:} paraphrase $\tilde{x}$

\STATE $u^\star \gets -\infty$; \ $\tilde{x}^\star \gets \emptyset$
\STATE $\text{prompt} \gets$ base prompt
\FOR{$n = 1,2,\ldots,N$}

  \STATE {\small\emph{Note: one LLM call returns an XML document with \textsc{structure} (identical) and
  \textsc{narrative} (paraphrased) sections.}}
  \STATE $\tilde{x}^{(n)} \gets \textsc{Paraphrase}(x,\text{prompt})$

  \STATE $\mathrm{SPS} \gets 0$; $\mathrm{WordSim} \gets 0$; $c \gets 0$
  \FOR{each narrative section $(r_i,\tilde r_i)$ in $\tilde{x}^{(n)}$}
    \STATE $\mathrm{SPS} \gets \mathrm{SPS} + \mathrm{SPS}(r_i,\tilde r_i)$ \hfill {\small(SPS; Def.~\ref{def:sps})}
    \STATE $\mathrm{WordSim} \gets \mathrm{WordSim} + \mathrm{WordSim}(r_i,\tilde r_i)$ \hfill {\small($\mathrm{WordSim}$; Def.~\ref{def:wordsim})}
    \STATE $c \gets c + 1$
  \ENDFOR
  \IF{$c=0$}
    \STATE \textbf{continue}
  \ENDIF
  \STATE $\mathrm{SPS} \gets \mathrm{SPS}/c$; \ $\mathrm{WordSim} \gets \mathrm{WordSim}/c$
  \STATE $u \gets \mathrm{SPS} - \mathrm{WordSim}$

  \IF{$u > u^\star$}
    \STATE $u^\star \gets u$; \ $\tilde{x}^\star \gets \tilde{x}^{(n)}$
  \ENDIF
  \IF{$\mathrm{SPS} \ge \tau_{\mathrm{sps}}$ \AND $\mathrm{WordSim} \le \tau_{\mathrm{ov}}$}
    \STATE \textbf{return} $\tilde{x}^{(n)}$ \hfill {\small(early stop)}
  \ENDIF

  \STATE $\text{prompt} \gets \textsc{Update}(\text{prompt}, \mathrm{SPS}, \mathrm{WordSim}, \tau_{\mathrm{sps}}, \tau_{\mathrm{ov}})$
\ENDFOR
\STATE \textbf{return} $\tilde{x}^\star$ \hfill {\small(fallback: best utility over attempts)}
\end{algorithmic}
\end{algorithm}

Generation proceeds iteratively, as illustrated in Algorithm~\ref{alg:sageapp}. For each input document $x$, SAGE generates a sequence of paraphrasing attempts $\{\tilde{x}^{(n)}\}_{n=1}^N$, where each attempt consists of a single paraphrase produced by a large language model operating under explicit semantic and surface-form constraints. Each paraphrase preserves structural sections verbatim while rewriting narrative sections only.

The structural--narrative decomposition is performed by the same language model used for paraphrasing, following a fixed instruction format. This decomposition relies on deterministic instructions rather than learned classifiers or external heuristics, and the resulting section labels are returned directly by the model as part of the generation output. Structural sections are preserved verbatim, while only narrative sections are eligible for rewriting.

After each paraphrasing attempt, SAGE evaluates the candidate using the semantic persistence score $\mathrm{SPS}(x,\tilde{x}^{(n)})$ and the word similarity score $\mathrm{WordSim}(x,\tilde{x}^{(n)})$, aggregated over narrative sections. If a candidate satisfies the target thresholds $\mathrm{SPS}(x,\tilde{x}) \ge 0.60$ and $\mathrm{WordSim}(x,\tilde{x}) \le 0.35$, the procedure terminates early and returns the candidate. Here, $\mathrm{WordSim}$ is computed over word sets extracted using a simple regex-based tokenizer that retains alphanumeric spans and apostrophes. We emphasize that these thresholds are heuristic choices used solely for early stopping and computational convenience. They should not be interpreted as universal thresholds for semantic equivalence or robustness, nor do they carry any formal guarantees beyond guiding the paraphrase generation process.

If the thresholds are not met, SAGE injects metric-based feedback into the next generation prompt. This feedback explicitly instructs the language model to either increase semantic fidelity or further reduce surface-form overlap, depending on which criterion failed. The process continues for a fixed number of iterations, after which SAGE selects the candidate maximizing the utility $\mathrm{SPS}(x,\tilde{x}) - \mathrm{WordSim}(x,\tilde{x})$. 

Between iterations, prompt updates encode only coarse feedback signals derived from the evaluation metrics, rather than exposing the metrics themselves. This design ensures that the paraphrasing process remains model-agnostic and does not rely on access to the audited model’s internals. The resulting paraphrases therefore reflect realistic, low-cost semantic obfuscations that an accused party could plausibly apply prior to training.

\subsection{SAGE-R: Factual Anchor Identification and Removal}
\label{app:factual-anchor-method}

Both the \emph{SAGE-R} defense and the \emph{FT-F} ablation (see Appendix~\ref{app:facts-ablation}) rely on identifying and removing \emph{factual anchors}, defined as concrete entity mentions and literals that can act as stable lexical hooks for membership inference. These include named entities (e.g., people, organizations, locations), numeric quantities, and date expressions.

\paragraph{LLM-based tagging.}
Factual anchors are first identified using a fixed LLM-prompting procedure. Given the original document and its narrative representation, the model is prompted to extract a structured list of factual items in a strict JSON format. Each item consists of a textual value, a coarse type (e.g., entity, number, date), and optional notes. We use a fixed system prompt and a small set of few-shot examples, with decoding performed at a fixed temperature. To ensure robustness to formatting errors, outputs are validated against a predefined schema and re-prompted if necessary. The code and prompts can be found in Appendix~\ref{app:prompts}.

\paragraph{Deterministic placeholder substitution.}
After factual anchors are identified, replacements are applied using a deterministic, rule-based procedure. Factual values are ordered by first occurrence in the original prose-only text and assigned canonical placeholders of the form \texttt{<<FACT\_1>>}, \texttt{<<FACT\_2>>}, etc. All exact occurrences of each factual value are then replaced using regex-based substitution, ensuring that the same anchor is consistently mapped to the same placeholder within a document. This step is entirely deterministic and does not involve additional model calls. We emphasize that this replacement is used solely for analysis and statistical control; during fine-tuning, the corresponding placeholder tokens are masked to prevent the model from learning or exploiting these anchors.

Qualitative examples illustrating factual anchor identification and replacement are provided in Appendix~\ref{app:qualitative-sage}.

\subsection{LLM-as-a-Judge Utility Evaluation}
\label{app:llm-judge}

The degradation of membership inference attacks under paraphrasing does not imply that fine-tuned models lose or forget the underlying training knowledge. To verify that models retain usable semantic and conceptual information, we evaluate downstream utility using an \emph{LLM-as-a-Judge} framework~\cite{zheng2023judging}, consistent with prior work on model evaluation beyond perplexity~\cite{DBLP:conf/uss/00020GCSA0FK0L25}. This approach replaces surface-level likelihood metrics with a semantic assessment of model outputs.

\paragraph{Question Generation.}
For each dataset, we randomly sample documents from the fine-tuning split of MIMIR (we use the \texttt{ngram\_13\_0.8} subset and the \texttt{member} field). We prompt GPT-4o-mini~\cite{openai2023gpt} to generate a fixed set of evaluation questions in structured JSON, using only the first 4{,}000 characters of each sampled document to control context length. Questions are designed to be conceptual and self-contained (e.g., asking for explanations, implications, or high-level relationships), and the prompt discourages reliance on verbatim identifiers (names, exact phrasing, or brittle constants) so that the evaluation targets semantic competence rather than string memorization. We include robust parsing and retry logic to handle occasional malformed JSON generations.

\paragraph{Question Generation and Coverage.}
To probe different notions of training-data familiarity, we generate a \emph{comprehensive} set of evaluation questions using GPT-4o-mini, conditioned on randomly sampled documents from each dataset. Question generation is prompt-driven and dataset-aware, with explicit constraints that discourage generic domain knowledge and instead target signals plausibly induced by exposure during training. Depending on the dataset, questions may include short excerpts (2–3 sentences) from the source text to ensure answerability without revealing verbatim identifiers. All questions are generated once and reused across models to ensure comparability.

\subsection{Question Set Composition}
\label{app:llm-judge-questions}

For each dataset, we generate a fixed set of \textbf{325 evaluation questions} to probe different forms of training-data familiarity beyond brittle surface memorization. All questions are generated once and reused across models to ensure controlled comparisons. The question set combines multiple complementary types, summarized as follows:

\begin{itemize}
    \item \textbf{Standard specific-content questions (50)}: Target concrete methods, findings, or claims unique to a document and are designed to fail under purely general domain knowledge.
    \item \textbf{Closed-book topic questions (25)}: Probe semantic topic recognition and method awareness without relying on exact values or formulas, ensuring compatibility with SAGE-R.
    \item \textbf{Tiered questions (60 total)}: Comprising 20 easy (topic recognition), 20 medium (core semantic content), and 20 hard (contextual understanding) questions to separate pretrained and fine-tuned behavior across difficulty regimes.
    \item \textbf{Contrastive pairs (25 pairs / 50 questions)}: Matched in-domain and out-of-domain questions with identical structure, testing whether a model discriminates between training-related and unrelated content.
    \item \textbf{Cloze tasks (40)}: Fill-in-the-blank questions over key semantic concepts, enabling objective evaluation independent of LLM judges.
    \item \textbf{Dataset-specific topic questions (100)}: Tailored to corpus characteristics (e.g., ArXiv field classification, Wikipedia article type, HackerNews discussion patterns), focusing on domain and style recognition rather than factual recall.
    \item \textbf{Passage completion tasks (30)}: Evaluate stylistic and register learning by asking models to continue a passage in a manner consistent with the source corpus.
\end{itemize}

\paragraph{Response Evaluation.}
Given a question, the subject model generates an answer with stochastic decoding (temperature $0.7$, up to 512 new tokens) to reflect typical usage rather than deterministic test-time decoding. Each answer is then graded by an LLM judge against a reference answer derived from the source document. Concretely, we use a panel of judges (GPT-4o-mini~\cite{openai2023gpt}, DeepSeek-V3.2~\cite{deepseek2024}, and Grok-4.1-fast~\cite{xai2025grok4}) and average their scores to reduce single-judge idiosyncrasies; judges output a scalar score (on a $0$--$100$ scale, normalized to $[0,1]$) with deterministic decoding (temperature $0.0$). To further mitigate run-to-run variance, we repeat the judging process three times per question and average the resulting scores. The final utility score for each model is the mean across all questions after these averages. Prompts and generated question sets are provided in Appendix~\ref{app:prompts}.

\paragraph{Implications for Semantic Equivalence.}
The LLM-as-a-Judge results complement SPS by validating semantic preservation at the level of functional utility. High SPS values demonstrate representational similarity between original and paraphrased texts, while the judge-based evaluation confirms that this similarity translates into preserved downstream performance. Together, these metrics operationalize semantic equivalence in a manner aligned with the requirements of the judge--prosecutor--accused protocol: transformations that preserve meaning and utility cannot be distinguished from the original copyrighted material, even when membership signals are eliminated.

\paragraph{Why can SAGE and SAGE-R outperform standard fine-tuning?}
Interestingly, SAGE, and in some cases SAGE-R achieves higher LLM-as-a-Judge scores than standard fine-tuning (Table~\ref{tab:utility-judge-lora}). We attribute this effect not to increased memorization, but to a regularization-like benefit induced by semantic-preserving paraphrasing. By rewriting training data while preserving meaning, SAGE reduces lexical correlations and discourages the model from overfitting to surface-level phrasing. This encourages representations that are more semantically grounded and better aligned with downstream prompting, which can improve judged utility.

This effect is most pronounced in structured or technical domains such as ArXiv and PubMed, where paraphrasing promotes abstraction over formulaic repetition. In contrast, gains are smaller or absent in noisier corpora such as HackerNews, where limited semantic structure constrains the benefits of paraphrasing. Overall, these results suggest that SAGE can act as a form of data-level regularization that improves semantic generalization without degrading utility.

Importantly, these effects do not impact our conclusions: utility comparisons are only interpreted within each dataset, and our core claim is the absence of systematic utility degradation under semantic-preserving paraphrasing, rather than absolute performance improvements.

\section{Ablation and Analysis on SPS \& WordSim}
\label{app:sps-anal}
\subsection{Paraphrase Quality}
\label{app:paraphrase-quality}
Table~\ref{tab:paraphrase_by_model_multicolumn} report paraphrase quality across datasets and paraphrasing models for SAGE and SAGE-R. Across all settings, SAGE consistently achieves higher $\mathrm{SPS}$ than SAGE-R, indicating stronger semantic persistence when structural and factual anchors are preserved. This gap is visible across domains and paraphrasers, including technical corpora (e.g., \texttt{arxiv}, \texttt{pubmed}) as well as more informal data (e.g., \texttt{hackernews}).

At the same time, absolute $\mathrm{SPS}$ values vary substantially across datasets and paraphrasing models. For instance, technical datasets exhibit higher $\mathrm{SPS}$ overall, reflecting the presence of stable factual anchors, while informal domains show lower and more variable scores. No single paraphraser uniformly dominates across datasets, and differences between models (e.g., DeepSeek v3.2~\cite{deepseek2024} vs.\ Gemini 2.5 Flash~\cite{gemini2025}) are often comparable in magnitude to differences induced by the dataset itself.

Importantly, all configurations maintain low $\mathrm{WordSim}$ values, confirming that semantic preservation is achieved alongside substantial surface-level rewriting. Taken together, these results indicate that while $\mathrm{SPS}$ reliably distinguishes between semantically richer (SAGE) and more aggressively obfuscated (SAGE-R) paraphrases, its absolute scale is dataset- and model-dependent, motivating a more detailed analysis of how $\mathrm{SPS}$ behaves across semantic observers.
\begin{table*}[htbp]
\centering
\small
\caption{\textbf{SAGE vs. SAGE-R:} Complete paraphrase quality metrics of datasets generated by each model. Higher $\mathrm{SPS}$ indicates better semantic preservation; lower $\mathrm{WordSim}$ indicates greater surface divergence.}
\begin{tabular}{llcccc}
\toprule
 &  & \multicolumn{2}{c}{\textbf{SAGE}} & \multicolumn{2}{c}{\textbf{SAGE-R}} \\
\cmidrule(lr){3-4} \cmidrule(lr){5-6}
Dataset & Model & $\mathrm{SPS}$ $\uparrow$ & $\mathrm{WordSim}$ $\downarrow$ & $\mathrm{SPS}$ $\uparrow$ & $\mathrm{WordSim}$ $\downarrow$ \\
\midrule

\multirow{3}{*}{\textbf{\texttt{arxiv}}}
 & DeepSeek v3.2     & 0.761 & 0.329 & 0.677 & 0.271 \\
 & Gemini 2.5 Flash  & 0.699 & 0.360 & 0.651 & 0.318 \\
 & Grok 4.1 Fast     & 0.703 & 0.332 & 0.644 & 0.278 \\
\midrule

\multirow{3}{*}{\textbf{\texttt{wikipedia}}}
 & DeepSeek v3.2     & 0.697 & 0.340 & 0.504 & 0.211 \\
 & Gemini 2.5 Flash  & 0.672 & 0.375 & 0.492 & 0.250 \\
 & Grok 4.1 Fast     & 0.692 & 0.350 & 0.488 & 0.201 \\
\midrule

\multirow{3}{*}{\textbf{\texttt{pile\_cc}}}
 & DeepSeek v3.2     & 0.652 & 0.230 & 0.556 & 0.181 \\
 & Gemini 2.5 Flash  & 0.576 & 0.241 & 0.499 & 0.195 \\
 & Grok 4.1 Fast     & 0.675 & 0.253 & 0.567 & 0.190 \\
\midrule

\multirow{3}{*}{\textbf{\texttt{pubmed}}}
 & DeepSeek v3.2     & 0.767 & 0.296 & 0.602 & 0.242 \\
 & Gemini 2.5 Flash  & 0.666 & 0.329 & 0.597 & 0.291 \\
 & Grok 4.1 Fast     & 0.753 & 0.306 & 0.631 & 0.246 \\
\midrule

\multirow{3}{*}{\textbf{\texttt{hackernews}}}
 & DeepSeek v3.2     & 0.555 & 0.195 & 0.501 & 0.165 \\
 & Gemini 2.5 Flash  & 0.528 & 0.187 & 0.488 & 0.162 \\
 & Grok 4.1 Fast     & 0.638 & 0.226 & 0.565 & 0.190 \\
\bottomrule
\end{tabular}
\label{tab:paraphrase_by_model_multicolumn}
\end{table*}

\subsection{SAE Implementation Details \& Ablation}
\label{app:sae_ablation}

We next analyze how the $\mathrm{SPS}$ depends on the choice of semantic observer by varying both the probe model and the SAE layer used for feature extraction. Tables~\ref{tab:ablation_merged_gemma}, \ref{tab:ablation_merged_pythia}, and~\ref{tab:ablation_merged_gpt2} report $\mathrm{SPS}$ under SAGE and SAGE-R for three representative probe models (Gemma-2B~\cite{team2024gemma}, Pythia-70M-deduped~\cite{pythia2023}, and GPT-2 Small~\cite{radford2019language}), evaluated at multiple transformer layers. All results are averaged across paraphrasing models to isolate the effect of the semantic representation itself.

\paragraph{No universal SPS threshold.}
Across all probe models and layers, we observe substantial variation in absolute $\mathrm{SPS}$ values. Earlier layers tend to yield lower scores, while deeper layers often produce higher $\mathrm{SPS}$, reflecting increasing semantic abstraction in the underlying representations. Moreover, the scale of $\mathrm{SPS}$ differs markedly between probe models (e.g., Gemma-2B vs.\ Pythia-70M), indicating that $\mathrm{SPS}$ is inherently tied to the robustness and expressivity of the semantic observer. As a result, there is no single, model-independent threshold that cleanly separates ``semantic preservation'' from ``semantic drift.''

This observation suggests that $\mathrm{SPS}$ should be interpreted relatively rather than absolutely: either calibrated to the specific probe model and layer, or aggregated (e.g., averaged) across multiple observers to mitigate idiosyncrasies of any single representation.

\paragraph{Consistent separation between SAGE and SAGE-R.}
Despite this variability, there is a consistent pattern across all ablations: SAGE achieves higher $\mathrm{SPS}$ than SAGE-R for every dataset, probe model, and layer considered. This gap persists even when absolute $\mathrm{SPS}$ values fluctuate substantially, and remains visible in both shallow and deep representations. In particular, SAGE-R exhibits a systematic reduction in $\mathrm{SPS}$ relative to SAGE, reflecting the deliberate removal of structural and factual anchors that contribute to stable semantic representations.

\paragraph{Implications for SPS as a semantic signal.}
Taken together, these results validate $\mathrm{SPS}$ as a meaningful measure of semantic persistence rather than a similarity heuristic. While $\mathrm{SPS}$ does not admit a universal threshold applicable across models, it robustly captures relative semantic degradation induced by stronger obfuscation. The fact that the SAGE vs.\ SAGE-R ordering is invariant to the choice of probe model and layer indicates that $\mathrm{SPS}$ reflects a genuine semantic property shared across representations, rather than an artifact of a particular SAE or architecture.

\begin{table}[htbp]
\centering
\caption{Ablation (SAGE vs SAGE-R): SPS by SAE layer for the Gemma-2B probe model at hook blocks.Lk.hook\_resid\_post. Here Lk denotes the transformer block index used for feature extraction; higher SPS indicates better semantic preservation. Results are averaged over paraphraser models (DeepSeek v3.2, Gemini 2.5 Flash, Grok 4.1 Fast).}
\begin{tabular}{lcccccc}
\toprule
\textbf{Dataset} & \multicolumn{2}{c}{\textbf{SPS L6 $\uparrow$}} & \multicolumn{2}{c}{\textbf{SPS L12 $\uparrow$}} & \multicolumn{2}{c}{\textbf{SPS L17 $\uparrow$}}\\
\cmidrule(lr){2-3}\cmidrule(lr){4-5}\cmidrule(lr){6-7}
 & \textbf{SAGE} & \textbf{SAGE-R} & \textbf{SAGE} & \textbf{SAGE-R} & \textbf{SAGE} & \textbf{SAGE-R}\\
\midrule
arXiv & 0.754 & 0.679 & 0.721 & 0.657 & 0.812 & 0.775\\
Wikipedia & 0.749 & 0.514 & 0.687 & 0.495 & 0.764 & 0.599\\
Pile-CC & 0.687 & 0.590 & 0.634 & 0.541 & 0.683 & 0.580\\
PubMed & 0.772 & 0.626 & 0.729 & 0.610 & 0.797 & 0.733\\
HackerNews & 0.644 & 0.589 & 0.574 & 0.518 & 0.598 & 0.519\\
\midrule
\textbf{Average} & 0.721 & 0.600 & 0.669 & 0.564 & 0.731 & 0.641\\
\bottomrule
\end{tabular}

\label{tab:ablation_merged_gemma}
\end{table}

\begin{table}[htbp]
\centering
\caption{Ablation (SAGE vs SAGE-R): SPS by SAE layer for the Pythia-70M-deduped probe model at hook blocks.Lk.hook\_resid\_post. Here Lk denotes the transformer block index used for feature extraction; higher SPS indicates better semantic preservation. Results are averaged over paraphraser models (DeepSeek v3.2, Gemini 2.5 Flash, Grok 4.1 Fast).}
\begin{tabular}{lcccc}
\toprule
\textbf{Dataset} & \multicolumn{2}{c}{\textbf{SPS L3 $\uparrow$}} & \multicolumn{2}{c}{\textbf{SPS L4 $\uparrow$}}\\
\cmidrule(lr){2-3}\cmidrule(lr){4-5}
 & \textbf{SAGE} & \textbf{SAGE-R} & \textbf{SAGE} & \textbf{SAGE-R}\\
\midrule
arXiv & 0.658 & 0.564 & 0.655 & 0.568\\
Wikipedia & 0.622 & 0.363 & 0.631 & 0.386\\
Pile-CC & 0.584 & 0.477 & 0.587 & 0.473\\
PubMed & 0.693 & 0.448 & 0.694 & 0.468\\
HackerNews & 0.571 & 0.514 & 0.570 & 0.503\\
\midrule
\textbf{Average} & 0.626 & 0.473 & 0.628 & 0.480\\
\bottomrule
\end{tabular}

\label{tab:ablation_merged_pythia}
\end{table}
\begin{table}[htbp]
\centering
\caption{Ablation (SAGE vs SAGE-R): SPS by SAE layer for the GPT-2 Small probe model at hook blocks.Lk.hook\_resid\_pre. Here Lk denotes the transformer block index used for feature extraction; higher SPS indicates better semantic preservation. Results are averaged over paraphraser models (DeepSeek v3.2, Gemini 2.5 Flash, Grok 4.1 Fast).}
\begin{tabular}{lcccccc}
\toprule
\textbf{Dataset} & \multicolumn{2}{c}{\textbf{SPS L6 $\uparrow$}} & \multicolumn{2}{c}{\textbf{SPS L8 $\uparrow$}} & \multicolumn{2}{c}{\textbf{SPS L10 $\uparrow$}}\\
\cmidrule(lr){2-3}\cmidrule(lr){4-5}\cmidrule(lr){6-7}
 & \textbf{SAGE} & \textbf{SAGE-R} & \textbf{SAGE} & \textbf{SAGE-R} & \textbf{SAGE} & \textbf{SAGE-R}\\
\midrule
arXiv & 0.672 & 0.562 & 0.675 & 0.582 & 0.685 & 0.622\\
Wikipedia & 0.637 & 0.348 & 0.651 & 0.378 & 0.664 & 0.419\\
Pile-CC & 0.559 & 0.456 & 0.555 & 0.450 & 0.559 & 0.449\\
PubMed & 0.732 & 0.521 & 0.727 & 0.553 & 0.735 & 0.630\\
HackerNews & 0.529 & 0.472 & 0.507 & 0.449 & 0.504 & 0.440\\
\midrule
\textbf{Average} & 0.626 & 0.472 & 0.623 & 0.483 & 0.629 & 0.512\\
\bottomrule
\end{tabular}

\label{tab:ablation_merged_gpt2}
\end{table}

\section{Qualitative Examples of SAGE variants}
\label{app:qualitative-sage}
Tables~\ref{tab:qual:qualitative-examples1}--\ref{tab:qual:qualitative-examples4} present qualitative examples of different SAGE variants, illustrating their behavior across representative samples.

\begin{table}[htbp]
\centering
\setlength{\tabcolsep}{5pt}
\renewcommand{\arraystretch}{1.2}
\footnotesize
\begin{tabular}{@{}>{\raggedright\arraybackslash}p{0.32\textwidth}>{\raggedright\arraybackslash}p{0.32\textwidth}>{\raggedright\arraybackslash}p{0.32\textwidth}@{}}
\toprule
\textbf{Original} & \textbf{SAGE} & \textbf{SAGE-R}\\
\midrule
\scriptsize\emph{arxiv / deepseek\_v3.2 / sample\_idx=70}\\--- abstract: | The families of braid groups, surface braid groups, mapping class groups and loop braid groups have a representation theory of “wild type”, so it is very useful to be able to construct such representations *topologically*, so that they may be understood by topological or geometric methods. For the braid groups \$\textbackslash\{\}mathbf\{B\}\_n\$, Lawrence and Bigelow have constructed families of representations starting from actions of \$\textbackslash\{\}mathbf\{B\}\_n\$ on the twisted homology of configuration spaces. These were then used by Bigelow and Krammer to prove that the braid groups are linear. We develop a general underlying procedure to build homological representations of families of groups, encompass... & --- abstract: | The representation theory for groups such as braid groups, surface braid groups, mapping class groups, and loop braid groups is of a "wild type," making topological constructions particularly valuable. Such topological representations facilitate comprehension through topological and geometric frameworks. In the case of the braid groups \$\textbackslash\{\}mathbf\{B\}\_n\$, Lawrence and Bigelow developed a series of representations. Their method originated from actions of \$\textbackslash\{\}mathbf\{B\}\_n\$ on the twisted homology groups of configuration spaces. Subsequently, Bigelow and Krammer utilized these constructions to establish the linearity of the braid groups. We introduce a universal, foundational proced... & The representation theory for groups such as braid groups, surface braid groups, mapping class groups, and loop braid groups is of a "\textless{}\textless{}FACT\_1\textgreater{}\textgreater{}," making topological constructions particularly valuable. Such topological representations facilitate comprehension through topological and geometric frameworks. In the case of the braid groups \textless{}\textless{}FACT\_2\textgreater{}\textgreater{}, \textless{}\textless{}FACT\_3\textgreater{}\textgreater{} and \textless{}\textless{}FACT\_4\textgreater{}\textgreater{} developed a series of representations. Their method originated from actions of \textless{}\textless{}FACT\_2\textgreater{}\textgreater{} on the twisted homology groups of configuration spaces. Subsequently, \textless{}\textless{}FACT\_4\textgreater{}\textgreater{} and \textless{}\textless{}FACT\_5\textgreater{}\textgreater{} utilized these constructions to establish the linearity of the braid groups. We introduce a universal, foundational procedure for constru...\\
\midrule
\scriptsize\emph{hackernews / grok\_4.1\_fast / sample\_idx=91}\\America’s Highest Minimum Wage Sparks Fight in Small California City - JumpCrisscross https://www.wsj.com/articles/americas-highest-minimum-wage-sparks-fight-in-small-california-city-11563960603?mod=rsswn ====== vgoh1 Typical WSJ rubbish. Focus on anecdotes, like the guy who sacks over half of his staff because of a \textless{} 15\% increase in minimum wage. Having grown up solidly in the class of the "working poor", I can tell you a side of the story that is not often told - there is not enough difference in quality of life between "working poor" and just being on public assistance and unemployed. When you are unemployed, you get government benefits, and most people either have a side hustle (odd j... & America’s Highest Minimum Wage Sparks Fight in Small California City - JumpCrisscross https://www.wsj.com/articles/americas-highest-minimum-wage-sparks-fight-in-small-california-city-11563960603?mod=rsswn ====== vgoh1 Standard WSJ nonsense. It highlights anecdotes, such as the man who lays off more than half his employees due to a minimum wage hike under 15\%. Raised firmly within the "working poor" class, I can share a seldom-discussed aspect: the quality-of-life disparity between the "working poor" and those unemployed on public assistance is quite small. The unemployed receive government benefits, with most engaging in side hustles—such as under-the-table odd jobs or occasionally outrig... & Standard \textless{}\textless{}FACT\_1\textgreater{}\textgreater{} nonsense. It highlights anecdotes, such as the man who lays off more than half his employees due to a minimum wage hike under \textless{}\textless{}FACT\_2\textgreater{}\textgreater{}. Raised firmly within the "working poor" class, I can share a seldom-discussed aspect: the quality-of-life disparity between the "working poor" and those unemployed on public assistance is quite small. The unemployed receive government benefits, with most engaging in side hustles—such as under-the-table odd jobs or occasionally outright illegal work—or perpetrating fraud, like claiming solo residency while actually sharing rent among several people. No judgment here; that's simply how it is. Thus, minimum-wage workers...\\
\midrule
\scriptsize\emph{pubmed\_central / gemini\_2.5\_flash / sample\_idx=405}\\Introduction \{\#Sec1\} ============ While meditation is often viewed and employed as a technique to reduce stress (Chiesa and Seretti [@CR8]), its potential to increase cognitive abilities has been emphasized from its beginnings (e.g., Luk [@CR33]). For instance, some Buddhist meditation techniques explicitly and intentionally target the training and improvement of concentration (manasikara) and insight (vipassanā; see Santucci [@CR45]), which tap into attentional control and higher-order cognition that according to Lutz et al. ([@CR34]) are systematically affected and improved by meditation. These possible links to cognitive improvement have attracted the attention of researchers, who inde... & Introduction \{\#Sec1\} ============ Meditation is frequently perceived and utilized as a method for stress reduction (Chiesa and Seretti [@CR8]). Nevertheless, its capacity to enhance cognitive functions has been highlighted since its earliest discussions (e.g., Luk [@CR33]). For example, certain Buddhist meditation practices, as noted by Santucci [@CR45], are specifically designed to cultivate and refine concentration (manasikara) and insight (vipassanā). These practices engage attentional control and higher-order cognition, which, according to Lutz et al. ([@CR34]), are systematically influenced and improved through meditation. The potential connections to cognitive enhancement have garne... & Meditation is frequently perceived and utilized as a method for stress reduction (\textless{}\textless{}FACT\_1\textgreater{}\textgreater{}). Nevertheless, its capacity to enhance cognitive functions has been highlighted since its earliest discussions (e.g., \textless{}\textless{}FACT\_2\textgreater{}\textgreater{}). For example, certain Buddhist meditation practices, as noted by \textless{}\textless{}FACT\_3\textgreater{}\textgreater{}, are specifically designed to cultivate and refine concentration (manasikara) and insight (vipassanā). These practices engage attentional control and higher-order cognition, which, according to \textless{}\textless{}FACT\_4\textgreater{}\textgreater{}, are systematically influenced and improved through meditation. The potential connections to cognitive enhancement have garnered considerable interest from the research community. Studies have in...\\
\bottomrule
\end{tabular}
\caption{Qualitative examples (truncated to 700 chars).}
\label{tab:qual:qualitative-examples1}
\end{table}

\begin{table}[htbp]
\centering
\setlength{\tabcolsep}{5pt}
\renewcommand{\arraystretch}{1.2}
\footnotesize
\begin{tabular}{@{}>{\raggedright\arraybackslash}p{0.32\textwidth}>{\raggedright\arraybackslash}p{0.32\textwidth}>{\raggedright\arraybackslash}p{0.32\textwidth}@{}}
\toprule
\textbf{Original} & \textbf{SAGE} & \textbf{SAGE-R}\\
\midrule
\scriptsize\emph{hackernews / grok\_4.1\_fast / sample\_idx=97}\\Homestead High School Newspaper, May 22, 1977 - timdierks http://imgur.com/st7eyiR ====== nealabq The paper was, and still is, called The Epitaph. Chris at the time was also one of the publishers/writers of an "underground" school paper called The Road Apple. ------ DrScump That McDonald's still exists in the same location, but I bet they won't accept the coupon anyway. \textasciitilde{}\textasciitilde{}\textasciitilde{} pikachu\_is\_cool Especially since it expires on May 29, 1XXX \textasciitilde{}\textasciitilde{}\textasciitilde{} DrScump Theoretically, it \_could\_ be A.D. 10,000+ ------ ashwinl Part of the ethos back then too: "Apple Computer Inc. will explain how to use any of the equipment you have or give demonstrations before you buy one" ------ coldcode If you invested the pric... & Homestead High School Newspaper, May 22, 1977 - timdierks http://imgur.com/st7eyiR ====== nealabq The newspaper was known as, and continues to be named, The Epitaph. ------ DrScump That McDonald's location remains operational to this day, though they probably won't honor the coupon regardless. \textasciitilde{}\textasciitilde{}\textasciitilde{} pikachu\_is\_cool Particularly given that it expires on May 29, 1XXX. \textasciitilde{}\textasciitilde{}\textasciitilde{} DrScump In theory, it \_could\_ refer to A.D. 10,000+. ------ ashwinl That era's philosophy included: "Apple Computer Inc. will explain how to use any of the equipment you have or give demonstrations before you buy one". ------ coldcode Had you invested the burger's cost (shortly after their IPO) and retained it until now, you... & The newspaper was known as, and continues to be named, \textless{}\textless{}FACT\_1\textgreater{}\textgreater{}. That \textless{}\textless{}FACT\_2\textgreater{}\textgreater{} location remains operational to this day, though they probably won't honor the coupon regardless. Particularly given that it expires on \textless{}\textless{}FACT\_3\textgreater{}\textgreater{}. In theory, it \_could\_ refer to \textless{}\textless{}FACT\_4\textgreater{}\textgreater{}. That era's philosophy included: "\textless{}\textless{}FACT\_5\textgreater{}\textgreater{} will explain how to use any of the equipment you have or give demonstrations before you buy one". Had you invested the burger's cost (shortly after their IPO) and retained it until now, you might afford an entire franchise. I wonder what became of those two \textless{}\textless{}FACT\_6\textgreater{}\textgreater{} units shipped to \textless{}\textless{}FACT\_7\textgreater{}\textgreater{}? Uncertain about those specific units, but an \textless{}\textless{}FACT\_6\textgreater{}\textgreater{} was still operatio...\\
\midrule
\scriptsize\emph{arxiv / grok\_4.1\_fast / sample\_idx=170}\\--- abstract: 'We review recent findings that the universe on its largest scales shows hints of the violation of statistical isotropy, in particular alignment with the geometry and direction of motion of the solar system, and missing power at scales greater than 60 degrees. We present the evidence, attempts to explain it using astrophysical, cosmological or instrumental mechanisms, and prospects for future understanding.' address: | \$\textasciicircum{}1\$Kavli Institute for Cosmological Physics and Department of Astronomy and Astrophysics, University of Chicago, Chicago, IL 60637\textbackslash\{\} E-mail: dhuterer@kicp.uchicago.edu author: - Dragan Huterer\$\textasciicircum{}1\$ title: 'Mysteries on Universe’s Largest Observable Scales' ---... & --- abstract: ' Recent findings indicate that the universe, on its largest scales, exhibits signs of statistical isotropy violation—specifically, alignments with the solar system's geometry and motion direction, along with reduced power at scales exceeding 60 degrees. This review outlines the supporting evidence, explanatory efforts via astrophysical, cosmological, or instrumental causes, and future investigative prospects. ' address: | \$\textasciicircum{}1\$Kavli Institute for Cosmological Physics and Department of Astronomy and Astrophysics, University of Chicago, Chicago, IL 60637\textbackslash\{\} E-mail: dhuterer@kicp.uchicago.edu author: - Dragan Huterer\$\textasciicircum{}1\$ title: 'Mysteries on Universe’s Largest Observable Scales'... & Recent findings indicate that the universe, on its largest scales, exhibits signs of statistical isotropy violation—specifically, alignments with the solar system's geometry and motion direction, along with reduced power at scales exceeding \textless{}\textless{}FACT\_1\textgreater{}\textgreater{}. This review outlines the supporting evidence, explanatory efforts via astrophysical, cosmological, or instrumental causes, and future investigative prospects. The cosmological principle asserts that the universe appears homogeneous and isotropic at its largest scales. Typically presented at the start of cosmology courses, this principle forms an essential foundation for deriving key outcomes in quantitative cosmology. For instance, by invoking...\\
\bottomrule
\end{tabular}
\caption{Qualitative examples (truncated to 700 chars).}
\label{tab:qual:qualitative-examples2}
\end{table}
\begin{table}[htbp]
\centering
\setlength{\tabcolsep}{5pt}
\renewcommand{\arraystretch}{1.2}
\footnotesize
\begin{tabular}{@{}>{\raggedright\arraybackslash}p{0.32\textwidth}>{\raggedright\arraybackslash}p{0.32\textwidth}>{\raggedright\arraybackslash}p{0.32\textwidth}@{}}
\toprule
\textbf{Original} & \textbf{SAGE} & \textbf{SAGE-R}\\
\midrule
\scriptsize\emph{pubmed\_central / grok\_4.1\_fast / sample\_idx=29}\\INTRODUCTION ============ Exclusive breastfeeding (EBF) practice is influenced by many factors. Up to the age of 6 months it is positively associated with rural residence, Malay speaking, non-working, smoking and multiparous mothers, term infants, having a supportive husband and the mother practicing bed-sharing \textbackslash\{\}[[@B1]\textbackslash\{\}]. The working status of mothers influences EBF practice and a favorable working environment and support in the workplace promotes EBF practice \textbackslash\{\}[[@B2]\textbackslash\{\}]. EBF is usually high during maternity leave in Indonesia and a national health survey found that the prevalence was 40\% in 2003 \textbackslash\{\}[[@B3]\textbackslash\{\}], before falling to 32\% in 2012 \textbackslash\{\}[[@B4]\textbackslash\{\}]. The increasing number of working mothers... & INTRODUCTION ============ Numerous factors affect the practice of exclusive breastfeeding (EBF). For infants up to 6 months old, EBF shows positive links to rural living, Malay-speaking backgrounds, non-working status, smoking habits and multiparity among mothers, term births, supportive husbands, and bed-sharing by the mother \textbackslash\{\}[[@B1]\textbackslash\{\}]. Mothers' employment status impacts EBF adherence, with supportive workplace conditions and assistance enhancing EBF continuation \textbackslash\{\}[[@B2]\textbackslash\{\}]. In Indonesia, EBF rates tend to peak during maternity leave periods; a national survey reported 40\% prevalence in 2003 \textbackslash\{\}[[@B3]\textbackslash\{\}], dropping to 32\% by 2012 \textbackslash\{\}[[@B4]\textbackslash\{\}]. This decline is attributed to rising numbers of work... & Numerous factors affect the practice of exclusive breastfeeding (EBF). For infants up to \textless{}\textless{}FACT\_1\textgreater{}\textgreater{} old, EBF shows positive links to rural living, Malay-speaking backgrounds, non-working status, smoking habits and multiparity among mothers, term births, supportive husbands, and bed-sharing by the mother \textless{}\textless{}FACT\_2\textgreater{}\textgreater{}. Mothers' employment status impacts EBF adherence, with supportive workplace conditions and assistance enhancing EBF continuation \textless{}\textless{}FACT\_3\textgreater{}\textgreater{}. In Indonesia, EBF rates tend to peak during maternity leave periods; a national survey reported \textless{}\textless{}FACT\_4\textgreater{}\textgreater{} prevalence in \textless{}\textless{}FACT\_5\textgreater{}\textgreater{} \textless{}\textless{}FACT\_6\textgreater{}\textgreater{}, dropping to \textless{}\textless{}FACT\_7\textgreater{}\textgreater{} by \textless{}\textless{}FACT\_8\textgreater{}\textgreater{} \textless{}\textless{}FACT\_9\textgreater{}\textgreater{}. This decline is attributed to rising numbers of w...\\
\midrule
\scriptsize\emph{hackernews / deepseek\_v3.2 / sample\_idx=972}\\Lawmakers Prod FCC to Act on SIM Swapping - mikece https://krebsonsecurity.com/2020/01/senators-prod-fcc-to-act-on-sim-swapping/ ====== mrandish This seems like it's already a serious and growing problem. It might be best to take a two-pronged approach of quickly implementing process requirements that make such swaps harder for crooks while working to come up with a longer- term technological solution. It's crazy that a minimum-wage retail mobile store employee is the weak link in a system that protects almost everyone's financial assets from banking and investment accounts to credit cards and crypto-currencies. \textasciitilde{}\textasciitilde{}\textasciitilde{} throwawayatt I purchased an unlocked iPhone recently from Apple and decid... & Lawmakers Prod FCC to Act on SIM Swapping - mikece https://krebsonsecurity.com/2020/01/senators-prod-fcc-to-act-on-sim-swapping/ ====== mrandish The issue of SIM swapping appears to be a significant and expanding threat. An effective strategy could involve a dual effort: first, to rapidly establish procedural mandates that increase the difficulty for malicious actors to perform these swaps, and second, to concurrently develop a more robust, permanent technological fix. It is concerning that a low-paid retail mobile store clerk serves as the critical vulnerability in a security framework safeguarding the vast majority of financial holdings, spanning from traditional bank and investment acc... & The issue of \textless{}\textless{}FACT\_1\textgreater{}\textgreater{} swapping appears to be a significant and expanding threat. An effective strategy could involve a dual effort: first, to rapidly establish procedural mandates that increase the difficulty for malicious actors to perform these swaps, and second, to concurrently develop a more robust, permanent technological fix. It is concerning that a low-paid retail mobile store clerk serves as the critical vulnerability in a security framework safeguarding the vast majority of financial holdings, spanning from traditional bank and investment accounts to credit lines and digital currencies. After buying an unlocked \textless{}\textless{}FACT\_2\textgreater{}\textgreater{} directly from \textless{}\textless{}FACT\_3\textgreater{}\textgreater{}, I opted to transfer my \textless{}\textless{}FACT\_4\textgreater{}\textgreater{}...\\
\bottomrule
\end{tabular}
\caption{Qualitative examples (truncated to 700 chars).}
\label{tab:qual:qualitative-examples3}
\end{table}

\begin{table}[htbp]
\centering
\setlength{\tabcolsep}{5pt}
\renewcommand{\arraystretch}{1.2}
\footnotesize
\begin{tabular}{@{}>{\raggedright\arraybackslash}p{0.32\textwidth}>{\raggedright\arraybackslash}p{0.32\textwidth}>{\raggedright\arraybackslash}p{0.32\textwidth}@{}}
\toprule
\textbf{Original} & \textbf{SAGE} & \textbf{SAGE-R}\\
\midrule
\scriptsize\emph{pile\_cc / grok\_4.1\_fast / sample\_idx=448}\\How We Work How We Work - It's As Easy As 1-2-3 ! - " really pleased how quickly they found experienced candidates for us. Very happy !” REM Here's how we can help you... *Don't have a Job or Person Specification? No problem, we'll send you our quick-to-complete, one-page template. Once you've answered the questions, you (and we) will have a very clear understanding of what you're looking for. *Not sure how much you should pay your new employee? We'll research the market and advise you. * Skill shortages? In theses times of full employment and declining immigration from Europe, skill shortages are becoming an ever more serious problem. As an employer you need to be able to demonstrate cle... & How We Work How We Work - It's As Easy As 1-2-3 ! - "Really pleased with how quickly they identified experienced candidates for us. Very happy!" REM Here's how we can assist you... *No Job or Person Specification available? That's fine—we'll provide our simple one-page template that's quick to fill out. After responding to the questions, both you and we will gain a clear picture of your requirements. *Uncertain about salary for your new hire? We will investigate market rates and provide guidance. *Facing skill shortages? During these periods of full employment and reduced immigration from Europe, such shortages are intensifying as a major challenge. Employers must clearly show candidates... & "Really pleased with how quickly they identified experienced candidates for us. Very happy!" \textless{}\textless{}FACT\_1\textgreater{}\textgreater{} Here's how we can assist you... *No Job or Person Specification available? That's fine—we'll provide our simple one-page template that's quick to fill out. After responding to the questions, both you and we will gain a clear picture of your requirements. *Uncertain about salary for your new hire? We will investigate market rates and provide guidance. *Facing skill shortages? During these periods of full employment and reduced immigration from \textless{}\textless{}FACT\_2\textgreater{}\textgreater{}, such shortages are intensifying as a major challenge. Employers must clearly show candidates why they should choose YOU over competitors...\\
\midrule
\scriptsize\emph{pile\_cc / grok\_4.1\_fast / sample\_idx=136}\\Wines and Meads, from "both sides of the pond" plus a few bits of other fermented food/drink Saturday, November 06, 2010 Disheartening Results.......... Been a "busy bee" today, as there was about 10 or 11 gallons of various brews that needed racking etc.... So why the strange thread title ? Well, it's one of those days where I've got the "WTF" thoughts going through my head. Why ? Well as I was racking various brews, not many of them taste like I think they should. And I don't really know why that is. For instance, a good few years ago, a reasonably local brewery, not only made some excellent beers (real ale type), but they also made a range of country wines. I'd tasted about 5 or 6 of t... & Wines and Meads, from "both sides of the pond" plus a few bits of other fermented food/drink Saturday, November 06, 2010 Disheartening Results.......... Today I was a "busy bee," since approximately 10 or 11 gallons of different brews had to be racked, among other things. This explains the unusual thread title. It's simply one of those days with "WTF" thoughts running through my mind. The cause? While racking the various brews, not many tasted the way I expected them to. I truly have no idea why. As an example, a few years ago a nearby brewery crafted not just excellent real ale-style beers but also a variety of country wines. I sampled about 5 or 6 of those wines, and the best descriptio... & Today I was a "busy bee," since approximately \textless{}\textless{}FACT\_1\textgreater{}\textgreater{} of different brews had to be racked, among other things. This explains the unusual thread title. It's simply one of those days with "WTF" thoughts running through my mind. The cause? While racking the various brews, not many tasted the way I expected them to. I truly have no idea why. As an example, a few years ago a nearby brewery crafted not just excellent real ale-style beers but also a variety of country wines. I sampled about \textless{}\textless{}FACT\_2\textgreater{}\textgreater{} of those wines, and the best description of my memory of them remains "alcoholic fruit cordials." Much of my knowledge about how wines "should" be produced derives from experts at resources like \textless{}\textless{}...\\
\bottomrule
\end{tabular}
\caption{Qualitative examples (truncated to 700 chars).}
\label{tab:qual:qualitative-examples4}
\end{table}

\section{Ablation on Factual Anchors}
\label{app:facts-ablation}
A natural question is whether the observed suppression of membership inference attacks is primarily driven by the removal of explicit factual anchors (e.g., names, numbers, dates), rather than by semantic-preserving paraphrasing. To isolate this effect, we consider an ablation (\textbf{FT-F}) in which factual anchors are removed directly from the original training data, while leaving the surrounding linguistic structure unchanged. This setting preserves much of the original surface form and syntax, but eliminates potentially identifiable facts.

Tables~\ref{tab:results-nolora-ensemble-AUC-datasets-defenses-None-True-llama-ftf} and~\ref{tab:results-nolora-ensemble-TPR-datasets-defenses-None-True-llama-ftf} show that factual removal alone yields only a partial reduction in attack effectiveness. Across datasets and attacks, FT-F consistently reduces AUC and TPR@FPR relative to standard fine-tuning (FT), indicating that factual anchors do contribute to membership signals. However, the residual leakage remains substantial, and FT-F is uniformly weaker than both SAGE and SAGE-R.

In contrast, SAGE further suppresses MIAs by jointly reducing surface-form overlap while preserving semantic content, and SAGE-R achieves the strongest suppression by additionally removing structural and factual cues. The gap between FT-F and SAGE demonstrates that eliminating facts alone is insufficient:
membership inference attacks continue to exploit broader lexical and structural regularities that survive factual deletion. These results confirm that robust suppression of MIAs requires semantic-preserving rewriting that actively disrupts surface-level correlations, rather than isolated removal of factual tokens.

\begin{table*}[htbp]
\centering
\small
\footnotesize
\caption{{AUC (Avg: \texttt{deepseek/deepseek-v3.2}, \texttt{x-ai/grok-4.1-fast}, \texttt{google/gemini-2.5-flash}) performance of MIAs across datasets and defenses. Evaluated on \texttt{meta-llama/Llama-3.2-3B}. \textbf{FT-F} denotes an ablation in which factual anchors (e.g., names, numbers, dates) are removed directly from the original training data without paraphrasing. (Full Finetuning)}}
\renewcommand{\arraystretch}{1.35}
\resizebox{\textwidth}{!}{
\begin{tabular}{lcccccccccccccccccccc}
\toprule
& \multicolumn{4}{c}{\textbf{\texttt{arxiv}}} & \multicolumn{4}{c}{\textbf{\texttt{wikipedia}}} & \multicolumn{4}{c}{\textbf{\texttt{pile\_cc}}} & \multicolumn{4}{c}{\textbf{\texttt{hackernews}}} & \multicolumn{4}{c}{\textbf{\texttt{pubmed}}} \\
\cmidrule(lr){2-5}
\cmidrule(lr){6-9}
\cmidrule(lr){10-13}
\cmidrule(lr){14-17}
\cmidrule(lr){18-21}
& {{FT}} & {{FT-F}} & {{SAGE}} & {{SAGE-R}} & {{FT}} & {{FT-F}} & {{SAGE}} & {{SAGE-R}} & {{FT}} & {{FT-F}} & {{SAGE}} & {{SAGE-R}} & {{FT}} & {{FT-F}} & {{SAGE}} & {{SAGE-R}} & {{FT}} & {{FT-F}} & {{SAGE}} & {{SAGE-R}} \\
\midrule
Loss & $1.000 $ & $0.995 $ & $0.921 $ & $0.794 $ & $1.000 $ & $0.978 $ & $0.883 $ & $0.671 $ & $0.999 $ & $0.994 $ & $0.848 $ & $0.696 $ & $1.000 $ & $1.000 $ & $0.947 $ & $0.759 $ & $1.000 $ & $1.000 $ & $0.842 $ & $0.765 $ \\
Zlib & $1.000 $ & $0.996 $ & $0.912 $ & $0.786 $ & $1.000 $ & $0.987 $ & $0.890 $ & $0.676 $ & $0.999 $ & $0.997 $ & $0.870 $ & $0.726 $ & $1.000 $ & $1.000 $ & $0.955 $ & $0.785 $ & $1.000 $ & $1.000 $ & $0.837 $ & $0.761 $ \\
Lowercase & $0.999 $ & $0.974 $ & $0.876 $ & $0.617 $ & $0.999 $ & $0.871 $ & $0.847 $ & $0.568 $ & $0.999 $ & $0.925 $ & $0.794 $ & $0.541 $ & $0.999 $ & $0.966 $ & $0.860 $ & $0.576 $ & $1.000 $ & $0.966 $ & $0.759 $ & $0.632 $ \\
Min-K\% & $0.688 $ & $0.670 $ & $0.644 $ & $0.587 $ & $0.705 $ & $0.654 $ & $0.634 $ & $0.546 $ & $0.747 $ & $0.713 $ & $0.627 $ & $0.561 $ & $0.689 $ & $0.668 $ & $0.618 $ & $0.535 $ & $0.698 $ & $0.672 $ & $0.603 $ & $0.573 $ \\
Min-K\%++ & $1.000 $ & $0.963 $ & $0.747 $ & $0.614 $ & $1.000 $ & $0.834 $ & $0.775 $ & $0.583 $ & $0.999 $ & $0.936 $ & $0.696 $ & $0.577 $ & $0.999 $ & $0.913 $ & $0.711 $ & $0.585 $ & $1.000 $ & $0.983 $ & $0.657 $ & $0.608 $ \\
ReCall & $1.000 $ & $0.995 $ & $0.933 $ & $0.809 $ & $1.000 $ & $0.981 $ & $0.889 $ & $0.678 $ & $0.999 $ & $0.995 $ & $0.856 $ & $0.706 $ & $1.000 $ & $1.000 $ & $0.954 $ & $0.766 $ & $1.000 $ & $1.000 $ & $0.867 $ & $0.786 $ \\
CON-ReCall & $1.000 $ & $0.995 $ & $0.893 $ & $0.699 $ & $1.000 $ & $0.968 $ & $0.845 $ & $0.619 $ & $0.999 $ & $0.994 $ & $0.765 $ & $0.609 $ & $1.000 $ & $0.999 $ & $0.816 $ & $0.616 $ & $1.000 $ & $1.000 $ & $0.789 $ & $0.697 $ \\
Ratio & $1.000 $ & $0.995 $ & $0.986 $ & $0.934 $ & $1.000 $ & $0.993 $ & $0.984 $ & $0.733 $ & $1.000 $ & $1.000 $ & $0.967 $ & $0.815 $ & $1.000 $ & $1.000 $ & $0.997 $ & $0.877 $ & $1.000 $ & $1.000 $ & $0.976 $ & $0.915 $ \\
Bag-of-Words & $0.498 $ & $0.498 $ & $0.498 $ & $0.498 $ & $0.510 $ & $0.510 $ & $0.510 $ & $0.510 $ & $0.498 $ & $0.498 $ & $0.498 $ & $0.498 $ & $0.524 $ & $0.524 $ & $0.524 $ & $0.524 $ & $0.528 $ & $0.528 $ & $0.528 $ & $0.528 $ \\
\midrule
Average & $0.96$ & $0.95$ & $0.86$ & $0.73$ & $0.96$ & $0.91$ & $0.84$ & $0.63$ & $0.97$ & $0.94$ & $0.80$ & $0.65$ & $0.96$ & $0.94$ & $0.86$ & $0.69$ & $0.96$ & $0.95$ & $0.79$ & $0.72$ \\
\bottomrule
\end{tabular}}
\label{tab:results-nolora-ensemble-AUC-datasets-defenses-None-True-llama-ftf}
\end{table*}

\begin{table*}[htbp]
\centering
\small
\footnotesize
\caption{{TPR@FPR=0.01 (Avg: \texttt{deepseek/deepseek-v3.2}, \texttt{x-ai/grok-4.1-fast}, \texttt{google/gemini-2.5-flash}) performance of MIAs across datasets and defenses. Evaluated on \texttt{meta-llama/Llama-3.2-3B}. \textbf{FT-F} denotes an ablation in which factual anchors (e.g., names, numbers, dates) are removed directly from the original training data without paraphrasing. (Full Finetuning)}}
\renewcommand{\arraystretch}{1.35}
\resizebox{\textwidth}{!}{
\begin{tabular}{lcccccccccccccccccccc}
\toprule
& \multicolumn{4}{c}{\textbf{\texttt{arxiv}}} & \multicolumn{4}{c}{\textbf{\texttt{wikipedia}}} & \multicolumn{4}{c}{\textbf{\texttt{pile\_cc}}} & \multicolumn{4}{c}{\textbf{\texttt{hackernews}}} & \multicolumn{4}{c}{\textbf{\texttt{pubmed}}} \\
\cmidrule(lr){2-5}
\cmidrule(lr){6-9}
\cmidrule(lr){10-13}
\cmidrule(lr){14-17}
\cmidrule(lr){18-21}
& {{FT}} & {{FT-F}} & {{SAGE}} & {{SAGE-R}} & {{FT}} & {{FT-F}} & {{SAGE}} & {{SAGE-R}} & {{FT}} & {{FT-F}} & {{SAGE}} & {{SAGE-R}} & {{FT}} & {{FT-F}} & {{SAGE}} & {{SAGE-R}} & {{FT}} & {{FT-F}} & {{SAGE}} & {{SAGE-R}} \\
\midrule
Loss & $1.000 $ & $0.985 $ & $0.393 $ & $0.051 $ & $0.994 $ & $0.559 $ & $0.208 $ & $0.024 $ & $0.988 $ & $0.853 $ & $0.074 $ & $0.011 $ & $1.000 $ & $0.997 $ & $0.450 $ & $0.084 $ & $1.000 $ & $0.995 $ & $0.210 $ & $0.094 $ \\
Zlib & $1.000 $ & $0.989 $ & $0.429 $ & $0.057 $ & $0.997 $ & $0.646 $ & $0.257 $ & $0.017 $ & $0.996 $ & $0.970 $ & $0.192 $ & $0.034 $ & $1.000 $ & $0.999 $ & $0.550 $ & $0.113 $ & $1.000 $ & $0.999 $ & $0.277 $ & $0.121 $ \\
Lowercase & $0.986 $ & $0.786 $ & $0.430 $ & $0.024 $ & $0.980 $ & $0.252 $ & $0.257 $ & $0.017 $ & $0.982 $ & $0.452 $ & $0.124 $ & $0.018 $ & $0.995 $ & $0.739 $ & $0.314 $ & $0.031 $ & $0.990 $ & $0.734 $ & $0.130 $ & $0.034 $ \\
Min-K\% & $0.597 $ & $0.541 $ & $0.356 $ & $0.084 $ & $0.622 $ & $0.324 $ & $0.187 $ & $0.034 $ & $0.673 $ & $0.523 $ & $0.091 $ & $0.017 $ & $0.612 $ & $0.539 $ & $0.265 $ & $0.055 $ & $0.620 $ & $0.558 $ & $0.195 $ & $0.101 $ \\
Min-K\%++ & $0.997 $ & $0.571 $ & $0.115 $ & $0.018 $ & $0.993 $ & $0.118 $ & $0.127 $ & $0.014 $ & $0.986 $ & $0.508 $ & $0.052 $ & $0.014 $ & $0.994 $ & $0.395 $ & $0.094 $ & $0.022 $ & $0.996 $ & $0.769 $ & $0.057 $ & $0.024 $ \\
ReCall & $1.000 $ & $0.988 $ & $0.447 $ & $0.074 $ & $0.995 $ & $0.616 $ & $0.226 $ & $0.023 $ & $0.987 $ & $0.832 $ & $0.070 $ & $0.010 $ & $1.000 $ & $0.999 $ & $0.485 $ & $0.090 $ & $1.000 $ & $0.998 $ & $0.279 $ & $0.130 $ \\
CON-ReCall & $1.000 $ & $0.975 $ & $0.373 $ & $0.066 $ & $0.993 $ & $0.528 $ & $0.202 $ & $0.018 $ & $0.988 $ & $0.871 $ & $0.099 $ & $0.021 $ & $1.000 $ & $0.988 $ & $0.286 $ & $0.062 $ & $1.000 $ & $0.993 $ & $0.161 $ & $0.082 $ \\
Ratio & $1.000 $ & $0.992 $ & $0.893 $ & $0.497 $ & $1.000 $ & $0.966 $ & $0.887 $ & $0.184 $ & $1.000 $ & $0.998 $ & $0.750 $ & $0.255 $ & $1.000 $ & $1.000 $ & $0.968 $ & $0.366 $ & $1.000 $ & $1.000 $ & $0.794 $ & $0.600 $ \\
Bag-of-Words & $0.007 $ & $0.007 $ & $0.007 $ & $0.007 $ & $0.004 $ & $0.004 $ & $0.004 $ & $0.004 $ & $0.010 $ & $0.010 $ & $0.010 $ & $0.010 $ & $0.011 $ & $0.011 $ & $0.011 $ & $0.011 $ & $0.011 $ & $0.011 $ & $0.011 $ & $0.011 $ \\
\midrule
Average & $0.95$ & $0.85$ & $0.43$ & $0.11$ & $0.95$ & $0.50$ & $0.29$ & $0.04$ & $0.95$ & $0.75$ & $0.18$ & $0.05$ & $0.95$ & $0.83$ & $0.43$ & $0.10$ & $0.95$ & $0.88$ & $0.26$ & $0.15$ \\
\bottomrule
\end{tabular}}
\label{tab:results-nolora-ensemble-TPR-datasets-defenses-None-True-llama-ftf}
\end{table*}

\section{Detailed Results on \texttt{meta-llama/Llama-3.2-3B}}
\label{app:detailed-result}
Table~\ref{tab:results-lora-ensemble-auc-tpr-datasets-defenses-pretrained} reports membership inference performance on the pretrained (PT) model across datasets and attacks that is fine-tuned on \texttt{Llama-3.2-3B}~\citep{llama3models2024}. Since the model has not been exposed to any downstream fine-tuning data, all attacks perform at or near random guessing, with AUC values tightly concentrated around $0.5$ and low TPR@0.01\%FPR. This establishes a lower-bound baseline and validates the experimental setup. From the rest of the tables, we can see that once the model is trained on downstream data, membership inference becomes highly effective, and that the effect is consistent across datasets, attacks, and evaluator backends. In particular, under full fine-tuning without defenses (FT), most MIAs achieve near-perfect AUCs across all five datasets (Table~\ref{tab:results-nolora-ensemble-datasets-defenses-None-True}, Table~\ref{tab:results-nolora-ensemble-AUC-datasets-defenses-deepseek-False}, Table~\ref{tab:results-nolora-ensemble-AUC-datasets-defenses-grok-False}, Table~\ref{tab:results-nolora-ensemble-AUC-datasets-defenses-gemini-False}) and this is mirrored by TPR@1\%FPR values near 1.0 (Table~\ref{tab:results-nolora-ensemble-TPRFPR-datasets-defenses-None-True}, Table~\ref{tab:results-nolora-ensemble-TPRFPR-datasets-defenses-deepseek-False}, Table~\ref{tab:results-nolora-ensemble-TPRFPR-datasets-defenses-gemini-False}). When defenses are applied, SAGE reduces leakage substantially and SAGE-R reduces it further, producing large drops in both AUC and TPR across all datasets (e.g., compare FT vs. SAGE vs. SAGE-R columns within Table~\ref{tab:results-nolora-ensemble-datasets-defenses-None-True} and Table~\ref{tab:results-nolora-ensemble-TPRFPR-datasets-defenses-None-True}). Finally, the LoRA regime is systematically less vulnerable than full fine-tuning, with noticeably lower AUC/TPR in the undefended setting and further reductions under SAGE/SAGE-R (Table~\ref{tab:results-lora-ensemble-AUC-datasets-defenses-deepseek-False}, Table~\ref{tab:results-lora-ensemble-TPRdatasets-defenses-deepseek-False}, Table~\ref{tab:results-lora-ensemble-AUC-datasets-defenses-grok-False}, Table~\ref{tab:results-lora-ensemble-TPRdatasets-defenses-grok-False}, Table~\ref{tab:results-lora-ensemble-AUC-datasets-defenses-gemini-False}, Table~\ref{tab:results-lora-ensemble-TPRdatasets-defenses-gemini-False}). Overall, these detailed results reinforce the main takeaway: fine-tuning introduces strong membership signal, SAGE mitigates it, and SAGE-R provides the strongest and most consistent suppression, while patterns are stable across datasets and evaluator models.

\begin{table}[htbp]
\centering
\small
\caption{{Pretrained (PT) membership inference attack performance across datasets and attacks. Each cell reports AUC, with the corresponding TPR@1\%FPR shown in parentheses. Evaluated on \texttt{meta-llama/Llama-3.2-3B}.}}
\begin{tabular}{lccccc}
\toprule
& \multicolumn{5}{c}{\textbf{Pretrained (PT)}} \\
\cmidrule(lr){2-6}
& \textbf{\texttt{arxiv}} 
& \textbf{\texttt{wikipedia}} 
& \textbf{\texttt{pile\_cc}} 
& \textbf{\texttt{hackernews}} 
& \textbf{\texttt{pubmed}} \\
\midrule
Loss          
& $0.513\,(0.005)$ & $0.502\,(0.015)$ & $0.489\,(0.012)$ & $0.501\,(0.007)$ & $0.481\,(0.005)$ \\
Zlib          
& $0.518\,(0.008)$ & $0.501\,(0.014)$ & $0.498\,(0.016)$ & $0.490\,(0.005)$ & $0.482\,(0.006)$ \\
Lowercase    
& $0.485\,(0.009)$ & $0.507\,(0.016)$ & $0.486\,(0.012)$ & $0.506\,(0.005)$ & $0.513\,(0.003)$ \\
Min-K\%      
& $0.510\,(0.012)$ & $0.503\,(0.016)$ & $0.505\,(0.008)$ & $0.482\,(0.017)$ & $0.498\,(0.007)$ \\
Min-K\%++    
& $0.504\,(0.006)$ & $0.507\,(0.014)$ & $0.491\,(0.010)$ & $0.506\,(0.012)$ & $0.491\,(0.007)$ \\
ReCall       
& $0.508\,(0.005)$ & $0.506\,(0.014)$ & $0.497\,(0.010)$ & $0.494\,(0.006)$ & $0.484\,(0.006)$ \\
CON-ReCall   
& $0.514\,(0.001)$ & $0.515\,(0.016)$ & $0.498\,(0.010)$ & $0.515\,(0.008)$ & $0.501\,(0.007)$ \\
Ratio        
& $0.516\,(0.008)$ & $0.510\,(0.010)$ & $0.482\,(0.012)$ & $0.509\,(0.007)$ & $0.505\,(0.014)$ \\
Bag-of-Words 
& $0.498\,(0.007)$ & $0.510\,(0.004)$ & $0.498\,(0.010)$ & $0.524\,(0.011)$ & $0.528\,(0.011)$ \\
Ensemble     
& $0.562\,(0.023)$ & $0.533\,(0.030)$ & $0.574\,(0.026)$ & $0.522\,(0.024)$ & $0.547\,(0.042)$ \\
\midrule
Average      
& $0.51\,(0.01)$ & $0.51\,(0.01)$ & $0.50\,(0.01)$ & $0.50\,(0.01)$ & $0.50\,(0.01)$ \\
\bottomrule
\end{tabular}

\label{tab:results-lora-ensemble-auc-tpr-datasets-defenses-pretrained}
\end{table}

\begin{table*}[htbp]
\centering
\small
\footnotesize
\caption{{AUC (Avg: \texttt{deepseek/deepseek-v3.2}, \texttt{x-ai/grok-4.1-fast}, \texttt{google/gemini-2.5-flash}) performance of MIAs across datasets and defenses. Evaluated on \texttt{meta-llama/Llama-3.2-3B}. (Full Finetuning).}}
\renewcommand{\arraystretch}{1.35}
\resizebox{\textwidth}{!}{
\begin{tabular}{lccccccccccccccc}
\toprule
& \multicolumn{3}{c}{\textbf{\texttt{arxiv}}} & \multicolumn{3}{c}{\textbf{\texttt{wikipedia}}} & \multicolumn{3}{c}{\textbf{\texttt{pile\_cc}}} & \multicolumn{3}{c}{\textbf{\texttt{hackernews}}} & \multicolumn{3}{c}{\textbf{\texttt{pubmed}}} \\
\cmidrule(lr){2-4}
\cmidrule(lr){5-7}
\cmidrule(lr){8-10}
\cmidrule(lr){11-13}
\cmidrule(lr){14-16}
& {{FT}} & {{SAGE}} & {{SAGE-R}} & {{FT}} & {{SAGE}} & {{SAGE-R}} & {{FT}} & {{SAGE}} & {{SAGE-R}} & {{FT}} & {{SAGE}} & {{SAGE-R}} & {{FT}} & {{SAGE}} & {{SAGE-R}} \\
\midrule
Loss & $1.000 $ & $0.921 $ & $0.794 $ & $1.000 $ & $0.883 $ & $0.671 $ & $0.999 $ & $0.848 $ & $0.696 $ & $1.000 $ & $0.947 $ & $0.759 $ & $1.000 $ & $0.842 $ & $0.765 $ \\
Zlib & $1.000 $ & $0.912 $ & $0.786 $ & $1.000 $ & $0.890 $ & $0.676 $ & $0.999 $ & $0.870 $ & $0.726 $ & $1.000 $ & $0.955 $ & $0.785 $ & $1.000 $ & $0.837 $ & $0.761 $ \\
Lowercase & $0.999 $ & $0.876 $ & $0.617 $ & $0.999 $ & $0.847 $ & $0.568 $ & $0.999 $ & $0.794 $ & $0.541 $ & $0.999 $ & $0.860 $ & $0.576 $ & $1.000 $ & $0.759 $ & $0.632 $ \\
Min-K\% & $0.688 $ & $0.644 $ & $0.587 $ & $0.705 $ & $0.634 $ & $0.546 $ & $0.747 $ & $0.627 $ & $0.561 $ & $0.689 $ & $0.618 $ & $0.535 $ & $0.698 $ & $0.603 $ & $0.573 $ \\
Min-K\%++ & $1.000 $ & $0.747 $ & $0.614 $ & $1.000 $ & $0.775 $ & $0.583 $ & $0.999 $ & $0.696 $ & $0.577 $ & $0.999 $ & $0.711 $ & $0.585 $ & $1.000 $ & $0.657 $ & $0.608 $ \\
ReCall & $1.000 $ & $0.933 $ & $0.809 $ & $1.000 $ & $0.889 $ & $0.678 $ & $0.999 $ & $0.856 $ & $0.706 $ & $1.000 $ & $0.954 $ & $0.766 $ & $1.000 $ & $0.867 $ & $0.786 $ \\
\bottomrule
\end{tabular}}
\label{tab:results-nolora-ensemble-datasets-defenses-None-True}
\end{table*}

\begin{table*}[htbp]
\centering
\small
\footnotesize
\caption{{TPR@1\% FPR (Avg: \texttt{deepseek/deepseek-v3.2}, \texttt{x-ai/grok-4.1-fast}, \texttt{google/gemini-2.5-flash}) performance of MIAs across datasets and defenses. Evaluated on \texttt{meta-llama/Llama-3.2-3B}. (Full Finetuning).}}
\renewcommand{\arraystretch}{1.35}
\resizebox{\textwidth}{!}{
\begin{tabular}{lccccccccccccccc}
\toprule
& \multicolumn{3}{c}{\textbf{\texttt{arxiv}}} & \multicolumn{3}{c}{\textbf{\texttt{wikipedia}}} & \multicolumn{3}{c}{\textbf{\texttt{pile\_cc}}} & \multicolumn{3}{c}{\textbf{\texttt{hackernews}}} & \multicolumn{3}{c}{\textbf{\texttt{pubmed}}} \\
\cmidrule(lr){2-4}
\cmidrule(lr){5-7}
\cmidrule(lr){8-10}
\cmidrule(lr){11-13}
\cmidrule(lr){14-16}
& {{FT}} & {{SAGE}} & {{SAGE-R}} & {{FT}} & {{SAGE}} & {{SAGE-R}} & {{FT}} & {{SAGE}} & {{SAGE-R}} & {{FT}} & {{SAGE}} & {{SAGE-R}} & {{FT}} & {{SAGE}} & {{SAGE-R}} \\
\midrule
Loss & $1.000 $ & $0.393 $ & $0.051 $ & $0.994 $ & $0.208 $ & $0.024 $ & $0.988 $ & $0.074 $ & $0.011 $ & $1.000 $ & $0.450 $ & $0.084 $ & $1.000 $ & $0.210 $ & $0.094 $ \\
Zlib & $1.000 $ & $0.429 $ & $0.057 $ & $0.997 $ & $0.257 $ & $0.017 $ & $0.996 $ & $0.192 $ & $0.034 $ & $1.000 $ & $0.550 $ & $0.113 $ & $1.000 $ & $0.277 $ & $0.121 $ \\
Lowercase & $0.986 $ & $0.430 $ & $0.024 $ & $0.980 $ & $0.257 $ & $0.017 $ & $0.982 $ & $0.124 $ & $0.018 $ & $0.995 $ & $0.314 $ & $0.031 $ & $0.990 $ & $0.130 $ & $0.034 $ \\
Min-K\% & $0.597 $ & $0.356 $ & $0.084 $ & $0.622 $ & $0.187 $ & $0.034 $ & $0.673 $ & $0.091 $ & $0.017 $ & $0.612 $ & $0.265 $ & $0.055 $ & $0.620 $ & $0.195 $ & $0.101 $ \\
Min-K\%++ & $0.997 $ & $0.115 $ & $0.018 $ & $0.993 $ & $0.127 $ & $0.014 $ & $0.986 $ & $0.052 $ & $0.014 $ & $0.994 $ & $0.094 $ & $0.022 $ & $0.996 $ & $0.057 $ & $0.024 $ \\
ReCall & $1.000 $ & $0.447 $ & $0.074 $ & $0.995 $ & $0.226 $ & $0.023 $ & $0.987 $ & $0.070 $ & $0.010 $ & $1.000 $ & $0.485 $ & $0.090 $ & $1.000 $ & $0.279 $ & $0.130 $ \\
\bottomrule
\end{tabular}}
\label{tab:results-nolora-ensemble-TPRFPR-datasets-defenses-None-True}
\end{table*}

\begin{table*}[htbp]
\centering
\small
\footnotesize
\caption{{AUC performance of MIAs across datasets and defenses. Evaluated on \texttt{meta-llama/Llama-3.2-3B}. (LoRA, \texttt{deepseek/deepseek-v3.2}).}}
\renewcommand{\arraystretch}{1.35}
\resizebox{\textwidth}{!}{
\begin{tabular}{lccccccccccccccc}
\toprule
& \multicolumn{3}{c}{\textbf{\texttt{arxiv}}} & \multicolumn{3}{c}{\textbf{\texttt{wikipedia}}} & \multicolumn{3}{c}{\textbf{\texttt{pile\_cc}}} & \multicolumn{3}{c}{\textbf{\texttt{hackernews}}} & \multicolumn{3}{c}{\textbf{\texttt{pubmed}}} \\
\cmidrule(lr){2-4}
\cmidrule(lr){5-7}
\cmidrule(lr){8-10}
\cmidrule(lr){11-13}
\cmidrule(lr){14-16}
& {{FT}} & {{SAGE}} & {{SAGE-R}} & {{FT}} & {{SAGE}} & {{SAGE-R}} & {{FT}} & {{SAGE}} & {{SAGE-R}} & {{FT}} & {{SAGE}} & {{SAGE-R}} & {{FT}} & {{SAGE}} & {{SAGE-R}} \\
\midrule
Loss & $0.685 $ & $0.600 $ & $0.553 $ & $0.658 $ & $0.580 $ & $0.525 $ & $0.645 $ & $0.547 $ & $0.512 $ & $0.710 $ & $0.588 $ & $0.533 $ & $0.645 $ & $0.545 $ & $0.522 $ \\
Zlib & $0.682 $ & $0.601 $ & $0.556 $ & $0.663 $ & $0.583 $ & $0.526 $ & $0.667 $ & $0.563 $ & $0.525 $ & $0.724 $ & $0.589 $ & $0.528 $ & $0.645 $ & $0.545 $ & $0.523 $ \\
Lowercase & $0.667 $ & $0.591 $ & $0.510 $ & $0.699 $ & $0.609 $ & $0.519 $ & $0.687 $ & $0.576 $ & $0.498 $ & $0.657 $ & $0.577 $ & $0.520 $ & $0.669 $ & $0.577 $ & $0.534 $ \\
Min-K\% & $0.579 $ & $0.546 $ & $0.521 $ & $0.561 $ & $0.532 $ & $0.508 $ & $0.568 $ & $0.531 $ & $0.515 $ & $0.548 $ & $0.512 $ & $0.490 $ & $0.566 $ & $0.519 $ & $0.511 $ \\
Min-K\%++ & $0.634 $ & $0.564 $ & $0.531 $ & $0.657 $ & $0.576 $ & $0.525 $ & $0.624 $ & $0.537 $ & $0.509 $ & $0.628 $ & $0.549 $ & $0.523 $ & $0.626 $ & $0.527 $ & $0.514 $ \\
ReCall & $0.698 $ & $0.605 $ & $0.552 $ & $0.666 $ & $0.587 $ & $0.529 $ & $0.661 $ & $0.558 $ & $0.521 $ & $0.719 $ & $0.590 $ & $0.528 $ & $0.666 $ & $0.556 $ & $0.529 $ \\
\bottomrule
\end{tabular}}
\label{tab:results-lora-ensemble-AUC-datasets-defenses-deepseek-False}
\end{table*}

\begin{table*}[htbp]
\centering
\small
\footnotesize
\caption{{AUC performance of MIAs across datasets and defenses. Evaluated on \texttt{meta-llama/Llama-3.2-3B}. (Full Finetuning, \texttt{deepseek/deepseek-v3.2}).}}
\renewcommand{\arraystretch}{1.35}
\resizebox{\textwidth}{!}{
\begin{tabular}{lccccccccccccccc}
\toprule
& \multicolumn{3}{c}{\textbf{\texttt{arxiv}}} & \multicolumn{3}{c}{\textbf{\texttt{wikipedia}}} & \multicolumn{3}{c}{\textbf{\texttt{pile\_cc}}} & \multicolumn{3}{c}{\textbf{\texttt{hackernews}}} & \multicolumn{3}{c}{\textbf{\texttt{pubmed}}} \\
\cmidrule(lr){2-4}
\cmidrule(lr){5-7}
\cmidrule(lr){8-10}
\cmidrule(lr){11-13}
\cmidrule(lr){14-16}
& {{FT}} & {{SAGE}} & {{SAGE-R}} & {{FT}} & {{SAGE}} & {{SAGE-R}} & {{FT}} & {{SAGE}} & {{SAGE-R}} & {{FT}} & {{SAGE}} & {{SAGE-R}} & {{FT}} & {{SAGE}} & {{SAGE-R}} \\
\midrule
Loss & $1.000 $ & $0.908 $ & $0.757 $ & $1.000 $ & $0.870 $ & $0.641 $ & $0.999 $ & $0.841 $ & $0.664 $ & $1.000 $ & $0.943 $ & $0.736 $ & $1.000 $ & $0.813 $ & $0.721 $ \\
Zlib & $1.000 $ & $0.900 $ & $0.750 $ & $1.000 $ & $0.876 $ & $0.644 $ & $0.999 $ & $0.862 $ & $0.691 $ & $1.000 $ & $0.952 $ & $0.756 $ & $1.000 $ & $0.805 $ & $0.717 $ \\
Lowercase & $0.999 $ & $0.867 $ & $0.592 $ & $0.999 $ & $0.842 $ & $0.560 $ & $0.999 $ & $0.798 $ & $0.528 $ & $0.999 $ & $0.860 $ & $0.579 $ & $1.000 $ & $0.732 $ & $0.604 $ \\
Min-K\% & $0.688 $ & $0.638 $ & $0.578 $ & $0.705 $ & $0.628 $ & $0.538 $ & $0.747 $ & $0.621 $ & $0.550 $ & $0.689 $ & $0.613 $ & $0.530 $ & $0.698 $ & $0.588 $ & $0.555 $ \\
Min-K\%++ & $1.000 $ & $0.732 $ & $0.597 $ & $1.000 $ & $0.761 $ & $0.569 $ & $0.999 $ & $0.690 $ & $0.561 $ & $0.999 $ & $0.714 $ & $0.585 $ & $1.000 $ & $0.630 $ & $0.583 $ \\
ReCall & $1.000 $ & $0.921 $ & $0.772 $ & $1.000 $ & $0.877 $ & $0.648 $ & $0.999 $ & $0.847 $ & $0.674 $ & $1.000 $ & $0.949 $ & $0.740 $ & $1.000 $ & $0.837 $ & $0.740 $ \\
\bottomrule
\end{tabular}}
\label{tab:results-nolora-ensemble-AUC-datasets-defenses-deepseek-False}
\end{table*}

\begin{table*}[htbp]
\centering
\small
\footnotesize
\caption{{TPR@1\% FPR performance of MIAs across datasets and defenses. Evaluated on \texttt{meta-llama/Llama-3.2-3B}. Evaluated on \texttt{meta-llama/Llama-3.2-3B}. (LoRA, \texttt{deepseek/deepseek-v3.2}).}}
\renewcommand{\arraystretch}{1.35}
\resizebox{\textwidth}{!}{
\begin{tabular}{lccccccccccccccc}
\toprule
& \multicolumn{3}{c}{\textbf{\texttt{arxiv}}} & \multicolumn{3}{c}{\textbf{\texttt{wikipedia}}} & \multicolumn{3}{c}{\textbf{\texttt{pile\_cc}}} & \multicolumn{3}{c}{\textbf{\texttt{hackernews}}} & \multicolumn{3}{c}{\textbf{\texttt{pubmed}}} \\
\cmidrule(lr){2-4}
\cmidrule(lr){5-7}
\cmidrule(lr){8-10}
\cmidrule(lr){11-13}
\cmidrule(lr){14-16}
& {{FT}} & {{SAGE}} & {{SAGE-R}} & {{FT}} & {{SAGE}} & {{SAGE-R}} & {{FT}} & {{SAGE}} & {{SAGE-R}} & {{FT}} & {{SAGE}} & {{SAGE-R}} & {{FT}} & {{SAGE}} & {{SAGE-R}} \\
\midrule
Loss & $0.028 $ & $0.013 $ & $0.006 $ & $0.041 $ & $0.024 $ & $0.016 $ & $0.022 $ & $0.015 $ & $0.012 $ & $0.036 $ & $0.016 $ & $0.008 $ & $0.036 $ & $0.013 $ & $0.010 $ \\
Zlib & $0.031 $ & $0.019 $ & $0.011 $ & $0.051 $ & $0.035 $ & $0.015 $ & $0.039 $ & $0.021 $ & $0.018 $ & $0.036 $ & $0.016 $ & $0.005 $ & $0.037 $ & $0.011 $ & $0.007 $ \\
Lowercase & $0.058 $ & $0.035 $ & $0.008 $ & $0.048 $ & $0.031 $ & $0.016 $ & $0.036 $ & $0.023 $ & $0.013 $ & $0.028 $ & $0.016 $ & $0.010 $ & $0.045 $ & $0.022 $ & $0.006 $ \\
Min-K\% & $0.048 $ & $0.022 $ & $0.011 $ & $0.047 $ & $0.028 $ & $0.018 $ & $0.030 $ & $0.014 $ & $0.010 $ & $0.042 $ & $0.011 $ & $0.010 $ & $0.078 $ & $0.020 $ & $0.018 $ \\
Min-K\%++ & $0.018 $ & $0.008 $ & $0.003 $ & $0.057 $ & $0.029 $ & $0.015 $ & $0.024 $ & $0.014 $ & $0.010 $ & $0.032 $ & $0.020 $ & $0.013 $ & $0.039 $ & $0.017 $ & $0.011 $ \\
ReCall & $0.030 $ & $0.015 $ & $0.007 $ & $0.047 $ & $0.027 $ & $0.015 $ & $0.021 $ & $0.011 $ & $0.009 $ & $0.037 $ & $0.014 $ & $0.008 $ & $0.041 $ & $0.010 $ & $0.010 $ \\
\bottomrule
\end{tabular}}
\label{tab:results-lora-ensemble-TPRdatasets-defenses-deepseek-False}
\end{table*}

\begin{table*}[htbp]
\centering
\small
\footnotesize
\caption{{TPR@1\% FPR performance of MIAs across datasets and defenses. Evaluated on \texttt{meta-llama/Llama-3.2-3B}. (Full Finetuning, \texttt{deepseek/deepseek-v3.2}).}}
\renewcommand{\arraystretch}{1.35}
\resizebox{\textwidth}{!}{
\begin{tabular}{lccccccccccccccc}
\toprule
& \multicolumn{3}{c}{\textbf{\texttt{arxiv}}} & \multicolumn{3}{c}{\textbf{\texttt{wikipedia}}} & \multicolumn{3}{c}{\textbf{\texttt{pile\_cc}}} & \multicolumn{3}{c}{\textbf{\texttt{hackernews}}} & \multicolumn{3}{c}{\textbf{\texttt{pubmed}}} \\
\cmidrule(lr){2-4}
\cmidrule(lr){5-7}
\cmidrule(lr){8-10}
\cmidrule(lr){11-13}
\cmidrule(lr){14-16}
& {{FT}} & {{SAGE}} & {{SAGE-R}} & {{FT}} & {{SAGE}} & {{SAGE-R}} & {{FT}} & {{SAGE}} & {{SAGE-R}} & {{FT}} & {{SAGE}} & {{SAGE-R}} & {{FT}} & {{SAGE}} & {{SAGE-R}} \\
\midrule
Loss & $1.000 $ & $0.369 $ & $0.034 $ & $0.994 $ & $0.189 $ & $0.022 $ & $0.988 $ & $0.089 $ & $0.009 $ & $1.000 $ & $0.447 $ & $0.091 $ & $1.000 $ & $0.171 $ & $0.056 $ \\
Zlib & $1.000 $ & $0.394 $ & $0.043 $ & $0.997 $ & $0.247 $ & $0.014 $ & $0.996 $ & $0.190 $ & $0.023 $ & $1.000 $ & $0.515 $ & $0.116 $ & $1.000 $ & $0.231 $ & $0.070 $ \\
Lowercase & $0.986 $ & $0.413 $ & $0.021 $ & $0.980 $ & $0.237 $ & $0.017 $ & $0.982 $ & $0.150 $ & $0.015 $ & $0.995 $ & $0.321 $ & $0.032 $ & $0.990 $ & $0.120 $ & $0.026 $ \\
Min-K\% & $0.597 $ & $0.321 $ & $0.080 $ & $0.622 $ & $0.169 $ & $0.026 $ & $0.673 $ & $0.087 $ & $0.010 $ & $0.612 $ & $0.245 $ & $0.057 $ & $0.620 $ & $0.154 $ & $0.086 $ \\
Min-K\%++ & $0.997 $ & $0.116 $ & $0.019 $ & $0.993 $ & $0.121 $ & $0.010 $ & $0.986 $ & $0.057 $ & $0.012 $ & $0.994 $ & $0.109 $ & $0.022 $ & $0.996 $ & $0.045 $ & $0.019 $ \\
ReCall & $1.000 $ & $0.421 $ & $0.047 $ & $0.995 $ & $0.214 $ & $0.022 $ & $0.987 $ & $0.081 $ & $0.008 $ & $1.000 $ & $0.471 $ & $0.088 $ & $1.000 $ & $0.225 $ & $0.078 $ \\
\bottomrule
\end{tabular}}
\label{tab:results-nolora-ensemble-TPRFPR-datasets-defenses-deepseek-False}
\end{table*}

\begin{table*}[htbp]
\centering
\small
\footnotesize
\caption{{AUC performance of MIAs across datasets and defenses. Evaluated on \texttt{meta-llama/Llama-3.2-3B}. (LoRA, \texttt{x-ai/grok-4.1-fast}).}}
\renewcommand{\arraystretch}{1.35}
\resizebox{\textwidth}{!}{
\begin{tabular}{lccccccccccccccc}
\toprule
& \multicolumn{3}{c}{\textbf{\texttt{arxiv}}} & \multicolumn{3}{c}{\textbf{\texttt{wikipedia}}} & \multicolumn{3}{c}{\textbf{\texttt{pile\_cc}}} & \multicolumn{3}{c}{\textbf{\texttt{hackernews}}} & \multicolumn{3}{c}{\textbf{\texttt{pubmed}}} \\
\cmidrule(lr){2-4}
\cmidrule(lr){5-7}
\cmidrule(lr){8-10}
\cmidrule(lr){11-13}
\cmidrule(lr){14-16}
& {{FT}} & {{SAGE}} & {{SAGE-R}} & {{FT}} & {{SAGE}} & {{SAGE-R}} & {{FT}} & {{SAGE}} & {{SAGE-R}} & {{FT}} & {{SAGE}} & {{SAGE-R}} & {{FT}} & {{SAGE}} & {{SAGE-R}} \\
\midrule
Loss & $0.685 $ & $0.601 $ & $0.559 $ & $0.658 $ & $0.581 $ & $0.524 $ & $0.645 $ & $0.548 $ & $0.515 $ & $0.710 $ & $0.593 $ & $0.540 $ & $0.645 $ & $0.549 $ & $0.529 $ \\
Zlib & $0.682 $ & $0.601 $ & $0.561 $ & $0.663 $ & $0.585 $ & $0.525 $ & $0.667 $ & $0.563 $ & $0.527 $ & $0.724 $ & $0.595 $ & $0.538 $ & $0.645 $ & $0.550 $ & $0.528 $ \\
Lowercase & $0.667 $ & $0.592 $ & $0.516 $ & $0.699 $ & $0.605 $ & $0.518 $ & $0.687 $ & $0.573 $ & $0.496 $ & $0.657 $ & $0.577 $ & $0.520 $ & $0.669 $ & $0.584 $ & $0.542 $ \\
Min-K\% & $0.579 $ & $0.547 $ & $0.525 $ & $0.561 $ & $0.533 $ & $0.510 $ & $0.568 $ & $0.532 $ & $0.517 $ & $0.548 $ & $0.514 $ & $0.493 $ & $0.566 $ & $0.523 $ & $0.517 $ \\
Min-K\%++ & $0.634 $ & $0.566 $ & $0.533 $ & $0.657 $ & $0.576 $ & $0.522 $ & $0.624 $ & $0.540 $ & $0.512 $ & $0.628 $ & $0.553 $ & $0.527 $ & $0.626 $ & $0.534 $ & $0.521 $ \\
ReCall & $0.698 $ & $0.607 $ & $0.558 $ & $0.666 $ & $0.587 $ & $0.529 $ & $0.661 $ & $0.561 $ & $0.524 $ & $0.719 $ & $0.596 $ & $0.536 $ & $0.666 $ & $0.560 $ & $0.536 $ \\
\bottomrule
\end{tabular}}
\label{tab:results-lora-ensemble-AUC-datasets-defenses-grok-False}
\end{table*}

\begin{table*}[htbp]
\centering
\small
\footnotesize
\caption{{AUC performance of MIAs across datasets and defenses. Evaluated on \texttt{meta-llama/Llama-3.2-3B}. (Full Finetuning, \texttt{x-ai/grok-4.1-fast}).}}
\renewcommand{\arraystretch}{1.35}
\resizebox{\textwidth}{!}{
\begin{tabular}{lccccccccccccccc}
\toprule
& \multicolumn{3}{c}{\textbf{\texttt{arxiv}}} & \multicolumn{3}{c}{\textbf{\texttt{wikipedia}}} & \multicolumn{3}{c}{\textbf{\texttt{pile\_cc}}} & \multicolumn{3}{c}{\textbf{\texttt{hackernews}}} & \multicolumn{3}{c}{\textbf{\texttt{pubmed}}} \\
\cmidrule(lr){2-4}
\cmidrule(lr){5-7}
\cmidrule(lr){8-10}
\cmidrule(lr){11-13}
\cmidrule(lr){14-16}
& {{FT}} & {{SAGE}} & {{SAGE-R}} & {{FT}} & {{SAGE}} & {{SAGE-R}} & {{FT}} & {{SAGE}} & {{SAGE-R}} & {{FT}} & {{SAGE}} & {{SAGE-R}} & {{FT}} & {{SAGE}} & {{SAGE-R}} \\
\midrule
Loss & $1.000 $ & $0.922 $ & $0.789 $ & $1.000 $ & $0.887 $ & $0.657 $ & $0.999 $ & $0.858 $ & $0.705 $ & $1.000 $ & $0.957 $ & $0.788 $ & $1.000 $ & $0.843 $ & $0.762 $ \\
Zlib & $1.000 $ & $0.913 $ & $0.783 $ & $1.000 $ & $0.894 $ & $0.664 $ & $0.999 $ & $0.880 $ & $0.735 $ & $1.000 $ & $0.964 $ & $0.819 $ & $1.000 $ & $0.839 $ & $0.759 $ \\
Lowercase & $0.999 $ & $0.878 $ & $0.615 $ & $0.999 $ & $0.847 $ & $0.555 $ & $0.999 $ & $0.809 $ & $0.542 $ & $0.999 $ & $0.877 $ & $0.572 $ & $1.000 $ & $0.767 $ & $0.624 $ \\
Min-K\% & $0.688 $ & $0.647 $ & $0.587 $ & $0.705 $ & $0.637 $ & $0.543 $ & $0.747 $ & $0.635 $ & $0.565 $ & $0.689 $ & $0.623 $ & $0.542 $ & $0.698 $ & $0.606 $ & $0.576 $ \\
Min-K\%++ & $1.000 $ & $0.757 $ & $0.617 $ & $1.000 $ & $0.788 $ & $0.581 $ & $0.999 $ & $0.715 $ & $0.586 $ & $0.999 $ & $0.721 $ & $0.593 $ & $1.000 $ & $0.672 $ & $0.615 $ \\
ReCall & $1.000 $ & $0.934 $ & $0.805 $ & $1.000 $ & $0.892 $ & $0.665 $ & $0.999 $ & $0.866 $ & $0.715 $ & $1.000 $ & $0.964 $ & $0.796 $ & $1.000 $ & $0.869 $ & $0.786 $ \\
\bottomrule
\end{tabular}}
\label{tab:results-nolora-ensemble-AUC-datasets-defenses-grok-False}
\end{table*}

\begin{table*}[htbp]
\centering
\small
\footnotesize
\caption{{TPR@1\% FPR performance of MIAs across datasets and defenses. Evaluated on \texttt{meta-llama/Llama-3.2-3B}. (LoRA, \texttt{x-ai/grok-4.1-fast}).}}
\renewcommand{\arraystretch}{1.35}
\resizebox{\textwidth}{!}{
\begin{tabular}{lccccccccccccccc}
\toprule
& \multicolumn{3}{c}{\textbf{\texttt{arxiv}}} & \multicolumn{3}{c}{\textbf{\texttt{wikipedia}}} & \multicolumn{3}{c}{\textbf{\texttt{pile\_cc}}} & \multicolumn{3}{c}{\textbf{\texttt{hackernews}}} & \multicolumn{3}{c}{\textbf{\texttt{pubmed}}} \\
\cmidrule(lr){2-4}
\cmidrule(lr){5-7}
\cmidrule(lr){8-10}
\cmidrule(lr){11-13}
\cmidrule(lr){14-16}
& {{FT}} & {{SAGE}} & {{SAGE-R}} & {{FT}} & {{SAGE}} & {{SAGE-R}} & {{FT}} & {{SAGE}} & {{SAGE-R}} & {{FT}} & {{SAGE}} & {{SAGE-R}} & {{FT}} & {{SAGE}} & {{SAGE-R}} \\
\midrule
Loss & $0.028 $ & $0.012 $ & $0.006 $ & $0.041 $ & $0.026 $ & $0.012 $ & $0.022 $ & $0.015 $ & $0.012 $ & $0.036 $ & $0.015 $ & $0.008 $ & $0.036 $ & $0.015 $ & $0.011 $ \\
Zlib & $0.031 $ & $0.016 $ & $0.010 $ & $0.051 $ & $0.031 $ & $0.015 $ & $0.039 $ & $0.020 $ & $0.019 $ & $0.036 $ & $0.015 $ & $0.007 $ & $0.037 $ & $0.016 $ & $0.009 $ \\
Lowercase & $0.058 $ & $0.032 $ & $0.007 $ & $0.048 $ & $0.030 $ & $0.015 $ & $0.036 $ & $0.021 $ & $0.014 $ & $0.028 $ & $0.017 $ & $0.007 $ & $0.045 $ & $0.023 $ & $0.008 $ \\
Min-K\% & $0.048 $ & $0.024 $ & $0.016 $ & $0.047 $ & $0.027 $ & $0.017 $ & $0.030 $ & $0.015 $ & $0.012 $ & $0.042 $ & $0.016 $ & $0.020 $ & $0.078 $ & $0.020 $ & $0.016 $ \\
Min-K\%++ & $0.018 $ & $0.009 $ & $0.005 $ & $0.057 $ & $0.027 $ & $0.014 $ & $0.024 $ & $0.015 $ & $0.013 $ & $0.032 $ & $0.019 $ & $0.011 $ & $0.039 $ & $0.020 $ & $0.013 $ \\
ReCall & $0.030 $ & $0.014 $ & $0.006 $ & $0.047 $ & $0.029 $ & $0.014 $ & $0.021 $ & $0.010 $ & $0.009 $ & $0.037 $ & $0.014 $ & $0.007 $ & $0.041 $ & $0.016 $ & $0.012 $ \\
\bottomrule
\end{tabular}}
\label{tab:results-lora-ensemble-TPRdatasets-defenses-grok-False}
\end{table*}

\begin{table*}[htbp]
\centering
\small
\footnotesize
\caption{{TPR@1\% FPR performance of MIAs across datasets and defenses. Evaluated on \texttt{meta-llama/Llama-3.2-3B}. Evaluated on \texttt{meta-llama/Llama-3.2-3B}. (Full Finetuning, \texttt{x-ai/grok-4.1-fast}).}}
\renewcommand{\arraystretch}{1.35}
\resizebox{\textwidth}{!}{
\begin{tabular}{lcccccccccccccccccccc}
\toprule
& \multicolumn{4}{c}{\textbf{\texttt{arxiv}}} & \multicolumn{4}{c}{\textbf{\texttt{wikipedia}}} & \multicolumn{4}{c}{\textbf{\texttt{pile\_cc}}} & \multicolumn{4}{c}{\textbf{\texttt{hackernews}}} & \multicolumn{4}{c}{\textbf{\texttt{pubmed}}} \\
\cmidrule(lr){2-5}
\cmidrule(lr){6-9}
\cmidrule(lr){10-13}
\cmidrule(lr){14-17}
\cmidrule(lr){18-21}
& {{FT}} & {{SOFT}} & {{SAGE}} & {{SAGE-R}} & {{FT}} & {{SOFT}} & {{SAGE}} & {{SAGE-R}} & {{FT}} & {{SOFT}} & {{SAGE}} & {{SAGE-R}} & {{FT}} & {{SOFT}} & {{SAGE}} & {{SAGE-R}} & {{FT}} & {{SOFT}} & {{SAGE}} & {{SAGE-R}} \\
\midrule
Loss & $1.000 $ & $0.899 $ & $0.409 $ & $0.050 $ & $0.994 $ & $0.886 $ & $0.227 $ & $0.024 $ & $0.988 $ & $0.877 $ & $0.068 $ & $0.013 $ & $1.000 $ & $0.899 $ & $0.493 $ & $0.111 $ & $1.000 $ & $0.899 $ & $0.212 $ & $0.085 $ \\
Zlib & $1.000 $ & $0.903 $ & $0.457 $ & $0.054 $ & $0.997 $ & $0.894 $ & $0.271 $ & $0.018 $ & $0.996 $ & $0.897 $ & $0.211 $ & $0.040 $ & $1.000 $ & $0.902 $ & $0.608 $ & $0.149 $ & $1.000 $ & $0.902 $ & $0.265 $ & $0.109 $ \\
Lowercase & $0.986 $ & $0.892 $ & $0.430 $ & $0.021 $ & $0.980 $ & $0.886 $ & $0.272 $ & $0.016 $ & $0.982 $ & $0.883 $ & $0.120 $ & $0.020 $ & $0.995 $ & $0.896 $ & $0.344 $ & $0.044 $ & $0.990 $ & $0.895 $ & $0.127 $ & $0.028 $ \\
Min-K\% & $0.597 $ & $0.533 $ & $0.385 $ & $0.089 $ & $0.622 $ & $0.572 $ & $0.203 $ & $0.037 $ & $0.673 $ & $0.605 $ & $0.104 $ & $0.024 $ & $0.612 $ & $0.560 $ & $0.301 $ & $0.074 $ & $0.620 $ & $0.559 $ & $0.208 $ & $0.092 $ \\
Min-K\%++ & $0.997 $ & $0.890 $ & $0.116 $ & $0.018 $ & $0.993 $ & $0.890 $ & $0.131 $ & $0.020 $ & $0.986 $ & $0.878 $ & $0.052 $ & $0.014 $ & $0.994 $ & $0.888 $ & $0.079 $ & $0.024 $ & $0.996 $ & $0.897 $ & $0.072 $ & $0.024 $ \\
ReCall & $1.000 $ & $0.904 $ & $0.470 $ & $0.060 $ & $0.995 $ & $0.895 $ & $0.244 $ & $0.024 $ & $0.987 $ & $0.880 $ & $0.066 $ & $0.013 $ & $1.000 $ & $0.903 $ & $0.533 $ & $0.121 $ & $1.000 $ & $0.903 $ & $0.286 $ & $0.119 $ \\
\bottomrule
\end{tabular}}
\label{tab:results-nolora-ensemble-TPRFPR-datasets-defenses-grok-False}
\end{table*}

\begin{table*}[htbp]
\centering
\small
\footnotesize
\caption{{AUC performance of MIAs across datasets and defenses. Evaluated on \texttt{meta-llama/Llama-3.2-3B}. (LoRA, \texttt{google/gemini-2.5-flash}).}}
\renewcommand{\arraystretch}{1.35}
\resizebox{\textwidth}{!}{
\begin{tabular}{lccccccccccccccc}
\toprule
& \multicolumn{3}{c}{\textbf{\texttt{arxiv}}} & \multicolumn{3}{c}{\textbf{\texttt{wikipedia}}} & \multicolumn{3}{c}{\textbf{\texttt{pile\_cc}}} & \multicolumn{3}{c}{\textbf{\texttt{hackernews}}} & \multicolumn{3}{c}{\textbf{\texttt{pubmed}}} \\
\cmidrule(lr){2-4}
\cmidrule(lr){5-7}
\cmidrule(lr){8-10}
\cmidrule(lr){11-13}
\cmidrule(lr){14-16}
& {{FT}} & {{SAGE}} & {{SAGE-R}} & {{FT}} & {{SAGE}} & {{SAGE-R}} & {{FT}} & {{SAGE}} & {{SAGE-R}} & {{FT}} & {{SAGE}} & {{SAGE-R}} & {{FT}} & {{SAGE}} & {{SAGE-R}} \\
\midrule
Loss & $0.685 $ & $0.604 $ & $0.566 $ & $0.658 $ & $0.584 $ & $0.535 $ & $0.645 $ & $0.546 $ & $0.517 $ & $0.710 $ & $0.583 $ & $0.533 $ & $0.645 $ & $0.554 $ & $0.538 $ \\
Zlib & $0.682 $ & $0.605 $ & $0.568 $ & $0.663 $ & $0.589 $ & $0.536 $ & $0.667 $ & $0.561 $ & $0.530 $ & $0.724 $ & $0.583 $ & $0.529 $ & $0.645 $ & $0.554 $ & $0.538 $ \\
Lowercase & $0.667 $ & $0.597 $ & $0.522 $ & $0.699 $ & $0.606 $ & $0.526 $ & $0.687 $ & $0.567 $ & $0.504 $ & $0.657 $ & $0.566 $ & $0.517 $ & $0.669 $ & $0.586 $ & $0.550 $ \\
Min-K\% & $0.579 $ & $0.546 $ & $0.526 $ & $0.561 $ & $0.534 $ & $0.513 $ & $0.568 $ & $0.529 $ & $0.517 $ & $0.548 $ & $0.510 $ & $0.492 $ & $0.566 $ & $0.526 $ & $0.519 $ \\
Min-K\%++ & $0.634 $ & $0.565 $ & $0.536 $ & $0.657 $ & $0.581 $ & $0.532 $ & $0.624 $ & $0.533 $ & $0.510 $ & $0.628 $ & $0.545 $ & $0.520 $ & $0.626 $ & $0.534 $ & $0.523 $ \\
ReCall & $0.698 $ & $0.609 $ & $0.566 $ & $0.666 $ & $0.591 $ & $0.539 $ & $0.661 $ & $0.558 $ & $0.526 $ & $0.719 $ & $0.584 $ & $0.530 $ & $0.666 $ & $0.566 $ & $0.545 $ \\
\bottomrule
\end{tabular}}
\label{tab:results-lora-ensemble-AUC-datasets-defenses-gemini-False}
\end{table*}

\begin{table*}[htbp]
\centering
\small
\footnotesize
\caption{{AUC performance of MIAs across datasets and defenses. Evaluated on \texttt{meta-llama/Llama-3.2-3B}. (Full Finetuning, \texttt{google/gemini-2.5-flash}).}}
\renewcommand{\arraystretch}{1.35}
\resizebox{\textwidth}{!}{
\begin{tabular}{lccccccccccccccc}
\toprule
& \multicolumn{3}{c}{\textbf{\texttt{arxiv}}} & \multicolumn{3}{c}{\textbf{\texttt{wikipedia}}} & \multicolumn{3}{c}{\textbf{\texttt{pile\_cc}}} & \multicolumn{3}{c}{\textbf{\texttt{hackernews}}} & \multicolumn{3}{c}{\textbf{\texttt{pubmed}}} \\
\cmidrule(lr){2-4}
\cmidrule(lr){5-7}
\cmidrule(lr){8-10}
\cmidrule(lr){11-13}
\cmidrule(lr){14-16}
& {{FT}} & {{SAGE}} & {{SAGE-R}} & {{FT}} & {{SAGE}} & {{SAGE-R}} & {{FT}} & {{SAGE}} & {{SAGE-R}} & {{FT}} & {{SAGE}} & {{SAGE-R}} & {{FT}} & {{SAGE}} & {{SAGE-R}} \\
\midrule
Loss & $1.000 $ & $0.933 $ & $0.835 $ & $1.000 $ & $0.893 $ & $0.715 $ & $0.999 $ & $0.844 $ & $0.719 $ & $1.000 $ & $0.942 $ & $0.753 $ & $1.000 $ & $0.871 $ & $0.812 $ \\
Zlib & $1.000 $ & $0.924 $ & $0.825 $ & $1.000 $ & $0.899 $ & $0.719 $ & $0.999 $ & $0.869 $ & $0.751 $ & $1.000 $ & $0.950 $ & $0.781 $ & $1.000 $ & $0.866 $ & $0.808 $ \\
Lowercase & $0.999 $ & $0.882 $ & $0.645 $ & $0.999 $ & $0.853 $ & $0.589 $ & $0.999 $ & $0.775 $ & $0.552 $ & $0.999 $ & $0.842 $ & $0.578 $ & $1.000 $ & $0.778 $ & $0.668 $ \\
Min-K\% & $0.688 $ & $0.647 $ & $0.595 $ & $0.705 $ & $0.636 $ & $0.557 $ & $0.747 $ & $0.624 $ & $0.567 $ & $0.689 $ & $0.617 $ & $0.532 $ & $0.698 $ & $0.614 $ & $0.589 $ \\
Min-K\%++ & $1.000 $ & $0.752 $ & $0.628 $ & $1.000 $ & $0.777 $ & $0.598 $ & $0.999 $ & $0.683 $ & $0.585 $ & $0.999 $ & $0.699 $ & $0.577 $ & $1.000 $ & $0.670 $ & $0.627 $ \\
ReCall & $1.000 $ & $0.943 $ & $0.850 $ & $1.000 $ & $0.899 $ & $0.721 $ & $0.999 $ & $0.855 $ & $0.729 $ & $1.000 $ & $0.950 $ & $0.761 $ & $1.000 $ & $0.895 $ & $0.833 $ \\
\bottomrule
\end{tabular}}
\label{tab:results-nolora-ensemble-AUC-datasets-defenses-gemini-False}
\end{table*}

\begin{table*}[htbp]
\centering
\small
\footnotesize
\caption{{TPR@1\% FPR performance of MIAs across datasets and defenses. Evaluated on \texttt{meta-llama/Llama-3.2-3B}. Evaluated on \texttt{meta-llama/Llama-3.2-3B}. (LoRA, \texttt{google/gemini-2.5-flash}).}}
\renewcommand{\arraystretch}{1.35}
\resizebox{\textwidth}{!}{
\begin{tabular}{lccccccccccccccc}
\toprule
& \multicolumn{3}{c}{\textbf{\texttt{arxiv}}} & \multicolumn{3}{c}{\textbf{\texttt{wikipedia}}} & \multicolumn{3}{c}{\textbf{\texttt{pile\_cc}}} & \multicolumn{3}{c}{\textbf{\texttt{hackernews}}} & \multicolumn{3}{c}{\textbf{\texttt{pubmed}}} \\
\cmidrule(lr){2-4}
\cmidrule(lr){5-7}
\cmidrule(lr){8-10}
\cmidrule(lr){11-13}
\cmidrule(lr){14-16}
& {{FT}} & {{SAGE}} & {{SAGE-R}} & {{FT}} & {{SAGE}} & {{SAGE-R}} & {{FT}} & {{SAGE}} & {{SAGE-R}} & {{FT}} & {{SAGE}} & {{SAGE-R}} & {{FT}} & {{SAGE}} & {{SAGE-R}} \\
\midrule
Loss & $0.028 $ & $0.013 $ & $0.007 $ & $0.041 $ & $0.027 $ & $0.014 $ & $0.022 $ & $0.014 $ & $0.012 $ & $0.036 $ & $0.017 $ & $0.007 $ & $0.036 $ & $0.014 $ & $0.011 $ \\
Zlib & $0.031 $ & $0.019 $ & $0.010 $ & $0.051 $ & $0.035 $ & $0.019 $ & $0.039 $ & $0.020 $ & $0.019 $ & $0.036 $ & $0.013 $ & $0.004 $ & $0.037 $ & $0.017 $ & $0.011 $ \\
Lowercase & $0.058 $ & $0.032 $ & $0.008 $ & $0.048 $ & $0.033 $ & $0.014 $ & $0.036 $ & $0.021 $ & $0.013 $ & $0.028 $ & $0.014 $ & $0.008 $ & $0.045 $ & $0.021 $ & $0.010 $ \\
Min-K\% & $0.048 $ & $0.027 $ & $0.017 $ & $0.047 $ & $0.026 $ & $0.017 $ & $0.030 $ & $0.012 $ & $0.012 $ & $0.042 $ & $0.009 $ & $0.015 $ & $0.078 $ & $0.020 $ & $0.018 $ \\
Min-K\%++ & $0.018 $ & $0.011 $ & $0.005 $ & $0.057 $ & $0.032 $ & $0.016 $ & $0.024 $ & $0.015 $ & $0.012 $ & $0.032 $ & $0.021 $ & $0.011 $ & $0.039 $ & $0.017 $ & $0.010 $ \\
ReCall & $0.030 $ & $0.018 $ & $0.008 $ & $0.047 $ & $0.029 $ & $0.018 $ & $0.021 $ & $0.011 $ & $0.010 $ & $0.037 $ & $0.013 $ & $0.007 $ & $0.041 $ & $0.011 $ & $0.011 $ \\
\bottomrule
\end{tabular}}
\label{tab:results-lora-ensemble-TPRdatasets-defenses-gemini-False}
\end{table*}

\begin{table*}[htbp]
\centering
\small
\footnotesize
\caption{{TPR@1\% FPR performance of MIAs across datasets and defenses. Evaluated on \texttt{meta-llama/Llama-3.2-3B}\cite{llama3models2024}. (Full Finetuning, \texttt{google/gemini-2.5-flash}~\cite{gemini2025}).}}
\renewcommand{\arraystretch}{1.35}
\resizebox{\textwidth}{!}{
\begin{tabular}{lccccccccccccccc}
\toprule
& \multicolumn{3}{c}{\textbf{\texttt{arxiv}}} & \multicolumn{3}{c}{\textbf{\texttt{wikipedia}}} & \multicolumn{3}{c}{\textbf{\texttt{pile\_cc}}} & \multicolumn{3}{c}{\textbf{\texttt{hackernews}}} & \multicolumn{3}{c}{\textbf{\texttt{pubmed}}} \\
\cmidrule(lr){2-4}
\cmidrule(lr){5-7}
\cmidrule(lr){8-10}
\cmidrule(lr){11-13}
\cmidrule(lr){14-16}
& {{FT}} & {{SAGE}} & {{SAGE-R}} & {{FT}} & {{SAGE}} & {{SAGE-R}} & {{FT}} & {{SAGE}} & {{SAGE-R}} & {{FT}} & {{SAGE}} & {{SAGE-R}} & {{FT}} & {{SAGE}} & {{SAGE-R}} \\
\midrule
Loss & $1.000 $ & $0.400 $ & $0.068 $ & $0.994 $ & $0.209 $ & $0.025 $ & $0.988 $ & $0.064 $ & $0.011 $ & $1.000 $ & $0.409 $ & $0.049 $ & $1.000 $ & $0.246 $ & $0.141 $ \\
Zlib & $1.000 $ & $0.435 $ & $0.075 $ & $0.997 $ & $0.254 $ & $0.019 $ & $0.996 $ & $0.174 $ & $0.038 $ & $1.000 $ & $0.528 $ & $0.074 $ & $1.000 $ & $0.334 $ & $0.185 $ \\
Lowercase & $0.986 $ & $0.447 $ & $0.031 $ & $0.980 $ & $0.261 $ & $0.017 $ & $0.982 $ & $0.103 $ & $0.019 $ & $0.995 $ & $0.276 $ & $0.018 $ & $0.990 $ & $0.143 $ & $0.048 $ \\
Min-K\% & $0.597 $ & $0.361 $ & $0.083 $ & $0.622 $ & $0.190 $ & $0.038 $ & $0.673 $ & $0.083 $ & $0.018 $ & $0.612 $ & $0.250 $ & $0.035 $ & $0.620 $ & $0.223 $ & $0.125 $ \\
Min-K\%++ & $0.997 $ & $0.112 $ & $0.018 $ & $0.993 $ & $0.130 $ & $0.013 $ & $0.986 $ & $0.047 $ & $0.017 $ & $0.994 $ & $0.094 $ & $0.020 $ & $0.996 $ & $0.055 $ & $0.028 $ \\
ReCall & $1.000 $ & $0.451 $ & $0.114 $ & $0.995 $ & $0.220 $ & $0.023 $ & $0.987 $ & $0.064 $ & $0.010 $ & $1.000 $ & $0.450 $ & $0.061 $ & $1.000 $ & $0.325 $ & $0.194 $ \\
\bottomrule
\end{tabular}}
\label{tab:results-nolora-ensemble-TPRFPR-datasets-defenses-gemini-False}
\end{table*}

\section{Detailed Results on \texttt{EleutherAI/pythia-6.9b}}
\label{app:pythia}
Ablation on \texttt{EleutherAI/pythia-6.9b}~\citep{pythia2023} (Tables~\ref{tab:results-lora-ensemble-AUC-datasets-defenses-grok-False-pythia-pretrained}--~\ref{tab:results-nolora-ensemble-TPR@FPR=0.01-datasets-defenses-gemini-False-pythia}) shows the same ordering as our main results: attacks are near-chance on the pretrained (PT) model (Table~\ref{tab:results-lora-ensemble-AUC-datasets-defenses-grok-False-pythia-pretrained}), LoRA adaptation yields moderate leakage that is further reduced by SAGE and most strongly by SAGE-R, and full fine-tuning produces the strongest leakage while SAGE/SAGE-R substantially suppress it. Tables~\ref{tab:results-lora-ensemble-AUC-datasets-defenses-None-True-pythia} and~\ref{tab:results-lora-ensemble-TPR-datasets-defenses-None-True-pythia} summarizes the results by reporting the average over 3 paraphrasers. Overall, the ablation confirms that our conclusions are not specific to a single backbone and that the relative effectiveness of SAGE and SAGE-R is consistent across datasets and attacks.

\begin{table*}[htbp]
\centering
\small
\caption{{Pretrained (PT) membership inference attack performance across datasets and attacks. Each cell reports AUC, with the corresponding TPR@1\%FPR shown in parentheses. Evaluated on \texttt{EleutherAI/pythia-6.9b}.}} 
\begin{tabular}{lccccc}
\toprule
& \multicolumn{5}{c}{\textbf{Pretrained (PT)}} \\
\cmidrule(lr){2-6}
& \textbf{\texttt{arxiv}} 
& \textbf{\texttt{wikipedia}} 
& \textbf{\texttt{pile\_cc}} 
& \textbf{\texttt{hackernews}} 
& \textbf{\texttt{pubmed}} \\
\midrule
Loss          
& $0.531\,(0.007)$ & $0.526\,(0.005)$ & $0.498\,(0.008)$ & $0.525\,(0.008)$ & $0.501\,(0.008)$ \\
Zlib          
& $0.536\,(0.007)$ & $0.527\,(0.010)$ & $0.504\,(0.011)$ & $0.519\,(0.008)$ & $0.504\,(0.004)$ \\
Lowercase    
& $0.515\,(0.017)$ & $0.538\,(0.007)$ & $0.504\,(0.011)$ & $0.511\,(0.018)$ & $0.519\,(0.006)$ \\
Min-K\%      
& $0.518\,(0.010)$ & $0.517\,(0.013)$ & $0.503\,(0.006)$ & $0.491\,(0.013)$ & $0.503\,(0.014)$ \\
Min-K\%++    
& $0.519\,(0.003)$ & $0.535\,(0.009)$ & $0.505\,(0.012)$ & $0.517\,(0.011)$ & $0.506\,(0.010)$ \\
ReCall       
& $0.532\,(0.007)$ & $0.533\,(0.006)$ & $0.506\,(0.007)$ & $0.523\,(0.010)$ & $0.509\,(0.007)$ \\
CON-ReCall   
& $0.530\,(0.008)$ & $0.531\,(0.008)$ & $0.502\,(0.008)$ & $0.520\,(0.011)$ & $0.516\,(0.007)$ \\
Ratio        
& $0.539\,(0.006)$ & $0.551\,(0.006)$ & $0.516\,(0.010)$ & $0.547\,(0.015)$ & $0.549\,(0.011)$ \\
Bag-of-Words 
& $0.498\,(0.007)$ & $0.510\,(0.004)$ & $0.498\,(0.010)$ & $0.524\,(0.011)$ & $0.528\,(0.011)$ \\
\midrule
Average      
& $0.53\,(0.01)$ & $0.53\,(0.01)$ & $0.50\,(0.01)$ & $0.52\,(0.01)$ & $0.51\,(0.01)$ \\
\bottomrule
\end{tabular}
\label{tab:results-lora-ensemble-AUC-datasets-defenses-grok-False-pythia-pretrained}
\end{table*}

\begin{table*}[htbp]
\centering
\small
\footnotesize
\caption{{AUC (Avg: \texttt{deepseek/deepseek-v3.2}, \texttt{x-ai/grok-4.1-fast}, \texttt{google/gemini-2.5-flash}) performance of MIAs across datasets and defenses. Evaluated on \texttt{EleutherAI/pythia-6.9b}. (LoRA)}}
\renewcommand{\arraystretch}{1.35}
\resizebox{\textwidth}{!}{
\begin{tabular}{lccccccccccccccc}
\toprule
& \multicolumn{3}{c}{\textbf{\texttt{arxiv}}} & \multicolumn{3}{c}{\textbf{\texttt{wikipedia}}} & \multicolumn{3}{c}{\textbf{\texttt{pile\_cc}}} & \multicolumn{3}{c}{\textbf{\texttt{hackernews}}} & \multicolumn{3}{c}{\textbf{\texttt{pubmed}}} \\
\cmidrule(lr){2-4}
\cmidrule(lr){5-7}
\cmidrule(lr){8-10}
\cmidrule(lr){11-13}
\cmidrule(lr){14-16}
& {{FT}} & {{SAGE}} & {{SAGE-R}} & {{FT}} & {{SAGE}} & {{SAGE-R}} & {{FT}} & {{SAGE}} & {{SAGE-R}} & {{FT}} & {{SAGE}} & {{SAGE-R}} & {{FT}} & {{SAGE}} & {{SAGE-R}} \\
\midrule
Loss & $0.673 $ & $0.616 $ & $0.571 $ & $0.671 $ & $0.609 $ & $0.551 $ & $0.645 $ & $0.563 $ & $0.528 $ & $0.731 $ & $0.616 $ & $0.556 $ & $0.657 $ & $0.570 $ & $0.546 $ \\
Zlib & $0.670 $ & $0.616 $ & $0.573 $ & $0.675 $ & $0.611 $ & $0.551 $ & $0.664 $ & $0.575 $ & $0.539 $ & $0.740 $ & $0.617 $ & $0.556 $ & $0.657 $ & $0.571 $ & $0.548 $ \\
Lowercase & $0.655 $ & $0.615 $ & $0.536 $ & $0.687 $ & $0.621 $ & $0.549 $ & $0.668 $ & $0.582 $ & $0.515 $ & $0.691 $ & $0.598 $ & $0.527 $ & $0.654 $ & $0.578 $ & $0.538 $ \\
Min-K\% & $0.579 $ & $0.554 $ & $0.529 $ & $0.573 $ & $0.551 $ & $0.525 $ & $0.565 $ & $0.534 $ & $0.517 $ & $0.553 $ & $0.518 $ & $0.500 $ & $0.566 $ & $0.529 $ & $0.519 $ \\
Min-K\%++ & $0.626 $ & $0.571 $ & $0.544 $ & $0.684 $ & $0.611 $ & $0.552 $ & $0.631 $ & $0.554 $ & $0.522 $ & $0.636 $ & $0.563 $ & $0.534 $ & $0.637 $ & $0.549 $ & $0.535 $ \\
ReCall & $0.689 $ & $0.626 $ & $0.576 $ & $0.682 $ & $0.618 $ & $0.556 $ & $0.661 $ & $0.575 $ & $0.537 $ & $0.749 $ & $0.624 $ & $0.556 $ & $0.678 $ & $0.583 $ & $0.557 $ \\
\bottomrule
\end{tabular}}
\label{tab:results-lora-ensemble-AUC-datasets-defenses-None-True-pythia}
\end{table*}

\begin{table*}[htbp]
\centering
\small
\footnotesize
\caption{{TPR@1\% FPR (Avg: \texttt{deepseek/deepseek-v3.2}, \texttt{x-ai/grok-4.1-fast}, \texttt{google/gemini-2.5-flash}) performance of MIAs across datasets and defenses. Evaluated on \texttt{EleutherAI/pythia-6.9b}. (LoRA)}}
\renewcommand{\arraystretch}{1.35}
\resizebox{\textwidth}{!}{
\begin{tabular}{lccccccccccccccc}
\toprule
& \multicolumn{3}{c}{\textbf{\texttt{arxiv}}} & \multicolumn{3}{c}{\textbf{\texttt{wikipedia}}} & \multicolumn{3}{c}{\textbf{\texttt{pile\_cc}}} & \multicolumn{3}{c}{\textbf{\texttt{hackernews}}} & \multicolumn{3}{c}{\textbf{\texttt{pubmed}}} \\
\cmidrule(lr){2-4}
\cmidrule(lr){5-7}
\cmidrule(lr){8-10}
\cmidrule(lr){11-13}
\cmidrule(lr){14-16}
& {{FT}} & {{SAGE}} & {{SAGE-R}} & {{FT}} & {{SAGE}} & {{SAGE-R}} & {{FT}} & {{SAGE}} & {{SAGE-R}} & {{FT}} & {{SAGE}} & {{SAGE-R}} & {{FT}} & {{SAGE}} & {{SAGE-R}} \\
\midrule
Loss & $0.032 $ & $0.021 $ & $0.007 $ & $0.030 $ & $0.021 $ & $0.012 $ & $0.020 $ & $0.011 $ & $0.005 $ & $0.039 $ & $0.021 $ & $0.008 $ & $0.033 $ & $0.017 $ & $0.012 $ \\
Zlib & $0.029 $ & $0.019 $ & $0.008 $ & $0.034 $ & $0.026 $ & $0.013 $ & $0.028 $ & $0.017 $ & $0.010 $ & $0.063 $ & $0.027 $ & $0.009 $ & $0.028 $ & $0.011 $ & $0.009 $ \\
Lowercase & $0.052 $ & $0.037 $ & $0.016 $ & $0.026 $ & $0.014 $ & $0.007 $ & $0.026 $ & $0.018 $ & $0.012 $ & $0.070 $ & $0.040 $ & $0.015 $ & $0.019 $ & $0.016 $ & $0.013 $ \\
Min-K\% & $0.066 $ & $0.028 $ & $0.013 $ & $0.042 $ & $0.030 $ & $0.016 $ & $0.020 $ & $0.010 $ & $0.005 $ & $0.059 $ & $0.041 $ & $0.024 $ & $0.066 $ & $0.032 $ & $0.027 $ \\
Min-K\%++ & $0.014 $ & $0.008 $ & $0.003 $ & $0.037 $ & $0.027 $ & $0.011 $ & $0.022 $ & $0.016 $ & $0.009 $ & $0.034 $ & $0.022 $ & $0.014 $ & $0.032 $ & $0.016 $ & $0.014 $ \\
ReCall & $0.033 $ & $0.020 $ & $0.006 $ & $0.031 $ & $0.020 $ & $0.011 $ & $0.016 $ & $0.008 $ & $0.004 $ & $0.058 $ & $0.024 $ & $0.012 $ & $0.032 $ & $0.010 $ & $0.008 $ \\
\bottomrule
\end{tabular}}
\label{tab:results-lora-ensemble-TPR-datasets-defenses-None-True-pythia}
\end{table*}

\begin{table*}[htbp]
\centering
\small
\footnotesize
\caption{{AUC (Avg: \texttt{deepseek/deepseek-v3.2}, \texttt{x-ai/grok-4.1-fast}, \texttt{google/gemini-2.5-flash}) performance of MIAs across datasets and defenses. Evaluated on \texttt{EleutherAI/pythia-6.9b}. (Full Finetuning)}}
\renewcommand{\arraystretch}{1.35}
\resizebox{\textwidth}{!}{
\begin{tabular}{lccccccccccccccc}
\toprule
& \multicolumn{3}{c}{\textbf{\texttt{arxiv}}} & \multicolumn{3}{c}{\textbf{\texttt{wikipedia}}} & \multicolumn{3}{c}{\textbf{\texttt{pile\_cc}}} & \multicolumn{3}{c}{\textbf{\texttt{hackernews}}} & \multicolumn{3}{c}{\textbf{\texttt{pubmed}}} \\
\cmidrule(lr){2-4}
\cmidrule(lr){5-7}
\cmidrule(lr){8-10}
\cmidrule(lr){11-13}
\cmidrule(lr){14-16}
& {{FT}} & {{SAGE}} & {{SAGE-R}} & {{FT}} & {{SAGE}} & {{SAGE-R}} & {{FT}} & {{SAGE}} & {{SAGE-R}} & {{FT}} & {{SAGE}} & {{SAGE-R}} & {{FT}} & {{SAGE}} & {{SAGE-R}} \\
\midrule
Loss & $1.000 $ & $0.908 $ & $0.758 $ & $1.000 $ & $0.852 $ & $0.651 $ & $1.000 $ & $0.826 $ & $0.666 $ & $1.000 $ & $0.914 $ & $0.725 $ & $1.000 $ & $0.827 $ & $0.742 $ \\
Zlib & $1.000 $ & $0.898 $ & $0.750 $ & $1.000 $ & $0.857 $ & $0.652 $ & $1.000 $ & $0.844 $ & $0.688 $ & $1.000 $ & $0.920 $ & $0.740 $ & $1.000 $ & $0.822 $ & $0.741 $ \\
Lowercase & $0.999 $ & $0.859 $ & $0.638 $ & $0.998 $ & $0.827 $ & $0.606 $ & $1.000 $ & $0.796 $ & $0.569 $ & $1.000 $ & $0.861 $ & $0.611 $ & $0.999 $ & $0.726 $ & $0.632 $ \\
Min-K\% & $0.697 $ & $0.645 $ & $0.584 $ & $0.709 $ & $0.631 $ & $0.549 $ & $0.731 $ & $0.617 $ & $0.551 $ & $0.675 $ & $0.609 $ & $0.544 $ & $0.703 $ & $0.606 $ & $0.576 $ \\
Min-K\%++ & $1.000 $ & $0.735 $ & $0.627 $ & $1.000 $ & $0.781 $ & $0.599 $ & $1.000 $ & $0.691 $ & $0.575 $ & $1.000 $ & $0.688 $ & $0.589 $ & $1.000 $ & $0.670 $ & $0.619 $ \\
ReCall & $1.000 $ & $0.921 $ & $0.770 $ & $1.000 $ & $0.861 $ & $0.656 $ & $1.000 $ & $0.833 $ & $0.676 $ & $1.000 $ & $0.928 $ & $0.737 $ & $1.000 $ & $0.848 $ & $0.759 $ \\
\bottomrule
\end{tabular}}
\label{tab:results-nolora-ensemble-AUC-datasets-defenses-None-True-pythia}
\end{table*}

\begin{table*}[htbp]
\centering
\small
\footnotesize
\caption{{TPR@1\% FPR (Avg: \texttt{deepseek/deepseek-v3.2}, \texttt{x-ai/grok-4.1-fast}, \texttt{google/gemini-2.5-flash}) performance of MIAs across datasets and defenses. Evaluated on \texttt{EleutherAI/pythia-6.9b}. (Full Finetuning)}}
\renewcommand{\arraystretch}{1.35}
\resizebox{\textwidth}{!}{
\begin{tabular}{lccccccccccccccc}
\toprule
& \multicolumn{3}{c}{\textbf{\texttt{arxiv}}} & \multicolumn{3}{c}{\textbf{\texttt{wikipedia}}} & \multicolumn{3}{c}{\textbf{\texttt{pile\_cc}}} & \multicolumn{3}{c}{\textbf{\texttt{hackernews}}} & \multicolumn{3}{c}{\textbf{\texttt{pubmed}}} \\
\cmidrule(lr){2-4}
\cmidrule(lr){5-7}
\cmidrule(lr){8-10}
\cmidrule(lr){11-13}
\cmidrule(lr){14-16}
& {{FT}} & {{SAGE}} & {{SAGE-R}} & {{FT}} & {{SAGE}} & {{SAGE-R}} & {{FT}} & {{SAGE}} & {{SAGE-R}} & {{FT}} & {{SAGE}} & {{SAGE-R}} & {{FT}} & {{SAGE}} & {{SAGE-R}} \\
\midrule
Loss & $1.000 $ & $0.365 $ & $0.040 $ & $0.995 $ & $0.148 $ & $0.017 $ & $0.995 $ & $0.082 $ & $0.009 $ & $1.000 $ & $0.354 $ & $0.071 $ & $1.000 $ & $0.189 $ & $0.075 $ \\
Zlib & $1.000 $ & $0.371 $ & $0.035 $ & $0.996 $ & $0.176 $ & $0.015 $ & $0.998 $ & $0.157 $ & $0.025 $ & $1.000 $ & $0.473 $ & $0.105 $ & $1.000 $ & $0.182 $ & $0.067 $ \\
Lowercase & $0.991 $ & $0.379 $ & $0.029 $ & $0.956 $ & $0.128 $ & $0.004 $ & $0.993 $ & $0.158 $ & $0.019 $ & $1.000 $ & $0.365 $ & $0.060 $ & $0.993 $ & $0.126 $ & $0.051 $ \\
Min-K\% & $0.616 $ & $0.314 $ & $0.077 $ & $0.628 $ & $0.131 $ & $0.021 $ & $0.663 $ & $0.073 $ & $0.009 $ & $0.588 $ & $0.244 $ & $0.055 $ & $0.632 $ & $0.176 $ & $0.111 $ \\
Min-K\%++ & $0.998 $ & $0.078 $ & $0.009 $ & $0.996 $ & $0.112 $ & $0.020 $ & $0.995 $ & $0.043 $ & $0.008 $ & $0.999 $ & $0.087 $ & $0.027 $ & $1.000 $ & $0.052 $ & $0.029 $ \\
ReCall & $1.000 $ & $0.429 $ & $0.047 $ & $0.996 $ & $0.160 $ & $0.015 $ & $0.994 $ & $0.069 $ & $0.007 $ & $1.000 $ & $0.425 $ & $0.089 $ & $1.000 $ & $0.222 $ & $0.082 $ \\
\bottomrule
\end{tabular}}
\label{tab:results-nolora-ensemble-TPR-datasets-defenses-None-True-pythia}
\end{table*}

\begin{table*}[htbp]
\centering
\small
\footnotesize
\caption{{AUC performance of MIAs across datasets and defenses. Evaluated on \texttt{EleutherAI/pythia-6.9b}. (LoRA, \texttt{deepseek/deepseek-v3.2})}}
\renewcommand{\arraystretch}{1.35}
\resizebox{\textwidth}{!}{
\begin{tabular}{lccccccccccccccc}
\toprule
& \multicolumn{3}{c}{\textbf{\texttt{arxiv}}} & \multicolumn{3}{c}{\textbf{\texttt{wikipedia}}} & \multicolumn{3}{c}{\textbf{\texttt{pile\_cc}}} & \multicolumn{3}{c}{\textbf{\texttt{hackernews}}} & \multicolumn{3}{c}{\textbf{\texttt{pubmed}}} \\
\cmidrule(lr){2-4}
\cmidrule(lr){5-7}
\cmidrule(lr){8-10}
\cmidrule(lr){11-13}
\cmidrule(lr){14-16}
& {{FT}} & {{SAGE}} & {{SAGE-R}} & {{FT}} & {{SAGE}} & {{SAGE-R}} & {{FT}} & {{SAGE}} & {{SAGE-R}} & {{FT}} & {{SAGE}} & {{SAGE-R}} & {{FT}} & {{SAGE}} & {{SAGE-R}} \\
\midrule
Loss & $0.673 $ & $0.616 $ & $0.566 $ & $0.671 $ & $0.606 $ & $0.547 $ & $0.645 $ & $0.564 $ & $0.526 $ & $0.731 $ & $0.619 $ & $0.554 $ & $0.657 $ & $0.565 $ & $0.539 $ \\
Zlib & $0.670 $ & $0.616 $ & $0.569 $ & $0.675 $ & $0.608 $ & $0.547 $ & $0.664 $ & $0.576 $ & $0.536 $ & $0.740 $ & $0.619 $ & $0.553 $ & $0.657 $ & $0.567 $ & $0.541 $ \\
Lowercase & $0.655 $ & $0.615 $ & $0.532 $ & $0.687 $ & $0.621 $ & $0.547 $ & $0.668 $ & $0.586 $ & $0.516 $ & $0.691 $ & $0.601 $ & $0.529 $ & $0.654 $ & $0.571 $ & $0.529 $ \\
Min-K\% & $0.579 $ & $0.553 $ & $0.527 $ & $0.573 $ & $0.551 $ & $0.525 $ & $0.565 $ & $0.534 $ & $0.515 $ & $0.553 $ & $0.519 $ & $0.497 $ & $0.566 $ & $0.527 $ & $0.515 $ \\
Min-K\%++ & $0.626 $ & $0.570 $ & $0.539 $ & $0.684 $ & $0.607 $ & $0.550 $ & $0.631 $ & $0.555 $ & $0.520 $ & $0.636 $ & $0.565 $ & $0.533 $ & $0.637 $ & $0.546 $ & $0.530 $ \\
ReCall & $0.689 $ & $0.625 $ & $0.570 $ & $0.682 $ & $0.615 $ & $0.552 $ & $0.661 $ & $0.576 $ & $0.535 $ & $0.749 $ & $0.626 $ & $0.555 $ & $0.678 $ & $0.579 $ & $0.550 $ \\
\bottomrule
\end{tabular}}
\label{tab:results-lora-ensemble-AUC-datasets-defenses-deepseek-False-pythia}
\end{table*}

\begin{table*}[htbp]
\centering
\small
\footnotesize
\caption{{AUC performance of MIAs across datasets and defenses. Evaluated on \texttt{EleutherAI/pythia-6.9b}. (Full Finetuning, \texttt{deepseek/deepseek-v3.2})}}
\renewcommand{\arraystretch}{1.35}
\resizebox{\textwidth}{!}{
\begin{tabular}{lccccccccccccccc}
\toprule
& \multicolumn{3}{c}{\textbf{\texttt{arxiv}}} & \multicolumn{3}{c}{\textbf{\texttt{wikipedia}}} & \multicolumn{3}{c}{\textbf{\texttt{pile\_cc}}} & \multicolumn{3}{c}{\textbf{\texttt{hackernews}}} & \multicolumn{3}{c}{\textbf{\texttt{pubmed}}} \\
\cmidrule(lr){2-4}
\cmidrule(lr){5-7}
\cmidrule(lr){8-10}
\cmidrule(lr){11-13}
\cmidrule(lr){14-16}
& {{FT}} & {{SAGE}} & {{SAGE-R}} & {{FT}} & {{SAGE}} & {{SAGE-R}} & {{FT}} & {{SAGE}} & {{SAGE-R}} & {{FT}} & {{SAGE}} & {{SAGE-R}} & {{FT}} & {{SAGE}} & {{SAGE-R}} \\
\midrule
Loss & $1.000 $ & $0.897 $ & $0.724 $ & $1.000 $ & $0.843 $ & $0.629 $ & $1.000 $ & $0.822 $ & $0.640 $ & $1.000 $ & $0.915 $ & $0.704 $ & $1.000 $ & $0.800 $ & $0.700 $ \\
Zlib & $1.000 $ & $0.888 $ & $0.717 $ & $1.000 $ & $0.848 $ & $0.629 $ & $1.000 $ & $0.838 $ & $0.659 $ & $1.000 $ & $0.921 $ & $0.713 $ & $1.000 $ & $0.793 $ & $0.698 $ \\
Lowercase & $0.999 $ & $0.850 $ & $0.617 $ & $0.998 $ & $0.824 $ & $0.594 $ & $1.000 $ & $0.799 $ & $0.560 $ & $1.000 $ & $0.869 $ & $0.607 $ & $0.999 $ & $0.707 $ & $0.598 $ \\
Min-K\% & $0.697 $ & $0.640 $ & $0.573 $ & $0.709 $ & $0.629 $ & $0.546 $ & $0.731 $ & $0.615 $ & $0.542 $ & $0.675 $ & $0.608 $ & $0.536 $ & $0.703 $ & $0.596 $ & $0.561 $ \\
Min-K\%++ & $1.000 $ & $0.727 $ & $0.608 $ & $1.000 $ & $0.770 $ & $0.589 $ & $1.000 $ & $0.689 $ & $0.563 $ & $1.000 $ & $0.691 $ & $0.581 $ & $1.000 $ & $0.648 $ & $0.591 $ \\
ReCall & $1.000 $ & $0.910 $ & $0.737 $ & $1.000 $ & $0.852 $ & $0.634 $ & $1.000 $ & $0.827 $ & $0.649 $ & $1.000 $ & $0.927 $ & $0.712 $ & $1.000 $ & $0.821 $ & $0.716 $ \\
\bottomrule
\end{tabular}}
\label{tab:results-nolora-ensemble-AUC-datasets-defenses-deepseek-False-pythia}
\end{table*}

\begin{table*}[htbp]
\centering
\small
\footnotesize
\caption{{TPR@1\% FPR performance of MIAs across datasets and defenses. Evaluated on \texttt{EleutherAI/pythia-6.9b}. (LoRA, \texttt{deepseek/deepseek-v3.2})}}
\renewcommand{\arraystretch}{1.35}
\resizebox{\textwidth}{!}{
\begin{tabular}{lccccccccccccccc}
\toprule
& \multicolumn{3}{c}{\textbf{\texttt{arxiv}}} & \multicolumn{3}{c}{\textbf{\texttt{wikipedia}}} & \multicolumn{3}{c}{\textbf{\texttt{pile\_cc}}} & \multicolumn{3}{c}{\textbf{\texttt{hackernews}}} & \multicolumn{3}{c}{\textbf{\texttt{pubmed}}} \\
\cmidrule(lr){2-4}
\cmidrule(lr){5-7}
\cmidrule(lr){8-10}
\cmidrule(lr){11-13}
\cmidrule(lr){14-16}
& {{FT}} & {{SAGE}} & {{SAGE-R}} & {{FT}} & {{SAGE}} & {{SAGE-R}} & {{FT}} & {{SAGE}} & {{SAGE-R}} & {{FT}} & {{SAGE}} & {{SAGE-R}} & {{FT}} & {{SAGE}} & {{SAGE-R}} \\
\midrule
Loss & $0.032 $ & $0.021 $ & $0.007 $ & $0.030 $ & $0.021 $ & $0.012 $ & $0.020 $ & $0.012 $ & $0.005 $ & $0.039 $ & $0.023 $ & $0.009 $ & $0.033 $ & $0.015 $ & $0.010 $ \\
Zlib & $0.029 $ & $0.020 $ & $0.008 $ & $0.034 $ & $0.027 $ & $0.012 $ & $0.028 $ & $0.018 $ & $0.010 $ & $0.063 $ & $0.025 $ & $0.010 $ & $0.028 $ & $0.010 $ & $0.008 $ \\
Lowercase & $0.052 $ & $0.035 $ & $0.015 $ & $0.026 $ & $0.014 $ & $0.007 $ & $0.026 $ & $0.017 $ & $0.014 $ & $0.070 $ & $0.038 $ & $0.016 $ & $0.019 $ & $0.015 $ & $0.012 $ \\
Min-K\% & $0.066 $ & $0.030 $ & $0.013 $ & $0.042 $ & $0.029 $ & $0.016 $ & $0.020 $ & $0.010 $ & $0.005 $ & $0.059 $ & $0.035 $ & $0.022 $ & $0.066 $ & $0.035 $ & $0.023 $ \\
Min-K\%++ & $0.014 $ & $0.008 $ & $0.003 $ & $0.037 $ & $0.029 $ & $0.012 $ & $0.022 $ & $0.015 $ & $0.011 $ & $0.034 $ & $0.022 $ & $0.014 $ & $0.032 $ & $0.013 $ & $0.014 $ \\
ReCall & $0.033 $ & $0.022 $ & $0.006 $ & $0.031 $ & $0.020 $ & $0.011 $ & $0.016 $ & $0.008 $ & $0.004 $ & $0.058 $ & $0.026 $ & $0.011 $ & $0.032 $ & $0.009 $ & $0.007 $ \\
\bottomrule
\end{tabular}}
\label{tab:results-lora-ensemble-TPR@FPR=0.01-datasets-defenses-deepseek-False-pythia}
\end{table*}

\begin{table*}[htbp]
\centering
\small
\footnotesize
\caption{{TPR@1\% FPR performance of MIAs across datasets and defenses. Evaluated on \texttt{EleutherAI/pythia-6.9b}. (Full Finetuning, \texttt{deepseek/deepseek-v3.2})}}
\renewcommand{\arraystretch}{1.35}
\resizebox{\textwidth}{!}{
\begin{tabular}{lccccccccccccccc}
\toprule
& \multicolumn{3}{c}{\textbf{\texttt{arxiv}}} & \multicolumn{3}{c}{\textbf{\texttt{wikipedia}}} & \multicolumn{3}{c}{\textbf{\texttt{pile\_cc}}} & \multicolumn{3}{c}{\textbf{\texttt{hackernews}}} & \multicolumn{3}{c}{\textbf{\texttt{pubmed}}} \\
\cmidrule(lr){2-4}
\cmidrule(lr){5-7}
\cmidrule(lr){8-10}
\cmidrule(lr){11-13}
\cmidrule(lr){14-16}
& {{FT}} & {{SAGE}} & {{SAGE-R}} & {{FT}} & {{SAGE}} & {{SAGE-R}} & {{FT}} & {{SAGE}} & {{SAGE-R}} & {{FT}} & {{SAGE}} & {{SAGE-R}} & {{FT}} & {{SAGE}} & {{SAGE-R}} \\
\midrule
Loss & $1.000 $ & $0.348 $ & $0.023 $ & $0.995 $ & $0.144 $ & $0.016 $ & $0.995 $ & $0.099 $ & $0.008 $ & $1.000 $ & $0.373 $ & $0.084 $ & $1.000 $ & $0.157 $ & $0.047 $ \\
Zlib & $1.000 $ & $0.368 $ & $0.024 $ & $0.996 $ & $0.177 $ & $0.013 $ & $0.998 $ & $0.166 $ & $0.016 $ & $1.000 $ & $0.475 $ & $0.102 $ & $1.000 $ & $0.142 $ & $0.033 $ \\
Lowercase & $0.991 $ & $0.372 $ & $0.024 $ & $0.956 $ & $0.115 $ & $0.004 $ & $0.993 $ & $0.169 $ & $0.017 $ & $1.000 $ & $0.338 $ & $0.060 $ & $0.993 $ & $0.110 $ & $0.032 $ \\
Min-K\% & $0.616 $ & $0.292 $ & $0.060 $ & $0.628 $ & $0.133 $ & $0.028 $ & $0.663 $ & $0.078 $ & $0.007 $ & $0.588 $ & $0.259 $ & $0.054 $ & $0.632 $ & $0.160 $ & $0.081 $ \\
Min-K\%++ & $0.998 $ & $0.081 $ & $0.006 $ & $0.996 $ & $0.115 $ & $0.020 $ & $0.995 $ & $0.051 $ & $0.007 $ & $0.999 $ & $0.106 $ & $0.027 $ & $1.000 $ & $0.047 $ & $0.022 $ \\
ReCall & $1.000 $ & $0.417 $ & $0.029 $ & $0.996 $ & $0.160 $ & $0.015 $ & $0.994 $ & $0.084 $ & $0.006 $ & $1.000 $ & $0.427 $ & $0.089 $ & $1.000 $ & $0.190 $ & $0.052 $ \\
\bottomrule
\end{tabular}}
\label{tab:results-nolora-ensemble-TPR@FPR=0.01-datasets-defenses-deepseek-False-pythia}
\end{table*}

\begin{table*}[htbp]
\centering
\small
\footnotesize
\caption{{AUC performance of MIAs across datasets and defenses. Evaluated on \texttt{EleutherAI/pythia-6.9b}. (LoRA, \texttt{x-ai/grok-4.1-fast})}}
\renewcommand{\arraystretch}{1.35}
\resizebox{\textwidth}{!}{
\begin{tabular}{lccccccccccccccc}
\toprule
& \multicolumn{3}{c}{\textbf{\texttt{arxiv}}} & \multicolumn{3}{c}{\textbf{\texttt{wikipedia}}} & \multicolumn{3}{c}{\textbf{\texttt{pile\_cc}}} & \multicolumn{3}{c}{\textbf{\texttt{hackernews}}} & \multicolumn{3}{c}{\textbf{\texttt{pubmed}}} \\
\cmidrule(lr){2-4}
\cmidrule(lr){5-7}
\cmidrule(lr){8-10}
\cmidrule(lr){11-13}
\cmidrule(lr){14-16}
& {{FT}} & {{SAGE}} & {{SAGE-R}} & {{FT}} & {{SAGE}} & {{SAGE-R}} & {{FT}} & {{SAGE}} & {{SAGE-R}} & {{FT}} & {{SAGE}} & {{SAGE-R}} & {{FT}} & {{SAGE}} & {{SAGE-R}} \\
\midrule
Loss & $0.673 $ & $0.613 $ & $0.569 $ & $0.671 $ & $0.608 $ & $0.548 $ & $0.645 $ & $0.563 $ & $0.526 $ & $0.731 $ & $0.620 $ & $0.561 $ & $0.657 $ & $0.568 $ & $0.545 $ \\
Zlib & $0.670 $ & $0.613 $ & $0.571 $ & $0.675 $ & $0.610 $ & $0.548 $ & $0.664 $ & $0.575 $ & $0.537 $ & $0.740 $ & $0.622 $ & $0.563 $ & $0.657 $ & $0.569 $ & $0.547 $ \\
Lowercase & $0.655 $ & $0.611 $ & $0.533 $ & $0.687 $ & $0.618 $ & $0.547 $ & $0.668 $ & $0.584 $ & $0.512 $ & $0.691 $ & $0.601 $ & $0.527 $ & $0.654 $ & $0.575 $ & $0.535 $ \\
Min-K\% & $0.579 $ & $0.553 $ & $0.529 $ & $0.573 $ & $0.549 $ & $0.523 $ & $0.565 $ & $0.535 $ & $0.517 $ & $0.553 $ & $0.520 $ & $0.502 $ & $0.566 $ & $0.530 $ & $0.519 $ \\
Min-K\%++ & $0.626 $ & $0.572 $ & $0.545 $ & $0.684 $ & $0.613 $ & $0.548 $ & $0.631 $ & $0.556 $ & $0.524 $ & $0.636 $ & $0.565 $ & $0.537 $ & $0.637 $ & $0.549 $ & $0.535 $ \\
ReCall & $0.689 $ & $0.623 $ & $0.574 $ & $0.682 $ & $0.617 $ & $0.553 $ & $0.661 $ & $0.575 $ & $0.535 $ & $0.749 $ & $0.629 $ & $0.563 $ & $0.678 $ & $0.581 $ & $0.556 $ \\
\bottomrule
\end{tabular}}
\label{tab:results-lora-ensemble-AUC-datasets-defenses-grok-False-pythia}
\end{table*}

\begin{table*}[htbp]
\centering
\small
\footnotesize
\caption{{AUC performance of MIAs across datasets and defenses. Evaluated on \texttt{EleutherAI/pythia-6.9b}. (Full Finetuning, \texttt{x-ai/grok-4.1-fast})}}
\renewcommand{\arraystretch}{1.35}
\resizebox{\textwidth}{!}{
\begin{tabular}{lccccccccccccccc}
\toprule
& \multicolumn{3}{c}{\textbf{\texttt{arxiv}}} & \multicolumn{3}{c}{\textbf{\texttt{wikipedia}}} & \multicolumn{3}{c}{\textbf{\texttt{pile\_cc}}} & \multicolumn{3}{c}{\textbf{\texttt{hackernews}}} & \multicolumn{3}{c}{\textbf{\texttt{pubmed}}} \\
\cmidrule(lr){2-4}
\cmidrule(lr){5-7}
\cmidrule(lr){8-10}
\cmidrule(lr){11-13}
\cmidrule(lr){14-16}
& {{FT}} & {{SAGE}} & {{SAGE-R}} & {{FT}} & {{SAGE}} & {{SAGE-R}} & {{FT}} & {{SAGE}} & {{SAGE-R}} & {{FT}} & {{SAGE}} & {{SAGE-R}} & {{FT}} & {{SAGE}} & {{SAGE-R}} \\
\midrule
Loss & $1.000 $ & $0.908 $ & $0.752 $ & $1.000 $ & $0.854 $ & $0.639 $ & $1.000 $ & $0.839 $ & $0.676 $ & $1.000 $ & $0.927 $ & $0.757 $ & $1.000 $ & $0.826 $ & $0.736 $ \\
Zlib & $1.000 $ & $0.898 $ & $0.745 $ & $1.000 $ & $0.859 $ & $0.642 $ & $1.000 $ & $0.857 $ & $0.701 $ & $1.000 $ & $0.933 $ & $0.776 $ & $1.000 $ & $0.822 $ & $0.735 $ \\
Lowercase & $0.999 $ & $0.857 $ & $0.630 $ & $0.998 $ & $0.823 $ & $0.598 $ & $1.000 $ & $0.811 $ & $0.566 $ & $1.000 $ & $0.876 $ & $0.620 $ & $0.999 $ & $0.724 $ & $0.625 $ \\
Min-K\% & $0.697 $ & $0.648 $ & $0.582 $ & $0.709 $ & $0.633 $ & $0.545 $ & $0.731 $ & $0.624 $ & $0.552 $ & $0.675 $ & $0.615 $ & $0.552 $ & $0.703 $ & $0.610 $ & $0.578 $ \\
Min-K\%++ & $1.000 $ & $0.746 $ & $0.632 $ & $1.000 $ & $0.791 $ & $0.593 $ & $1.000 $ & $0.709 $ & $0.581 $ & $1.000 $ & $0.701 $ & $0.602 $ & $1.000 $ & $0.679 $ & $0.624 $ \\
ReCall & $1.000 $ & $0.922 $ & $0.764 $ & $1.000 $ & $0.862 $ & $0.646 $ & $1.000 $ & $0.847 $ & $0.686 $ & $1.000 $ & $0.940 $ & $0.770 $ & $1.000 $ & $0.847 $ & $0.753 $ \\
\bottomrule
\end{tabular}}
\label{tab:results-nolora-ensemble-AUC-datasets-defenses-grok-False-pythia}
\end{table*}

\begin{table*}[htbp]
\centering
\small
\footnotesize
\caption{{TPR@1\% FPR performance of MIAs across datasets and defenses. Evaluated on \texttt{EleutherAI/pythia-6.9b}. (LoRA, \texttt{x-ai/grok-4.1-fast})}}
\renewcommand{\arraystretch}{1.35}
\resizebox{\textwidth}{!}{
\begin{tabular}{lccccccccccccccc}
\toprule
& \multicolumn{3}{c}{\textbf{\texttt{arxiv}}} & \multicolumn{3}{c}{\textbf{\texttt{wikipedia}}} & \multicolumn{3}{c}{\textbf{\texttt{pile\_cc}}} & \multicolumn{3}{c}{\textbf{\texttt{hackernews}}} & \multicolumn{3}{c}{\textbf{\texttt{pubmed}}} \\
\cmidrule(lr){2-4}
\cmidrule(lr){5-7}
\cmidrule(lr){8-10}
\cmidrule(lr){11-13}
\cmidrule(lr){14-16}
& {{FT}} & {{SAGE}} & {{SAGE-R}} & {{FT}} & {{SAGE}} & {{SAGE-R}} & {{FT}} & {{SAGE}} & {{SAGE-R}} & {{FT}} & {{SAGE}} & {{SAGE-R}} & {{FT}} & {{SAGE}} & {{SAGE-R}} \\
\midrule
Loss & $0.032 $ & $0.021 $ & $0.007 $ & $0.030 $ & $0.021 $ & $0.012 $ & $0.020 $ & $0.011 $ & $0.005 $ & $0.039 $ & $0.020 $ & $0.008 $ & $0.033 $ & $0.019 $ & $0.013 $ \\
Zlib & $0.029 $ & $0.018 $ & $0.008 $ & $0.034 $ & $0.025 $ & $0.013 $ & $0.028 $ & $0.017 $ & $0.010 $ & $0.063 $ & $0.026 $ & $0.007 $ & $0.028 $ & $0.012 $ & $0.009 $ \\
Lowercase & $0.052 $ & $0.037 $ & $0.017 $ & $0.026 $ & $0.014 $ & $0.007 $ & $0.026 $ & $0.019 $ & $0.012 $ & $0.070 $ & $0.043 $ & $0.012 $ & $0.019 $ & $0.017 $ & $0.012 $ \\
Min-K\% & $0.066 $ & $0.026 $ & $0.012 $ & $0.042 $ & $0.029 $ & $0.016 $ & $0.020 $ & $0.009 $ & $0.004 $ & $0.059 $ & $0.051 $ & $0.020 $ & $0.066 $ & $0.025 $ & $0.026 $ \\
Min-K\%++ & $0.014 $ & $0.007 $ & $0.003 $ & $0.037 $ & $0.025 $ & $0.011 $ & $0.022 $ & $0.017 $ & $0.009 $ & $0.034 $ & $0.023 $ & $0.011 $ & $0.032 $ & $0.019 $ & $0.014 $ \\
ReCall & $0.033 $ & $0.017 $ & $0.006 $ & $0.031 $ & $0.019 $ & $0.012 $ & $0.016 $ & $0.008 $ & $0.004 $ & $0.058 $ & $0.024 $ & $0.012 $ & $0.032 $ & $0.012 $ & $0.008 $ \\
\bottomrule
\end{tabular}}
\label{tab:results-lora-ensemble-TPR@FPR=0.01-datasets-defenses-grok-False-pythia}
\end{table*}

\begin{table*}[htbp]
\centering
\small
\footnotesize
\caption{{TPR@1\% FPR performance of MIAs across datasets and defenses. Evaluated on \texttt{EleutherAI/pythia-6.9b}. (Full Finetuning, \texttt{x-ai/grok-4.1-fast})}}
\renewcommand{\arraystretch}{1.35}
\resizebox{\textwidth}{!}{
\begin{tabular}{lccccccccccccccc}
\toprule
& \multicolumn{3}{c}{\textbf{\texttt{arxiv}}} & \multicolumn{3}{c}{\textbf{\texttt{wikipedia}}} & \multicolumn{3}{c}{\textbf{\texttt{pile\_cc}}} & \multicolumn{3}{c}{\textbf{\texttt{hackernews}}} & \multicolumn{3}{c}{\textbf{\texttt{pubmed}}} \\
\cmidrule(lr){2-4}
\cmidrule(lr){5-7}
\cmidrule(lr){8-10}
\cmidrule(lr){11-13}
\cmidrule(lr){14-16}
& {{FT}} & {{SAGE}} & {{SAGE-R}} & {{FT}} & {{SAGE}} & {{SAGE-R}} & {{FT}} & {{SAGE}} & {{SAGE-R}} & {{FT}} & {{SAGE}} & {{SAGE-R}} & {{FT}} & {{SAGE}} & {{SAGE-R}} \\
\midrule
Loss & $1.000 $ & $0.379 $ & $0.041 $ & $0.995 $ & $0.151 $ & $0.015 $ & $0.995 $ & $0.076 $ & $0.010 $ & $1.000 $ & $0.376 $ & $0.092 $ & $1.000 $ & $0.192 $ & $0.062 $ \\
Zlib & $1.000 $ & $0.379 $ & $0.038 $ & $0.996 $ & $0.173 $ & $0.013 $ & $0.998 $ & $0.177 $ & $0.031 $ & $1.000 $ & $0.506 $ & $0.137 $ & $1.000 $ & $0.190 $ & $0.058 $ \\
Lowercase & $0.991 $ & $0.368 $ & $0.027 $ & $0.956 $ & $0.129 $ & $0.005 $ & $0.993 $ & $0.173 $ & $0.017 $ & $1.000 $ & $0.405 $ & $0.061 $ & $0.993 $ & $0.108 $ & $0.049 $ \\
Min-K\% & $0.616 $ & $0.340 $ & $0.085 $ & $0.628 $ & $0.132 $ & $0.020 $ & $0.663 $ & $0.072 $ & $0.008 $ & $0.588 $ & $0.272 $ & $0.069 $ & $0.632 $ & $0.164 $ & $0.108 $ \\
Min-K\%++ & $0.998 $ & $0.076 $ & $0.010 $ & $0.996 $ & $0.115 $ & $0.021 $ & $0.995 $ & $0.041 $ & $0.008 $ & $0.999 $ & $0.079 $ & $0.030 $ & $1.000 $ & $0.051 $ & $0.023 $ \\
ReCall & $1.000 $ & $0.428 $ & $0.046 $ & $0.996 $ & $0.166 $ & $0.013 $ & $0.994 $ & $0.064 $ & $0.008 $ & $1.000 $ & $0.476 $ & $0.122 $ & $1.000 $ & $0.206 $ & $0.070 $ \\
\bottomrule
\end{tabular}}
\label{tab:results-nolora-ensemble-TPR@FPR=0.01-datasets-defenses-grok-False-pythia}
\end{table*}

\begin{table*}[htbp]
\centering
\small
\footnotesize
\caption{{AUC performance of MIAs across datasets and defenses. Evaluated on \texttt{EleutherAI/pythia-6.9b}. (LoRA, \texttt{google/gemini-2.5-flash})}}
\renewcommand{\arraystretch}{1.35}
\resizebox{\textwidth}{!}{
\begin{tabular}{lccccccccccccccc}
\toprule
& \multicolumn{3}{c}{\textbf{\texttt{arxiv}}} & \multicolumn{3}{c}{\textbf{\texttt{wikipedia}}} & \multicolumn{3}{c}{\textbf{\texttt{pile\_cc}}} & \multicolumn{3}{c}{\textbf{\texttt{hackernews}}} & \multicolumn{3}{c}{\textbf{\texttt{pubmed}}} \\
\cmidrule(lr){2-4}
\cmidrule(lr){5-7}
\cmidrule(lr){8-10}
\cmidrule(lr){11-13}
\cmidrule(lr){14-16}
& {{FT}} & {{SAGE}} & {{SAGE-R}} & {{FT}} & {{SAGE}} & {{SAGE-R}} & {{FT}} & {{SAGE}} & {{SAGE-R}} & {{FT}} & {{SAGE}} & {{SAGE-R}} & {{FT}} & {{SAGE}} & {{SAGE-R}} \\
\midrule
Loss & $0.673 $ & $0.620 $ & $0.578 $ & $0.671 $ & $0.612 $ & $0.558 $ & $0.645 $ & $0.561 $ & $0.531 $ & $0.731 $ & $0.609 $ & $0.552 $ & $0.657 $ & $0.576 $ & $0.554 $ \\
Zlib & $0.670 $ & $0.620 $ & $0.580 $ & $0.675 $ & $0.614 $ & $0.558 $ & $0.664 $ & $0.574 $ & $0.543 $ & $0.740 $ & $0.610 $ & $0.551 $ & $0.657 $ & $0.578 $ & $0.556 $ \\
Lowercase & $0.655 $ & $0.618 $ & $0.542 $ & $0.687 $ & $0.625 $ & $0.554 $ & $0.668 $ & $0.577 $ & $0.517 $ & $0.691 $ & $0.592 $ & $0.526 $ & $0.654 $ & $0.587 $ & $0.550 $ \\
Min-K\% & $0.579 $ & $0.555 $ & $0.531 $ & $0.573 $ & $0.553 $ & $0.528 $ & $0.565 $ & $0.533 $ & $0.518 $ & $0.553 $ & $0.516 $ & $0.501 $ & $0.566 $ & $0.530 $ & $0.522 $ \\
Min-K\%++ & $0.626 $ & $0.571 $ & $0.547 $ & $0.684 $ & $0.612 $ & $0.558 $ & $0.631 $ & $0.551 $ & $0.523 $ & $0.636 $ & $0.558 $ & $0.533 $ & $0.637 $ & $0.551 $ & $0.539 $ \\
ReCall & $0.689 $ & $0.629 $ & $0.583 $ & $0.682 $ & $0.621 $ & $0.562 $ & $0.661 $ & $0.574 $ & $0.540 $ & $0.749 $ & $0.617 $ & $0.551 $ & $0.678 $ & $0.590 $ & $0.566 $ \\
\bottomrule
\end{tabular}}
\label{tab:results-lora-ensemble-AUC-datasets-defenses-gemini-False-pythia}
\end{table*}

\begin{table*}[htbp]
\centering
\small
\footnotesize
\caption{{AUC performance of MIAs across datasets and defenses. Evaluated on \texttt{EleutherAI/pythia-6.9b}. (Full Finetuning, \texttt{google/gemini-2.5-flash})}}
\renewcommand{\arraystretch}{1.35}
\resizebox{\textwidth}{!}{
\begin{tabular}{lccccccccccccccc}
\toprule
& \multicolumn{3}{c}{\textbf{\texttt{arxiv}}} & \multicolumn{3}{c}{\textbf{\texttt{wikipedia}}} & \multicolumn{3}{c}{\textbf{\texttt{pile\_cc}}} & \multicolumn{3}{c}{\textbf{\texttt{hackernews}}} & \multicolumn{3}{c}{\textbf{\texttt{pubmed}}} \\
\cmidrule(lr){2-4}
\cmidrule(lr){5-7}
\cmidrule(lr){8-10}
\cmidrule(lr){11-13}
\cmidrule(lr){14-16}
& {{FT}} & {{SAGE}} & {{SAGE-R}} & {{FT}} & {{SAGE}} & {{SAGE-R}} & {{FT}} & {{SAGE}} & {{SAGE-R}} & {{FT}} & {{SAGE}} & {{SAGE-R}} & {{FT}} & {{SAGE}} & {{SAGE-R}} \\
\midrule
Loss & $1.000 $ & $0.919 $ & $0.797 $ & $1.000 $ & $0.860 $ & $0.684 $ & $1.000 $ & $0.818 $ & $0.683 $ & $1.000 $ & $0.901 $ & $0.715 $ & $1.000 $ & $0.856 $ & $0.791 $ \\
Zlib & $1.000 $ & $0.909 $ & $0.787 $ & $1.000 $ & $0.865 $ & $0.685 $ & $1.000 $ & $0.837 $ & $0.705 $ & $1.000 $ & $0.905 $ & $0.732 $ & $1.000 $ & $0.852 $ & $0.790 $ \\
Lowercase & $0.999 $ & $0.869 $ & $0.666 $ & $0.998 $ & $0.833 $ & $0.627 $ & $1.000 $ & $0.779 $ & $0.580 $ & $1.000 $ & $0.837 $ & $0.606 $ & $0.999 $ & $0.748 $ & $0.674 $ \\
Min-K\% & $0.697 $ & $0.646 $ & $0.596 $ & $0.709 $ & $0.632 $ & $0.556 $ & $0.731 $ & $0.613 $ & $0.558 $ & $0.675 $ & $0.605 $ & $0.545 $ & $0.703 $ & $0.612 $ & $0.589 $ \\
Min-K\%++ & $1.000 $ & $0.733 $ & $0.642 $ & $1.000 $ & $0.782 $ & $0.616 $ & $1.000 $ & $0.674 $ & $0.581 $ & $1.000 $ & $0.671 $ & $0.583 $ & $1.000 $ & $0.682 $ & $0.642 $ \\
ReCall & $1.000 $ & $0.930 $ & $0.809 $ & $1.000 $ & $0.868 $ & $0.689 $ & $1.000 $ & $0.826 $ & $0.693 $ & $1.000 $ & $0.917 $ & $0.729 $ & $1.000 $ & $0.876 $ & $0.808 $ \\
\bottomrule
\end{tabular}}
\label{tab:results-nolora-ensemble-AUC-datasets-defenses-gemini-False-pythia}
\end{table*}

\begin{table*}[htbp]
\centering
\small
\footnotesize
\caption{{TPR@1\% FPR performance of MIAs across datasets and defenses. Evaluated on \texttt{EleutherAI/pythia-6.9b}. (LoRA, \texttt{google/gemini-2.5-flash})}}
\renewcommand{\arraystretch}{1.35}
\resizebox{\textwidth}{!}{
%
}
\label{tab:results-lora-ensemble-TPR@FPR=0.01-datasets-defenses-gemini-False-pythia}
\end{table*}

\begin{table*}[htbp]
\centering
\small
\footnotesize
\caption{{TPR@1\% FPR performance of MIAs across datasets and defenses. Evaluated on \texttt{EleutherAI/pythia-6.9b}. (Full Finetuning, \texttt{google/gemini-2.5-flash})}}
\renewcommand{\arraystretch}{1.35}
\resizebox{\textwidth}{!}{
%
}
\label{tab:results-nolora-ensemble-TPR@FPR=0.01-datasets-defenses-gemini-False-pythia}
\end{table*}

\end{document}